%% file: PowellSnakes_I_rev_v2.tex
\documentclass[useAMS,usenatbib]{mn2e}
\usepackage{color}
\input{rgb}
\usepackage{textcomp}
\usepackage{times}
\usepackage{graphicx}
\usepackage{subfigure}
\usepackage{psfrag}
\usepackage{amsmath}
\usepackage{amssymb}
\usepackage{stmaryrd}

\def\reff@jnl#1{{\rm#1\/}}
\def\apj{\reff@jnl{ApJ}}       
\def\apjs{\reff@jnl{ApJS}}     
\def\aaps{\reff@jnl{A\&AS}}    
\def\mnras{\reff@jnl{MNRAS}}   
\def\prd{\reff@jnl{Phys.\ Rev.\ D}}    

\newcommand{\va}{\mathbf{a}}

\newcommand{\vM}{\mathbf{M}}
\newcommand{\vN}{\mathbf{N}}

\newcommand{\vX}{\mathbf{X}}

\newcommand{\vd}{\mathbf{d}}

\newcommand{\vs}{\mathbf{s}}
\newcommand{\vC}{\mathbf{C}}

\newcommand{\vt}{\mathbf{t}}

\newcommand{\vx}{\mathbf{x}}
\newcommand{\vphi}{\mathbf{\phi}}

\newcommand{\beq}{\begin{equation}}
\newcommand{\eeq}{\end{equation}}

\newcommand{\be}{\begin{equation}}
\newcommand{\ee}{\end{equation}}

{\begin{enumerate}\setlength{\itemsep}{0mm}}%
{\end{enumerate}}
{\begin{enumerate}\setlength{\itemsep}{0mm}}%
{\end{enumerate}}

\def\spd#1#2#3{{\upartial ^2 #1 \over \upartial #2 \upartial #3}}

\def\tfrac#1#2{{\textstyle\frac{#1}{#2}}}
\def\vect#1{{\mathbfit{#1}}}

\title[Fast Bayesian object detection]{A fast Bayesian approach to discrete object detection in astronomical datasets - PowellSnakes I}
\author[P.~Carvalho, G.~Rocha and M.P.~Hobson]
{Pedro Carvalho,$^1$\thanks{email: carvalho@mrao.cam.ac.uk}
 Gra\c{c}a Rocha,$^{2,3}$\thanks{email: graca@caltech.edu}
 M.P.~Hobson$^1$\thanks{email: mph@mrao.cam.ac.uk}
\\
  $^1$Astrophysics Group, Cavendish Laboratory,
  J.J.~Thomson Avenue, Cambridge CB3 0HE, UK\\
  $^2$ IPAC, Keith Spalding building, 11200 East California Blvd, Pasadena, California 91125\\
  $^3$  JPL, 4800 Oak Grove Drive, Pasadena, California 91109}

\date{Accepted - Version7 - V---. Received ---; in original form 25 July 2008}
\pagerange{\pageref{firstpage}--\pageref{lastpage}}
\pubyear{2007}

\onecolumn

\begin{document}

\label{firstpage}
\maketitle

\begin{abstract}
A new fast Bayesian approach is introduced for the detection of
discrete objects immersed in a diffuse background.  This new method,
called PowellSnakes, speeds up traditional Bayesian techniques by:
i) replacing the standard form of the likelihood for the parameters
characterizing the discrete objects by an alternative exact form
that is much quicker to evaluate; ii) using a simultaneous multiple
minimization code based on Powell's direction set algorithm to
locate rapidly the local maxima in the posterior; and iii) deciding
whether each located posterior peak corresponds to a real object by
performing a Bayesian model selection using an approximate evidence
value based on a local Gaussian approximation to the peak. The
construction of this Gaussian approximation also provides the
covariance matrix of the uncertainties in the derived parameter
values for the object in question. This new approach provides a
speed up in performance by a factor of `hundreds' as compared to
existing Bayesian source extraction methods that use MCMC to explore
the parameter space, such as that presented by Hobson \& McLachlan
\cite{mikecharlie}. The method can be implemented in either real or
Fourier space. In the case of objects embedded in a homogeneous
random field, working in Fourier space provides a further speed up
that takes advantage of the fact that the correlation matrix of the
background is circulant. We illustrate the capabilities of the
method by applying to some simplified toy models. Furthermore
PowellSnakes has the advantage of consistently defining  the
threshold for acceptance/rejection based on priors which cannot be
said of the frequentist methods.
We present here the first implementation of this technique (Version-I).
Further improvements to  this implementation are currently under investigation and will be published shortly.
The application of the method to realistic simulated Planck
observations will be presented in a forthcoming publication.

\end{abstract}
\begin{keywords}
Cosmology: observations -- methods: data analysis -- cosmic microwave
background
\end{keywords}

\section{Introduction}

The detection and characterisation of discrete objects is a generic
problem in many areas of astrophysics and cosmology.  Indeed, one of
the major challenges in
such analyses is to separate a localised signal from a diffuse
background. It is most common to face this problem in the analysis
of two dimensional images, where one wishes to detect discrete
objects in a diffuse background.

When performing this task it is often assumed that the background is
smoothly varying and has a characteristic length-scale much larger
than the scale of the discrete objects being sought.  For example,
SExtractor (Bertin \& Arnouts \cite{SExtractor}) approximates the
background emission by a low-order polynomial, which is subtracted
from the image. Object detection is then performed by finding sets
of connected pixels above some given threshold. Such methods run
into problems, however, when the diffuse background varies on length
scales and with amplitudes similar to those of the discrete objects
of interest. Moreover, difficulties also arise when the rms level of
instrumental noise is comparable to, or somewhat larger than, the
amplitude of the localised signal one is seeking.

A specific example illustrating the above difficulties is provided
by high-resolution observations of the cosmic microwave background
(CMB). In addition to the CMB emission, which varies on a
characteristic scale of order $\sim 10$ arcmin, one is often
interested in detecting emission from discrete objects such as
extragalactic `point' (i.e. beam-shaped) sources or the
Sunyaev-Zel'dovich (SZ) effect in galaxy clusters, which have
characteristic scales similar to that of the primordial CMB
emission. Moreover, the rms of the instrumental noise in CMB
observations can be greater than the amplitude of the discrete
sources. In such cases, it is not surprising that straightforward
methods, such as that outlined above, often fail to detect the
localised objects.

The standard approach for dealing with such difficulties is to apply a
linear filter $\psi(\vect{x})$ to the original image $d(\vect{x})$ and
instead analyse the resulting filtered field
\begin{equation}
d_f(\vect{x}) = \int \psi(\vect{x}-\vect{y}) d(\vect{y})\,d^2\vect{y}.
\label{dfilt}
\end{equation}
Suppose one is interested in detecting objects with some given spatial
template $t(\vect{x})$ (normalised for convenience to unit peak
amplitude). If the original image contains $N_{\rm obj}$ objects at
positions $\vect{X}_i$ with amplitudes $A_i$, together with
contributions from other astrophysical components and instrumental
noise, then
\begin{equation}
d(\vect{x}) \equiv s(\vect{x})+n(\vect{x}) = \sum_{k=1}^{N_{\rm
obj}} A_i t(\vect{x}-\vect{X}_k) + n(\vect{x}),
\end{equation}
where $s(\vect{x})$ is the signal of interest and $n(\vect{x})$ is
the generalised background `noise', defined as all contributions to
the image aside from the discrete objects.
It is straightforward to design an optimal filter function
$\psi(\vect{x})$ such that the filtered field (\ref{dfilt}) has the
following properties: (i) $d_f(\vect{X}_k)$ is an unbiassed
estimator of $A_i$; (ii) the variance of the filtered noise field
$n_f(\vect{x})$ is minimised;
The corresponding function $\psi(\vect{x})$
is the standard matched filter (see, for example, Haehnelt \&
Tegmark \cite{TegmarkMatchedFilter})
One may consider the filtering process as `optimally boosting' (in a
linear sense) the signal from discrete objects, with a given spatial
template, and simultaneously suppressing emission from the
background.

An additional subtlety in most practical applications is that the
set of objects are not all identical, but one can repeat the
filtering process with filter functions optimised for different
spatial templates to obtain several filtered fields, each of which
will optimally boost objects with that template. In the case of the
SZ effect, for example, one might assume that the functional form of
the template is the same for all clusters, but that the `core
radius' differs from one cluster to another. The functional form of
the filter function is then the same in each case, and one can
repeat the filtering for a number of different scales (Herranz et
al. \cite{HerranzSZ_MatchedFilter}). Moreover, one can trivially
extend the linear filtering approach to the simultaneous analysis of
a set of images. Once again, the SZ effect provides a good example.
Owing to the distinctive frequency dependence of the thermal SZ
effect, it is better to use the maps at all the observed frequencies
simultaneously when attempting to detect and characterise thermal SZ
clusters hidden in the emission from other astrophysical components
(Herranz et al. \cite{HerranzSZ_Multifreq}).

The approaches outlined above have been shown to produce good
results, but the filtering process is only optimal among the rather
limited class of linear filters and is logically separated from the
subsequent object detection step performed on the filtered map(s).
As a result, Hobson \& McLachlan (\cite{mikecharlie}; hereinafter
HM03) introduced a Bayesian approach to the detection and
characterisation of discrete objects in a diffuse background. As in
the filtering techniques, the method assumed a parameterised form
for the objects of interest, but the optimal values of these
parameters, and their associated errors, were obtained in a single
step by evaluating their full posterior distribution. If available,
one could also place physical priors on the parameters defining an
object and on the number of objects present. Although this approach
represents the theoretically-optimal method for performing
parametrised object detection, its implementation was performed
using Monte-Carlo Markov chain (MCMC) sampling from the posterior
which was extremely computationally intensive. Although considerable
progress has recently been made in increasing the efficiency of
sampling-based Bayesian object detection methods (Feroz \& Hobson
\cite{HobsonMCMC2007}), such approaches are still substantially
slower than simple linear filtering methods. Therefore, in this
paper, we explore a new, fast method for performing Bayesian object
detection in which sampling is replaced by multiple local
maximisation of the posterior, and the evaluation of errors and
Bayesian evidence values is performed by making a Gaussian
approximation to the posterior at each peak. This approach yields a
speed-up over sampling-based methods of many 100s, making the
computational complexity of the approach comparable to that of
linear filtering methods.

The outline of the paper is as follows. In section
\ref{bayesian_infer} we provide a brief outline of the essential
Bayesian logical framework which supports the employed methodology.
The previously defined entities (see section \ref{bayesian_infer})
are then characterized in terms of the source detection problem in
section \ref{sect:BayObjDetect}. A summary of the strategy
to fulfil the desire goals is then suggested in section
\ref{sect:ObjDetStrat}. In section \ref{maxpost} we develop the
first step of the proposed algorithm: The maximisation of the
posterior distribution and we give account of the errors bars on the
estimated parameters of the putative source under study. A two step
approach split between Fourier and real space is indicated. The
second phase of the detection algorithm: The Bayesian validation
procedure which we apply on the previously found likelihood maximum
is explained in section \ref{validation}.
A prior is suggested and a upper bound on the quality of the
detection is advanced. We end this section by considering a method
for correcting possible systematic deviations on the evidence
evaluation. In section \ref{sect:IntroColour} we give a detailed
study on the effects that develop when adding correlation into the
diffuse background that produces the substructure in which the
sources are imbedded. The results from a series of simulations
covering several typical scenarios are presented in section
\ref{results} and we close indicating our conclusions and possible
directions for further work in section \ref{sect:conclusions}.
%
%

\section{Bayesian Inference}
\label{bayesian_infer}

Bayesian inference methods provide a consistent approach to the
estimation of a set parameters $\mathbf{\Theta}$ in a model (or
hypothesis) $H$ for the data $\vect{d}$. Bayes' theorem states that
\begin{equation} \Pr(\mathbf{\Theta}|\vect{d}, H) =
\frac{\Pr(\vect{d}|\,\mathbf{\Theta},H)\Pr(\mathbf{\Theta}|H)}
{\Pr(\vect{d}|H)},
\end{equation}
where $\Pr(\mathbf{\Theta}|\vect{d}, H) \equiv P(\mathbf{\Theta})$
is the posterior probability
distribution of the parameters, $\Pr(\vect{d}|\mathbf{\Theta}, H)
\equiv L(\mathbf{\Theta})$ is the likelihood,
$\Pr(\mathbf{\Theta}|H) \equiv \pi(\mathbf{\Theta})$ is the prior, and
$\Pr(\vect{d}|H) \equiv E$ is the Bayesian evidence.

In parameter estimation, the normalising evidence factor is usually
ignored, since it is independent of the parameters
$\mathbf{\Theta}$. This (unnormalised) posterior constitutes the
complete Bayesian inference of the parameter values.  Inferences are
usually obtained either by taking samples from the (unnormalised)
posterior using MCMC methods, or by locating its maximum (or maxima)
and approximating the shape around the peak(s) by a multivariate
Gaussian.

In contrast to parameter estimation problems, in model selection the
evidence takes the central role and is simply the factor required to
normalize the posterior over $\mathbf{\Theta}$:
\begin{equation}
\mathcal{Z} =
\int{L(\mathbf{\Theta})\pi(\mathbf{\Theta})}d^D\mathbf{\Theta},
\label{eq:3}
\end{equation}
where $D$ is the dimensionality of the parameter space.  As the
average of the likelihood over the prior, the evidence is larger for a
model if more of its parameter space is likely and smaller for a model
with large areas in its parameter space having low likelihood values,
even if the likelihood function is very highly peaked. Thus, the
evidence automatically implements Occam's razor: a simpler
theory with compact parameter space will have a larger evidence than a
more complicated one, unless the latter is significantly better at
explaining the data.
The question of model selection between two
models $H_{0}$ and $H_{1}$ can then be decided by comparing their
respective posterior probabilities given the observed
data set $\vect{d}$, as follows
\begin{equation}
\frac{\Pr(H_{1}|\vect{d})}{\Pr(H_{0}|\vect{d})}
=\frac{\Pr(\vect{d}|H_{1})\Pr(H_{1})}{\Pr(\vect{d}|
H_{0})\Pr(H_{0})}
=\frac{\mathcal{Z}_1}{\mathcal{Z}_0}\frac{\Pr(H_{1})}{\Pr(H_{0})},
\label{eq:3.1}
\end{equation}
where $\Pr(H_{1})/\Pr(H_{0})$ is the a priori probability ratio for
the two models, which can often be set to unity but occasionally
requires further consideration.  Unfortunately, evaluation of the
multidimensional integral \eqref{eq:3} is a challenging numerical
task. Nonetheless, a fast approximate method for evidence evaluation
is to model the posterior as a multivariate Gaussian centred at its
peak(s) and apply the Laplace formula (see e.g. Hobson, Bridle \&
Lahav \cite{HBL01}, thereinafter HBL01).


\section{Bayesian object detection}\label{sect:BayObjDetect}

A Bayesian approach to detecting and characterizing discrete objects
hidden in some background noise was first presented in an astronomical
context by HM03, and our general framework follows this closely.  For
brevity, we will consider our data vector $\vect{d}$ to denote
the pixel values in a single patch in which we wish to search for discrete
objects, although $\vect{d}$ could equally well represent the
Fourier coefficients of the image, or coefficients in some other
basis.

\subsection{Discrete objects in a background}

Let us suppose that we are interested in detecting and
characterizing some set of (two-dimensional) discrete objects, each
of which is described by a template $\tau(\vect{x};\vect{a})$. This
template is defined in terms of a set of parameters $\vect{a}$ that
might typically denote (collectively) the position $(X,Y)$ of the
object, its amplitude $A$ and some measure $R$ of its spatial
extent. A particular example is the circularly-symmetric
Gaussian-shaped object defined by
\begin{equation}
\tau(\vect{x};\vect{a})=A\exp
\left[-\frac{(x-X)^2+(y-Y)^2}{2R^2}\right],
\label{objdef}
\end{equation}
so that $\vect{a} = \{X,Y,A,R\}$.  In what follows, we will consider
only circularly-symmetric objects, but our general approach
accommodates the templates that are, for example, elongated in one
direction, at the expense of increasing the dimensionality of the
parameter space.\footnote{Such an elongation can be caused by
asymmetry of the beam, source or both. If our model assumes a
circularly symmetric source but an asymmetric beam, the elongation
will be constant across the map, but not if the sources are
intrinsically asymmetric. In the latter case, one thus has to
introduce the source position angle as a parameter, in addition to
the elongation.} In the analysis of single-frequency maps, most
authors take the template $\tau(\vect{x};\vect{a})$ to be the
(pixelised) intrinsic shape of the object convolved with the beam
profile. For multi-frequency data, however, it is more natural to
treat the intrinsic object shape and the beam profiles separately.
\footnote{In its early stages PowellSnakes treated a point source
convolved with the beam as a single parameterized template. In the
current version these are separate entities.}

If $N_{\rm obj}$ objects are present in the map and the contribution
of each object to the data is additive, we may write
\begin{equation}
\vect{d} = \vect{n}+ \sum_{k=1}^{N_{\rm obj}} \vect{s}(\vect{a}_k),
\label{eq:sum_obj}
\end{equation}
where $\vect{s}(\vect{a}_k)$ denotes the contribution to the data
from the $k$th discrete object and $\vect{n}$ denotes the generalised
`noise' contribution to the data from other `background' emission and
instrumental noise. Clearly, we wish to use the data $\vect{d}$ to
place constraints on the values of the unknown parameters $N_{\rm
obj}$ and $\vect{a}_k$ $(k=1,\ldots,N_{\rm obj})$.

\subsection{Defining the posterior distribution}

As discussed in HM03, in analysing the data the Bayesian purist would
attempt to infer simultaneously the full set of parameters
$\mathbf{\Theta} \equiv (N_{\rm obj}, \vect{a}_1,\vect{a}_2,\ldots,
\vect{a}_{N_{\rm obj}})$. The crucial complication inherent to this
approach is that the length of the parameter vector $\mathbf{\Theta}$
is variable, since it depends on the unknown value $N_{\rm obj}$.
Some sampling-based approaches are able to move between spaces of
different dimensionality, and such techniques were investigated in
HM03.

An alternative approach, also discussed by HM03, is simply to set
$N_{\rm obj}=1$. In other words, the model for the data consists of
just a single object and so the full parameter space under
consideration is $\vect{a}=\{X,Y,A,R\}$, which is fixed and
only 4-dimensional.  Although we have fixed $N_{\rm obj}=1$, it is
important to understand that this does {\em not} restrict us to
detecting just a single object in the map.  Indeed, by modelling
the data in this way, we would expect the posterior distribution to
possess numerous local maxima in the 4-dimensional parameter space,
each corresponding to the location in this space of one of the
objects present in the image. HM03 show this vastly simplified
approach is indeed reliable when the objects of interest are spatially
well-separated, and we adopt this method here.

\subsection{Likelihood}

In this case, if the background `noise' $\vect{n}$ is a
statistically homogeneous Gaussian random field with covariance
matrix $\vect{N} = \langle \vect{n}\vect{n}^{\rm T} \rangle$, then
the likelihood function takes the form
\begin{equation}
L(\vect{a})=\frac{\exp\left\{
-\frac{1}{2}\left[\vect{d}-\vect{s}(\vect{a})\right]^{\rm T}
\vect{N}^{-1}\left[\vect{d}-\vect{s}(\vect{a})\right]\right\}
}{\left(2\pi\right)^{\vect{N}_{\rm
pix}/2}\left|\vect{N}\right|^{1/2}}. \label{eq:22}
\end{equation}
Moreover, if the background is just independent pixel noise, then
$\vect{N} = \sigma^2 \vect{I}$, where $\sigma$ is the noise rms.

In the general case, the log-likelihood thus takes the form
\begin{equation}
\ln L (\vect{a}) =
c-\tfrac{1}{2}\left[\vect{d}-\vect{s}(\vect{a})\right]^{\rm T}
\vect{N}^{-1}\left[\vect{d}-\vect{s}(\vect{a})\right]
\label{oldlike}
\end{equation}
where $c$ is an unimportant constant. The first innovation in our new
method is to re-cast the log-likelihood in such a way that the
computational cost of evaluating it is considerably reduced. This is
achieved by instead writing the log-likelihood as
\begin{equation}
\ln L (\vect{a}) = c'-\tfrac{1}{2}\vect{s}(\vect{a})^{\rm T}
\vect{N}^{-1}\vect{s}(\vect{a}) + \vect{d}^{\rm
T}\vect{N}^{-1}\vect{s}(\vect{a}), \label{newlike}
\end{equation}
where $c'=c - \tfrac{1}{2}\vect{d}^{\rm T} \vect{N}^{-1}\vect{d}$
is again independent of the parameters $\vect{a}$. The advantage of
this formulation is that the part of the log-likelihood dependent on
the parameters $\vect{a}$ consists only of products involving the
data and the signal. Since the signal from a discrete source with
parameters $\vect{a}$ is only (significantly) non-zero in a limited
region centred on its putative position one need only calculate the
quadratic forms in (\ref{newlike}) over a very limited number of
pixels, whereas as the quadratic form in the standard version
(\ref{oldlike}) must be calculated over all the pixels in the map.

In the case of independent pixel noise, the covariance matrix
$\vect{N}$ in (\ref{newlike}) is diagonal, and so the quadratic
forms can be calculated very rapidly. For correlated noise, however,
$\vect{N}$ is no longer diagonal and so evaluating the necessary
elements of its inverse can be costly for a large map. Nonetheless,
if the background is a statistically homogeneous random field
$\vect{N}^{-1}$ can be computed in Fourier space, where it is
diagonal. As a result, we perform our analysis of sufficiently small
patches of sky that the assumption of statistical homogeneity is
reasonable.
 An alternative procedure would be to transform to a set of
basis functions, such as wavelets, in which the data are readily
compress, hence reducing the effective dimensionality of the problem;
we will not, however, pursue this route.

\subsection{Prior}

Having determined the likelihood function, it remains only to assign a
prior on the parameters $\vect{a}=(X,Y,A,R)$.
Although not a formal requirement of our method,
the simplest choice is to assume the prior is separable, so that
\begin{equation}
\pi(\vect{a})=\pi(X)\pi(Y)\pi(A)\pi(R).
\label{prior}
\end{equation}
Moreover, we will assume that the priors on $X$ and $Y$ are uniform
within the range of the image, and that priors on $A$ and $R$ are the
uniform distributions within some ranges $[A_{\rm min},A_{\rm max}]$
and $[R_{\rm min},R_{\rm max}]$ respectively.
This is true for our first implementation of PowellSnakes (version
I). Our general approach can, in fact, accommodate more general
priors. Indeed, the assumption of uniform priors on $A$ and $R$ is
not a good one for most astrophysical problems.  Typically one
expects the number of discrete sources to decrease with amplitude,
and their angular sizes are unlikely to be uniformly distributed.
The best prior to adopt for $A$ is one derived from existing source
counts. These are typically power laws over a wide range (although
care must be taken at the ends of the range to obtain a properly
normalised prior). In our approach presented below, however, we will
assume the log-posterior is at most quadratic in the source
amplitude $A$, which restricts $\pi(A)$ to be of uniform,
exponential or Gaussian form.  (The second implementation (version
II) of PowellSnakes currently under development adopts a Power law.)
Turning to $\pi(R)$, existing knowledge of the angular sizes of the
objects sought can be use to construct an appropriate prior, and no
restriction is placed by our method on this.  Of course, in the case
of point sources, the overall template is simply the beam profile,
the angular size of which is (usually) known in advance.
\footnote{Ongoing work on an improved implementation of PowellSnakes
(Version II) uses the following priors: uniform priors on the
position,  $X$ and $Y$, and scale, $R$, while the brightness, source
amplitude $A$, is modelled by a Power-law: $N(s) \propto s^{-
\beta}$ with $\beta$  in $[2,3)$, where $s$ is the source flux. The
range for parameter $R$ is taken from a parameter file.}

\subsection{Optimal patch size. The patch border}
The patches are divided into two different areas: A central area
where the detection takes place and a surrounding border.  The
border is used to prevent the sources from being truncated by the
edges of the patch since this usually impacts severely on the
process of their detection and characterization. In order to avoid
truncation, the border size must be at least half the largest source
radial size. When enforcing this rule we are assuring that a source
profile which has been cut by the patch edge is left undisturbed at
least on one of its neighbour patches. When defining an optimal
patch size several somewhat opposing criteria must be taken into
account:
\begin{itemize}
\item The assumption of statistical homogeneity which always favours
small patches
\item Execution speed which tends to exhibit a bias towards somewhat larger
patches.
\begin{itemize}
\item The greater the number of patches the greater the number of
border overlaps which occur. This increases the number of pixels
processed more than once.
\item There is a non-negligible amount of time preparing a new patch for detection.
\item The pre-filtering stage which involves an FFT requires a power of 2 array size in order to be efficient
\end{itemize}
\end{itemize}

Our work shows that the optimal patch size is always the one which
minimises the effects of the in-homogeneities of the background.
However, (128 x 128) pixels patches on the NSide=1024 Healpix maps
($\sim$ 7.33\textdegree~x 7.33\textdegree) and (256 x 256) pixels
patches on NSide=2048 Healpix maps ($\sim$ 7.33\textdegree~ x
7.33\textdegree) are always good choices in terms of being a good
balance between statistical homogeneity and execution speed.

\section{Object detection strategy}\label{sect:ObjDetStrat}

Once the likelihood and prior have been defined, the problem of
object identification and characterization reduces simply to
exploring the resulting posterior distribution
\begin{equation}
\ln P(\vect{a}) = \ln L(\vect{a}) + \ln \pi(\vect{a}),
\end{equation}
where we have omitted an unimportant additive constant term on the
right-hand side.  Rather than using MCMC methods, as advocated by
HM03 and Feroz \& Hobson \cite{HobsonMCMC2007}, here instead we
locate the local maxima of the posterior in the 4-dimensional
parameter space $\vect{a}=\{X,Y,A,R\}$ and perform a Gaussian
approximation about each peak. The former provides the estimates of
the object parameters associated with that posterior peak and, as
outlined below, the latter allows one to assign uncertainties to the
derived parameter values. The Gaussian approximation also allows one
to estimate the Bayesian evidence associated with each posterior
peak. The Bayesian evidence is used to decide whether the detected
peak corresponds to a real object or is just a `conspiracy' of the
background noise field (see Section~\ref{validation}).

Such an approach was also advocated by HM03, in which a simulated
annealing downhill simplex minimiser was used in an iterative object
detection scheme. At each iteration, the minimiser located the
global maximum of the posterior and an object with the optimal
parameter values was subtracted from the map before commencing the
next iteration.
The process was repeated until the first rejected detection.

Here we adopt a different strategy that obviates the need to attempt
to locate the global maximum at each iteration, since this is very
computationally expensive. Instead, our approach consists of
launching a series of simple downhill minimisations. The choice of
starting point for each minimisation, the minimiser employed and the
number of minimisations performed are discussed in detail in
Section~\ref{maxpost}. The end-point of each minimisation launched
will be a local maximum in the posterior, giving the optimal
parameter values for some putative detected object.  A Gaussian
approximation to the posterior is constructed about the peak and the
detection is either accepted or rejected based on an evidence
criterion (see Section~\ref{validation}). If the detection is
accepted, then an object with the optimal parameter values is
subtracted from the map before the next downhill minimisation is
launched.
Although our method does leave open the possibility of repeatedly
detecting the same spurious sources, the speed of our algorithm
assures that this is not a problem (see Section ~\ref{results}).

It should be noted, however, that the only reason for subtracting
accepted objects from the map is to avoid multiple real detections.
Moreover, when attempting to detect sources at very low
signal-to-noise ratios, one might occasionally accept a spurious
detection as a real source, and its subtraction from the map may
damage subsequent downhill minimisations. Ideally, therefore, one
should avoid subtracting sources altogether. One such approach is
simply to accept that real objects may be detected repeatedly, and
check for duplications in the optimal sets of parameters obtained in
each accepted detection (e.g. parameters sets that are similar to well
within the derived uncertainties). We have not implemented such a
method for this paper, but this is under investigation.

As discussed in Section \ref{maxpost} determining the number of the
downhill minimisations to perform is a key part of our method.
However once we perform these minimizations , we ensure that no real
object has been missed by performing a further set of downhill
minimisations as follows. Since the spatial extent of any object in
the data map must be greater than that of the beam, we divide the
map into patches of area equal to the beam size. For each such
patch, we then launch a downhill minimiser with initial values of
$X$ and $Y$ equal to the coordinates of the centre of the patch, and
with $A$ and $R$ values equal to the midpoints of their respective
prior ranges . This is the case for our first implementation of
PowellSnakes (version I, for white noise only).
This approach is unlikely to miss any remaining peaks in the
posterior associated with real objects, provided the number density
of sources is sufficiently low that they are spatially
well-separated.  Clearly, such an approach will perform less well if
two or more sources (partially) overlap, in which case a single
source of large spatial extent may be fitted to them.

In the earlier stages of PowellSnakes implementation we performed
the analysis completely in real space, rather than pre-filtering the
map. However this procedure slowed down the execution of the
algorithm. The evaluation of the objective function for each set of
parameters is a very expensive operation which unfortunately lies
deep in the inner loop of the code, ending up being called several
hundreds of times for each ``\emph{snake}''. Even a small
optimization of its execution has a gigantic impact on the overall
performance of the code.
We initialize the Powell minimizer, (ie the tail of each of the
``\emph{snakes}''), with the central points of the grid defined in
subsection \ref{PowMinStep} and the midpoints of the {[}A,R]
parameters. The first step of the minimization process is always to
search the position sub-space keeping R and A constant. As we have
to repeat this step each time we start a new ``\emph{snake}'',
finding a fast way of evaluating the likelihood in all the
positional sub-space available is very advantageous. This is where
the filtering process steps in. Filtering with the matched filter
automatically provides us with a fast evaluation of the likelihood
restricted to the positional subspace. Furthermore it gives a very
good estimate of the starting value for the $A$ parameter, using the
classical formula of the amplitude of a field
 filtered by a matched filter (\ref{mfahat}). Next, we
need to find out the optimal value of $R$ parameter which is no
longer considered constant.


As a final note, one must take care when a source is located near
the edge of the map, since it will be partially truncated. This is
not, however, a severe limitation as one can allow an overlap of at
least the large expected size when dividing the full map into small
patches; a truncate source in one patch will appear intact in a
neighbouring one. This strategy does result in a small extra
processing overhead, since several pixels will be processed more
than once, but this is a minor consideration.


\section{Maximising the posterior distribution}\label{maxpost}

In this section, we outline the method employed to locate the multiple
local maxima of the posterior distribution.

\subsection{Pre-filtering step}

Let us begin by writing the template (\ref{objdef}) as
\begin{equation}
\tau(\vect{x};\vect{a}) = At(\vect{x}-\vect{\underline{X}};R),
\end{equation}
where $A$, $\vect{\underline{X}}$ and $R$ are, respectively, the
amplitude, vector position and size of the object, and
$t(\vect{x};R)$ is the spatial shape of a unit height reference
object of size $R$ centred on the origin. Then the signal vector in
the log-likelihood (\ref{newlike}) is simply
\begin{equation}
\vect{s}(\vect{a}) = A\vect{t}(\vect{\underline{X}},R),
\end{equation}
where the vector $\vect{t}$ has components
$t_i=t(\vect{x}_i-\vect{\underline{X}};R)$, in which $\vect{x}_i$ is
the position of the $i$th pixel in the map. Substituting the above
expression into the log-likelihood, assuming a constant prior and
differentiating with respect to $A$ gives
\begin{equation}
\frac{\partial \ln P}{\partial A} = \vect{t}^{\rm
T}(\vect{\underline{X}},R)
\vect{N}^{-1}[\vect{d}-A\vect{t}(\vect{\underline{X}},R)],
\end{equation}
which on setting equal to zero yields an analytic estimate for the
amplitude:
\begin{equation}
\hat{A}(\vect{\underline{X}},R) = \frac{\vect{t}^{\rm
T}(\vect{\underline{X}},R)\vect{N}^{-1}\vect{d}} {\vect{t}^{\rm
T}(\vect{\underline{X}},R)\vect{N}^{-1}\vect{t}(\vect{\underline{X}},R)}.
\label{mfahat}
\end{equation}
Note that, under the assumption of statistical homogeneity, the
denominator in the above expression does not depend on the source
position $\vect{\underline{X}}$, and thus need only be calculated
once (say for a source at the origin) for any given value of $R$.
Generalising the notation of Savage \& Oliver \cite{SavageOliver},
we denote this denominator by $\alpha(R)$ and the numerator in
(\ref{mfahat}) by $\gamma(\vect{\underline{X}},R)$. It is worth
noting that the quantity $\sqrt{\alpha(R)}$ is merely a generalised
signal-to-noise ratio for a unit amplitude object integrated over
its spatial template.

More importantly the estimator (\ref{mfahat}) is precisely the filtered field produced
by a classical matched filter, in which one assumes a value for $R$.
Moreover, it is straightforward to show that the corresponding
log-likelihood (\ref{newlike}) can be written
\begin{equation}
\ln
L(\vect{\underline{X}},\hat{A},R)=c'+\tfrac{1}{2}\alpha(R)[\hat{A}(\vect{\underline{X}},R)]^2.
\label{newlike2}
\end{equation}
Thus, we show here that for a given value of $R$, peaks in the
filtered field correspond precisely to peaks in the log-likelihood
considered as a function of $\vect{\underline{X}}$. If the objects
sought all have the same size $R$ and this is known in advance, then
the local maxima of the now 3-dimensional posterior in the space
$(X,Y,A)$ are all identified by this method. This scenario
corresponds to point sources observed by an experiment with a known
beam profile.
However the assumption that the sizes of the objects are known in
advance might lead us astray. To show that this is indeed the case,
consider the following:
\begin{itemize}
\item Suppose two sources are so close they cannot be resolved by the optical device:
The intensity profile usually still matches fairly well the of beam
profile but with a much enlarged radial scale. Taking the radius of
the beam profile as an unknown parameter, PowellSnakes will still
identify them wrongly as a single source, but the total estimated
flux will much closely match the true value of the total flux coming
from both sources.
\item We are assuming a statistical homogeneous background: this is never strictly true, but keeping the patches small enough
this is usually a good assumption. Nevertheless, small changes in
the beam profile across the patch, which in terms of the noise
components (CMB, Galactic foreground, etc.) may be perfectly
ignored, when estimating the sources fluxes might lead us to
severely biased values.
\item We are assuming  point sources: this limits the range of application of this method.
One of the driving motivations behind PowellSnakes is the detection
of SZ clusters. This implies the capability of multi-frequency
operation and a source profile with a well defined shape. This
source shape template is only useful in real situations if we are to
allow different spatial scales. In order to achieve this extra
degree of freedom we introduced the Radius [$R$] parameter.
\end{itemize}

When the source size $R$ is not known in advance, or the sources
have different sizes, most matched filter analyses (see e.g. Herranz
et al. \cite{HerranzMultifreq04}) filter the field at several
predetermined values of $R$ and combine the results in some way to
obtain estimates of the parameters $(X,Y,A,R)$ for each identified
source. This approach is rather ad hoc, however, and so we will
pursue a different strategy based on locating the local maxima of
the posterior using multiple downhill minimisations. At first sight
it might seem tempting to use the result (\ref{mfahat}) to reduce
the parameter space to the 3 dimensions $(X,Y,R)$, since with
knowledge of these parameters one can immediately calculate
$\hat{A}$. Such an approach can, however, lead to numerical
instabilities when the denominator $\alpha(R)$ in (\ref{mfahat}) is
very small (which occurs in the low signal-to-noise regime) as our
numerical tests confirmed.
Thus we do not pursue this approach.

\begin{figure}
\begin{center}
\includegraphics[width=7cm]{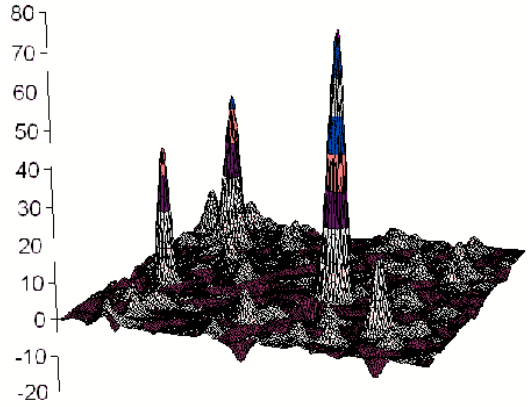}
\caption{Log-likelihood manifold, equation (\ref{newlike}) using
template (\ref{objdef}) ($A$ and $R$ dimensions suppressed); white
noise background, $R$ parameter = 4.0 pixels; a constant has been
subtracted} \label{fig:LikeWhiteNoise}
\end{center}
\end{figure}

We do, however, make use of the result (\ref{mfahat}) by first
pre-filtering the map assuming a value for $R$ equal to the mean of
its prior range.
The advantage of this pre-filtering stage is that, even in the
presence of objects of different sizes, the filter field often has a
pronounced local maximum near many real objects in case the $R$
parameter range is not too wide and the value of $R$ employed in the
filtering process stays close to 0 (figure
\ref{fig:LikeWhiteNoise}).

\subsection{2-dimensional Powell minimisation step}
\label{PowMinStep}

In the next step of our analysis, we launch $N=100$ (see below)
 downhill minimisations in the two-dimensions
$(X,Y)$ using the Powell algorithm (Press et al. \cite{NR}) with
starting points chosen as described below.
This method is a simple direction-set algorithm that requires only
function values. The method makes repeated use of a 1-dimensional
minimisation algorithm, which in our case is Brent's method; this
is an interpolation scheme that alternates between parabolic steps
and golden sections. In addition, we `enhance' the basic Brent line
minimizer algorithm by introducing a `\emph{jump}' procedure
designed to avoid small local minima (such as those resulting in the
posterior from the noise background). A description of this
`\emph{jump}' procedure is given below.
The reason for this 2-dimensional minimisation step is simply to
obtain a good estimate of the $(X,Y)$-positions of the sources. As a
byproduct, using formula (\ref{mfahat}), we end up obtaining an
approximate estimate of the putative source amplitude $A$ as well.

Assuming that sensitivity and experiment resolution are such that we
should only expect a very small number of partially resolved
sources, one may assume that in each patch of beam size area we
should not find more than one real peak. The likelihood peaks in the
positional sub-space (keeping $A$ and $R$ constant) are not expected
narrower than the original beam at least when we are dealing with
the white noise only case. This is so because the likelihood
positional surface is nothing but the correlation of the beam shape
with, on average, another beam shape eventually with a different
$R$. Thus the total area where the correlation will have values
significantly different from 0 is expected to be larger than the
largest area of each one of the original beams. These assumptions
naturally define the coarseness one should use when exploring the
patch, namely, the beam size. As we are assuming that beam shapes
with different radii may actually occur, the largest possible
resolution
happens when the smallest beam is considered. So far there is no
prior information to  eventually distinguish one part of the patch
from the others, therefore we assume that a uniform rectangular grid
with as many cells as the total number of ``\emph{beams in the
patch}'', would be perfectly adequate to define the starting points
of the ``\emph{snakes}''. Thus:

\begin{equation}
\label{eq:PwM_NBeams} N_{\text{\textrm{Snakes}}}=\frac{\text{Patch
total number of pixels inside the `\emph{core}' area\ }}{\text{Size
of the smallest beam (in pixels)}}
\end{equation}

\begin{figure}
\begin{center}
\includegraphics[width=7cm]{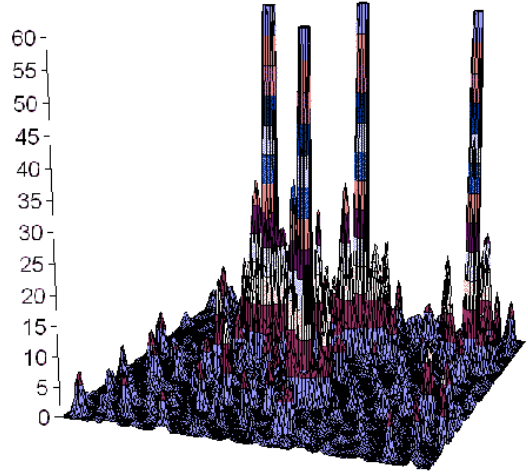}
\caption{The likelihood ratio manifold of the models for the
presence of a source over its absence $\frac{L_{H_1}}{L_{H_0}}$ ($A$
and $R$ dimensions suppressed): radius $R=4.0$ pix; Background
parameters $s_0$ = 11.57 pix, $w_I=0.01$, $w_B=0.99$ (see section
\ref{sect:IntroColour})} \label{fig:LikelihoodColourBaseDetail}
\end{center}
\end{figure}

We add an extra two dimensional random offset from the central
points of each grid cell in order to avoid some unwanted correlation
effects due to the perfect spacing of the grid. Some may argue that,
in addition to the likelihood real peaks (those resulting from the
sources in the map), we should expect a large number of fake ones
resulting from the interaction of the filter with the noise
fluctuations (figure \ref{fig:LikeWhiteNoise}). We considered such
possibility by devising a ``\emph{jump}'' procedure (see below) to
avoid the ``\emph{snakes}'' from being ``\emph{caught}'' on those
fake peaks. Nevertheless, as in the case of a CMB background, when
the number of these spurious peaks largely outnumbers the real ones,
one should increase the density of ``\emph{snakes}'' (figure
\ref{fig:LikelihoodColourBaseDetail}). For the white noise detection
examples presented in section~\ref{results}, we have taken all of
the above into consideration. The number of snakes chosen,
$N_{\text{\textrm{Snakes}}}=100$, meets these requirements, for both
cases (ie LISNR and HMcL, see section~\ref{results}).  As we assumed
that no sources are placed into the patch borders we consequently
did not include those pixels.

\subsubsection{The jump procedure}
\label{jump}

The Brent line minimizer keeps at any time a set of three pairs of
values, argument and function value, where the middle pair is said
to be ``\emph{bracket}''. This is because the argument part of the
second pair lies in the interval defined by the other two and its
function value is lower than the function values of the interval
limiting pairs. We call them

\begin{equation}
\label{eq:JP_Bounds}
\left\{ \begin{array}{ll}
[a,f(a)] & \text{\textrm{Left bound}}\\
{}[b,f(b)] & \text{\textrm{Current "particle" solution}}\\
{}[c,f(c)] & \text{\textrm{Right bound}}\end{array}\right.
\end{equation}

Like the simulated annealing example previously referred, the
rational behind the jump procedure was borrowed from natural world,
but this time from the realm of quantum mechanics. We imagine our
current best guess, the bracketed pair, $[b,f(b)]$, as the particle
inside a potential well defined by the limiting pairs (left and
right bound). If we adopt the ``classical'' point of view there is
no way the particle could possibly escape from the bracketing
barriers. This is exactly what happens with the traditional Brent
procedure. Instead if we use a quantum approach it is still possible
for the particle to tunnel out from the barrier.

\begin{figure}
\begin{center}
\includegraphics[width=7cm,keepaspectratio]{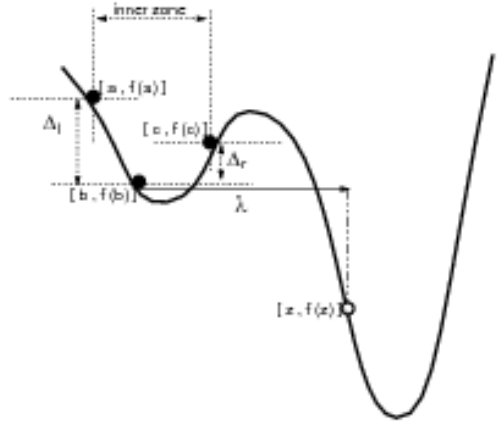}
\caption{The jump procedure} \label{fig:jump}
\end{center}
\end{figure}

The transmission coefficient $T$ of a particle with energy $E$ which
faces a potential barrier with a height equal to $V_{0}$($V_{0}>E$),
and a width $l$, is given by:

\begin{equation}
\label{eq:JP_TransCoef}
T\simeq\frac{16E(V_{0}-E)}{V_{0}^{2}}\:
e^{-2\rho_{2}l},
\end{equation}
where $\rho_{2}$ is $=\sqrt{\frac{2m(V_{0}-E)}{\hbar^{2}}}$,
$m$ is the particle mass and $\hbar$ is Dirac constant, where we
assume $\rho_{2}l\ll1$ (Claude Cohen-Tannoudji et al.
\cite{ClaudeCohen}). From the quantum mechanical example we have
only retained the negative exponential dependence with the barrier
length (the jump distance in our jump model) and the characteristic
constant depending on the difference between the particle energy and
the barrier maximum potential. Following this example we define a
random variable $\Lambda$ exponentially distributed

\begin{equation}\label{eq:JP_Distrib}
p(\Lambda=\lambda)=\frac{1}{\theta}\, e^{-\lambda / \theta},
\end{equation}
with average $\theta$. The next step is to define $\theta$
dependence on the parameters, bearing in mind
the tunnelling effect behaviour. We chose:

\begin{equation}\label{eq:JP_AverageVal}
\theta=L\left(\frac{f_{s}-\Delta}{f_{s}}\right)
\end{equation}
where $L$  is the ``maximum average jump length'', $f_{s}$ is the
``function values scale'', and $\Delta$ is either the ``right
barrier'' $\equiv\Delta_r$ when we jump forward or  ~``left
barrier'' $\equiv\Delta_l$ when we are trying a jump along the
opposite direction (figure~\ref{fig:jump}).

The procedure has two tuning parameters: one (whose value is
given as input from the parameter file) to divide the maximum allowed range
for the current jump in order to produce the
$L$
value, and
another one which  purpose is to define a scale for the function
values. This second value:
$f_{s}$
is the maximum range of function
values evaluated from the ``\emph{snake's tails}'' locations (initial
values drawn from the grid as explained above). When we start the
minimization process by defining the bracketing interval, the
well walls (defined by the differences of the function values,
$\Delta_{r}$ and $\Delta_{l}$) are usually large and consequently
only very small tunnels will be allowed. If
$\Delta>f_{s}$ no jumps will be tempted. As the minimizer converges
and approaches the minimum, the barriers will become smaller and
consequently bigger tunnels will become more and more probable. This
jump procedure has the advantage of exploring the function domain
progressively as the algorithm approaches the bottom of the well. If
by accident we have first hit the global minimum when the solution
is still far from the bottom only small jumps are allowed and the
probability of tunnelling to a local peak is small. As soon as we
get to the bottom, $\Delta\approx0$, large tunnels become probable
($\theta\approx L$),
but as the function is now close to its
lowest value, a successful jump is now improbable. But if instead we have
started within a local peak instead of becoming
``\emph{stuck}'', as the minimizer approaches its last steps,
large jumps become common and the probability of ``\emph{plunging}''
into a deeper well is now very high. One of the main reasons for
choosing the exponential distribution, apart its close relationship
with the quantum tunnelling, is the simplicity and
efficiency of a random deviates generator whose output will follow
it  (provided we already have a routine which produces random
deviates with a uniform probability distribution inside a certain
interval). We proceed as follows:

\paragraph*{For each Brent iteration:}

\begin{itemize}
\item Draw a sample $u$ from an uniform distribution between $\left[0,1\right)$.
If $u\geq0.5$ choose a positive jump, otherwise move
backwards.
\item Draw from the tunnelling distribution. If the jump ends up inside
the ``\emph{inner zone}'' we reject it because we are on the same
well we have started from. If the jump is
outside the ''\emph{inner zone}'' we evaluate the function at that
point $[z,f(z)]$. If the new function value is  lower
than the one that initiated the jump, $f(z)<f(b)$, we accept it and
restart the Brent minimizer with this new value.
\end{itemize}

\subsubsection{The effectiveness of the jump procedure in recovering "lost
snakes"}

Our extensive set of simulations covering a typical assemblage of
different scenarios shows that the use of the jump procedure has a
fairly good effectiveness. We have achieved equal levels of quality
on the obtained results using $\sim$ 20-25 \% fewer snakes. However,
within our current setup its importance is far from critical since
speed is not a concern to us. The extra level of complexity which
results from the introduction of the procedure, could have been
avoided by launching extra "snakes". However, when dealing with such
a scenario where the cost of starting additional "snakes" is much
higher or the problem cannot just be circumvented by setting up more
starting points for the minimization procedure, we believe that the
jump procedure might have an important role to play.

\subsection{4-dimensional Powell minimisation step}

Instead of just repeating the filtering procedure using different
values for the $R$ parameter, we proceed by following a rationale
which tries to mimic that behind the FFT immense speed up. FFT is
fast because it reduces, using factorization, the number of
operations needed to compute the ``\emph{complete set}'' of Fourier
coefficients. If we only need one or two (if the number of
coefficients we want to compute is less than $\ln(n)$; $n$ being the
size of the vector) then it would be more advantageous to use the
traditional formula directly. Hence, we now search the full
parameter space ($X$, $Y$, $A$ and $R$) with the help of the Powell
minimizer, but this time using the previously estimated parameters
(from the 2-dimensional step) as starting values. We are not
expecting the ``\emph{Powell snake}'' to unfold significantly
outside a small neighbourhood of the initial value. Thus, as we do
not need to evaluate the likelihood outside this small positional
range anymore (as in the FFT case where we just want to compute one
or two coefficients) using the real space proves to be more
advantageous.

In the next step, we perform multiple Powell minimisations in the
full 4-dimensional space, starting each minimisation from the $(X,Y,
A,R)$ - positions found in the previous step. In this step the
`\emph{jump}' procedure described above is not used. This is one of
the main reasons why we have first performed the 2-dimensional step
(and the pre-filtering). From our experience we could verify that
after performing the 2-dimensional step, the neighbourhood of the
likelihood manifold around the vector solution $\left\{X, Y,
A,R\right\}$ we had so far obtained was rather smooth and ``well
behaved''. Thus, the necessity for a procedure in order to avoid
local maxima is no longer required. The resulting end-point of each
minimisation is a local maximum of the posterior, giving the optimal
parameter values for some putative detected object.  A Gaussian
approximation to the posterior is constructed about the peak and the
detection is either accepted or rejected based on an evidence
criterion (see Section \ref{validation}). If the detection is
accepted, then an object with the optimal parameter values is
subtracted from the map before the next downhill minimisation is
launched.

\subsection{Error estimation - Fisher
analysis}\label{subsect:FisherAnalysis}

For each accepted object, estimates of the uncertainties on the
derived parameters are obtained from the Gaussian approximation to
the posterior around that peak. In detail the covariance matrix of the
parameter uncertainties has elements
$C_{ij} \equiv \langle \delta a_i \delta a_j \rangle$, which can be
obtained by inverting (minus) the Hessian matrix at the peak. In the
case of uniform priors, one has simply
\begin{equation}
(C^{-1})_{ij} = -\left\langle\spd{\ln
L}{\theta_i}{\theta_j}\right\rangle \label{eq:error}
\end{equation}

\noindent

Here we present an estimation of the putative source parameters'
error bars using the Fisher analysis over the model $H_{1}$. We
present two different cases:

\[
\left\{ \begin{array}{ll}
\mbox{White noise} & \mbox{$\rightarrow$ The analysis is carried out in real space}\\
\mbox{``\emph{Coloured}'' noise} & \mbox{$\rightarrow$ Fourier space
is used instead (see subsection
\ref{subsect:FisherColour})}\end{array}\right.\]

\noindent
We start by defining $F_{ij}$ as the negative expectation
of the  log-likelihood Hessian matrix, in other words the
{}``\emph{Fisher information matrix}''. We are assuming, as we have
already stated, a Gaussian approximation of the likelihood around
its maximum in parameter space.

Let
$\mathit{L}_{H_{1}}\equiv\mathrm{Pr}(\boldsymbol{D}|\boldsymbol{\mathbf{\Theta}}\,H_{1})$
be the likelihood of model $H_{1}$, then

\begin{equation}
F_{ij}=-\left\langle
\frac{\partial^{2}\ln\left(\mathit{L}_{H_{1}}\right)}{\partial\theta_{i}\,\partial\theta_{j}}\right\rangle
_{\hat{\boldsymbol{\Theta}}}\label{eq:1.1}\end{equation}
Assuming a Gaussian model for the diffuse background we have:

\begin{equation}
-\ln\left(\mathit{L}_{H_{1}}\right)=
c+\sum_{i,j}\boldsymbol{N}_{ij}^{-1}d_{i}\tau_{j}-\frac{1}{2}\sum_{i,j}\boldsymbol{N}_{ij}^{-1}\tau_{i}\tau_{j}-\frac{1}{2}\sum_{i,j}\boldsymbol{N}_{ij}^{-1}d_{i}d_{j},\label{eq:1.2}\end{equation}
where $\boldsymbol{N}_{ij}^{-1}$ is the $(ij)$ element of the
inverse correlation matrix of the diffuse background, $p_{i}$ is the
$i$ map pixel, $\tau_{i}$ is the $i$ pixel of the object template
$\boldsymbol{\tau}(\mathbf{r},A,\boldsymbol{\bar{\Theta}},\boldsymbol{r}_{0})$
and $c$ is an unimportant constant.
The parameters of the object template are: $A$ the amplitude of the
object, $\mathbf{r}_{0}$ the positional parameters
and $\mathbf{\bar{\Theta}}$ which represents the ensemble of the remaining
parameters. In this analysis we are assuming a circular symmetric
source. Thus, $\boldsymbol{\bar{\Theta}}$ contains only a single parameter,
$R$ which provides a scale for the radial dimension of the object.
ie, $R\equiv\mbox{"Radius of the object"}$. We will assume, from now
on, the following functional dependence of the source template on
its parameters:

\begin{equation}
\boldsymbol{\tau}(\mathbf{r},A,\boldsymbol{\bar{\Theta}},\boldsymbol{r}_{0})=A\, t\left(\frac{|\boldsymbol{r}-\boldsymbol{r}_{0}|^{2}}{R^{2}}\right)\label{eq:1.3}\end{equation}

\subsubsection{White noise}

For a background diffuse component generated from a Gaussian
stationary process with coherence length much smaller than the pixel
size, ie  Gaussian white noise, the following condition
applies:

\begin{equation}
\boldsymbol{N}_{ij}^{-1}=\frac{1}{\sigma_{i}^{2}}\delta_{ij}\label{eq:2.1}
\end{equation}
This is the situation when dealing with a system where the
instrumental pixel noise is the dominant source of the background
component. Adding the condition for statistical homogeneity :

\begin{equation}
\frac{1}{\sigma_{i}^{2}}=\frac{1}{\sigma^{2}},\label{eq:2.2}\end{equation}
and substituting both conditions into the log-likelihood expression
this becomes

\begin{equation}
-\ln\left(\mathit{L}_{H_{1}}\right)=c+\frac{1}{\sigma^{2}}\left(\sum_{i}d_{i}\tau_{i}-\frac{1}{2}\sum_{i}\tau_{i}^{2}-\frac{1}{2}\sum_{i}d_{i}^{2}\right),\label{eq:2.3}\end{equation}
Evaluating expression \eqref{eq:1.1} for all the possible
parameters' combinations and using the following sort criterion
$\left\{A,R,X,Y\right\} $, one obtains

\begin{equation}
F_{ij}=\left[\begin{array}{cccc}
\frac{\alpha \hat{R}^{2}}{\sigma^{2}} & \frac{-2\delta \hat{A}\hat{R}}{\sigma^{2}} & 0 & 0\\
\frac{-2\delta \hat{A}\hat{R}}{\sigma^{2}} & \frac{4\beta \hat{A}^{2}}{\sigma^{2}} & 0 & 0\\
0 & 0 & \frac{4\gamma \hat{A}^{2}}{\sigma^{2}} & 0\\
0 & 0 & 0 & \frac{4\gamma
\hat{A}^{2}}{\sigma^{2}}\end{array}\right],\label{eq:2.4}
\end{equation}
where $\left\{\alpha,\beta,\delta,\gamma\right\} $ are pure
numerical constants

\begin{equation}
\left\{ \begin{array}{ll}
\alpha=2\pi\,\intop_{0}^{\infty}r\, t\left(r^{2}\right)^{2}\, dr, & \beta=2\pi\,\intop_{0}^{\infty}r^{5}\, t'\left(r^{2}\right)^{2}\, dr\\
\gamma=\pi\,\intop_{0}^{\infty}r^{3}\, t'\left(r^{2}\right)^{2}\,
dr, & \delta=2\pi\,\intop_{0}^{\infty}r^{3}\,
t\left(r^{2}\right)t'\left(r^{2}\right)\,
dr\end{array}\right.,\label{eq:2.5}
\end{equation}
$t'\left(x\right)\equiv\frac{d\, t(x)}{d\, x}$ and
$\hat{\boldsymbol{\Theta}}\equiv\left\{\hat{A},\hat{R},\hat{X},\hat{Y}\right\}
$ are the values of the template parameters which maximize the
likelihood. We have used pixel metrics together with the assumption
that the sums can be evaluated using the continuous limit (this
introduces only minor errors). When considering a source template
with exactly the same functional dependence of
equation~\eqref{objdef}, the numerical constants become:

\begin{equation}
\left\{ \begin{array}{ll}
\alpha=\pi, & \beta=\frac{\pi}{2}\\
\gamma=\frac{\pi}{8}, & \delta=-\frac{\pi}{2}\end{array}\right.\label{eq:2.7}
\end{equation}
Now inverting the $\boldsymbol{F}$ matrix

\begin{equation}
F_{ij}^{-1}=\left[\begin{array}{cccc}
\frac{\beta\sigma^{2}}{4\hat{R}^{2}\mathcal{D}} & \frac{\delta\sigma^{2}}{2\hat{R}\hat{A}\mathcal{D}} & 0 & 0\\
\frac{\delta\sigma^{2}}{2\hat{R}\hat{A}\mathcal{D}} & \frac{\alpha\sigma^{2}}{4\hat{A}^{2}\mathcal{D}} & 0 & 0\\
0 & 0 & \frac{\sigma^{2}}{4\hat{A}^{2}\gamma} & 0\\
0 & 0 & 0 &
\frac{\sigma^{2}}{4\hat{A}^{2}\gamma}\end{array}\right],\label{eq:2.8}\end{equation}
where $\mathcal{D}\equiv\alpha\beta-\delta^{2}$ and using the
Cram\`{e}r-Rao inequality, which in this case reduces to an
equality:

\begin{equation}
\Delta\theta_{i}=\sqrt{F_{\theta_{i}\theta_{i}}^{-1}},\label{eq:2.9}\end{equation}
since we are dealing with a max-likelihood estimator which is
efficient. Writing explicitly the above expression \eqref{eq:2.9}
for each parameter, one finally obtains the desired parameters'
error bars:

\begin{equation}
\begin{array}{llll}
\Delta A=\frac{\sigma}{\hat{R}}\,\sqrt{\frac{\beta}{4\mathcal{D}}},
& \Delta
R=\frac{\sigma}{\hat{A}}\,\sqrt{\frac{\alpha}{4\mathcal{D}}}, &
\Delta X=\frac{\sigma}{\hat{A}}\,\frac{1}{\sqrt{4\gamma}}, & \Delta
Y=\frac{\sigma}{\hat{A}}\,\frac{1}{\sqrt{4\gamma}}\end{array}\label{eq:2.10}
\end{equation}
Defining $PSNR\equiv\frac{\hat{A}}{\sigma},\mbox{ "Peak Signal to
Noise Ratio"}$ , and substituting into \eqref{eq:2.10}:

\begin{equation}
\begin{array}{llll}
\Delta A=\frac{\sigma}{\hat{R}}\,\sqrt{\frac{\beta}{4\mathcal{D}}},
& \Delta R=\frac{1}{PSNR}\,\sqrt{\frac{\alpha}{4\mathcal{D}}}, &
\Delta X=\frac{1}{PSNR}\,\frac{1}{\sqrt{4\gamma}}, & \Delta
Y=\frac{1}{PSNR}\,\frac{1}{\sqrt{4\gamma}}\end{array},\label{eq:2.11}\end{equation}
we get another way of expressing the error bars, where we have
emphasized their dependence on the ``\emph{quality}'' of the signal.


\section{Bayesian object validation}
\label{validation}

It is clear that a key component of our approach is the step for
accepting/rejecting as a real object each posterior peak. Indeed, a
reliable means for performing this step is crucial for the method to
function properly.  The decision whether to accept or reject a
detection is based on the evaluation of the Bayesian evidence for
two competing models for the data given by equation \eqref{eq:3.1}.
PowellSnakes  evaluates the models evidence using a Gaussian
approximation to the logarithm of the likelihood ratio of the
competing models

\begin{equation}
\ln\left(\frac{\mathrm{Pr}(\boldsymbol{D}|\boldsymbol{\mathbf{\Theta}}\,H_{1})}{\mathrm{Pr}(\boldsymbol{D}|H_{0})}\right),\label{eq:A1}
\end{equation}
It does so by expanding Equation (\ref{eq:A1}) around its maxima in
the parameter space and rejecting all the terms equal and above the
third order.
As the null model ($H_{0}\equiv$ ``\emph{There is no source centred
in region S}'') doesn't have any parameter,
 thus behaving like a constant when considering the parameters space,
maximizing the logarithm of likelihood ratio holds the same results
as the max-likelihood estimator of the source parameters of the
$H_{1}$ model only.  Therefore finding the peaks of the logarithm of
likelihood ratio, corresponds to computing  the max-likelihood
estimator of the putative source parameters.

The precise definition of the models $H_0$ and $H_1$ are given below.
In general, however, we shall adopt a formulation of the model
selection problem in which each model $H_i$ ($i=0,1$) is parameterised
by the same parameters $\vect{a}=(X,Y,A,R)$, but with different
priors. In this case the evidence for each model is:
\begin{equation}
Z_i = \int L(\vect{a})\pi_i(\vect{a})d\vect{a},
\end{equation}
where the likelihood function $L(\vect{a})$ is the same for both
models and is given by \eqref{newlike}. Although not required by the method, we
will assume that for each model the prior is separable, as in equation
\eqref{prior}, so that for $i=0,1$:
\begin{equation}
\pi_i(\vect{a})=\pi_i(\vect{\underline{X}})\pi_i(A)\pi_i(R).
\end{equation}
where $\vect{\underline{X}}$ is the vector position of the source.

Moreover, we shall assume uniform priors within specified ranges on
each parameter, although the method is easily generalised to
non-uniform priors.
The general form of our models for the data are:
\begin{eqnarray*}
H_0 & = & \mbox{`there is no source centred in the region $S$',} \\
H_1 & = & \mbox{`there is one source centred in the region $S$'.}
\end{eqnarray*}
If the region $S$ corresponds to the coordinate ranges
$[X_{\rm min},X_{\rm max}]$ and $[Y_{\rm min},Y_{\rm max}]$, then
\begin{equation}
\pi_0(\vect{\underline{X}})=\pi_1(\vect{\underline{X}})\equiv \pi
(\vect{\underline{X}}) =
\begin{cases}
1/|S| & \mbox{if $\vect{\underline{X}} \in S$} \\
0 & \mbox{otherwise}
\end{cases},
\end{equation}
where $|S|$ is the area of the region $S$. We assume the prior on
$R$ is also the same for both models, so
$\pi_0(R)=\pi_1(R)\equiv\pi(R)$, but the priors on $A$ are
substantially different. Guided by the forms for $H_0$ and $H_1$
given above, we take $\pi_0(A)=\delta(A)$ (which forces $A=0$) and
\begin{equation}
\pi_1(A)=
\begin{cases}
1/\Delta A & \mbox{if $A \in [A_{\rm min},A_{\rm max}]$} \\
0 & \mbox{otherwise}
\end{cases},
\end{equation}
One could, of course, consider alternative definitions of these
hypotheses, such as setting $H_0$: $A \le A_{\rm lim}$ and $H_1$: $A >
A_{\rm lim}$, where $A_{\rm lim}$ is some (non-zero) cut-off value
below which one is not interested in the identified object. We shall
not, however, pursue this further here.

Given the above forms of the priors, the evidence for
model $H_0$ is simply:
\begin{eqnarray}
Z_0
& = & \int d\vect{\underline{X}}\,\,dA\,\,dR\,\, \pi(\vect{\underline{X}})\,\delta(A)\,\pi(R) \,L(\vect{\underline{X}},A,R)\\
& = & L_0\int
d\vect{\underline{X}}\,\,dR\,\,\pi(\vect{\underline{X}})\,\pi(R) =
L_0,
\end{eqnarray}
since $L_0\equiv L(\vect{\underline{X}},A=0,R)$ is a constant and
the priors are normalised. For model $H_1$, the evidence reads
\begin{eqnarray}
Z_1(S) & = & \int_S d\vect{\underline{X}}\,\pi(\vect{\underline{X}})
\int dA\,dR\,\, \pi(A)\,\pi(R)\,
L(\vect{\underline{X}},A,R) \\
& = & \int_S d\vect{\underline{X}}\,\pi(\vect{\underline{X}})\, \bar{P}_1(\vect{\underline{X}}),\\
& = & \frac{1}{|S|} \int_S
\bar{P}_1(\vect{\underline{X}})\,\,d\vect{\underline{X}}\,
\label{eq:e1}
\end{eqnarray}
where we wrote explicitly the dependence of the evidence on the
chosen spatial region $S$; we have also defined the (unnormalised)
two-dimensional marginal posterior $\bar{P}_1(\vect{\underline{X}})$
and $|S|$ is the area of the region $S$.

So far we have not addressed the prior ratio $\Pr(H_1)/Pr(H_0)$ in
\eqref{eq:3.1}. Although this factor is often set to unity in model
selection problems, one must be more careful in the current setting.
For the sake of illustration and simplicity, let us assume that the
objects we seek are randomly distributed in spatial position, i.e.
they are not clustered. In that case, if $\mu_S$ is the (in general
non-integer) expected number of objects per region of size $|S|$,
then the probability of there being $N$ objects in such a region
follows a Poissonian distribution:
\begin{equation}
\Pr(N|\mu_s) = \frac{e^{-\mu_S} \mu_S^N}{N!}.
\end{equation}
Thus, bearing in mind the above definitions of $H_0$ and $H_1$, we
have:
\begin{equation}
\frac{\Pr(H_1)}{\Pr(H_0)} = \mu_S.
\end{equation}
Hence, the key equation \eqref{eq:3.1} for model selection becomes:
\begin{equation}\label{eq:ModelSelKeyEq}
\rho \equiv \frac{\Pr(H_1|\vect{d})}{\Pr(H_0|\vect{d})} =
\mu_S\frac{Z_1(S)}{L_0} = \mu \, \frac{\int_S
\bar{P}_1(\vect{\underline{X}}) \, d\vect{\underline{X}}  }{L_0},
\end{equation}
where $\mu = \frac{\mu_S}{|S|}$  is the expected number sources
(centres) per unit area.

There is a certain degree of freedom when choosing the level $\rho$
must reach in order to consider that model $H_{1}$ is present (a
detection) or not. This ambiguity may only be overcome by advocating
decision theory (see Jaynes \cite{jaynes}). From now on, we will
always assume a criterion of
\begin{equation}\label{eq:SymmLossDef}
\mbox{symmetric loss} \equiv \mbox{``an undetected source is as bad
as a spurious one.''}
\end{equation}
When introducing the criterion of symmetric loss the condition for
detection immediately becomes uniquely defined. One accepts the
detection if

\begin{equation}\label{eq:SymmLossCond}
\rho > 1,
\end{equation}
and rejects it otherwise. One can change our decision criteria, for
instance suppose we want to put emphasis on reliability (which means
we are willing to accept an increase in the number of undetected
sources by a significant decrease in the fake ones), than one must
change the threshold for acceptance/rejection. This means that one
must change the value that $\rho$ must reach.

The only remaining issue is the choice of the region $S$. Our method
for locating peaks in the posterior is direct (local) maximisation,
which yields the parameter values
$\hat{\vect{a}}=(\hat{\vect{\underline{X}}}\, ,\hat{A},\hat{R})$ at
the peak.

In this peak-based approach, $S$ is taken to be a region enclosing
the entire posterior peak in the $(\vect{\underline{X}})$ - subspace
(to a good approximation). This, in fact, requires a little care. If
$S$ were taken to be the full mapped region, then the evidence
$Z_1(S)$ would have contributions from many local peaks in the
posterior. For each putative detection, however, we are only
interested in the contribution to $Z_1(S)$ resulting from the
posterior peak of interest. In practice, our Gaussian approximation
to the posterior around this peak means that other peaks will not
contribute to our estimate of $Z_1(S)$, thus making a virtue out of
a necessity. Moreover, as we show below, provided the region $S$
does enclose the posterior peak of interest in the
$(\vect{\underline{X}})$ - subspace, the resulting model selection
ratio $\rho$ will be independent of the region size $|S|$.

In the present case, using (\ref{eq:e1}) and assuming uniform priors on $A$ and
$R$, one has:
\begin{equation}
Z_1(S) = \frac{1}{|S|}\int_S
\bar{P}(\vect{\underline{X}})\,d\vect{\underline{X}} = \frac{1}
{|S|\Delta  A\Delta R} \int L({\vect{a}}) \,d^4\vect{a}.
\label{eq:E1PeakPrior}
\end{equation}
Making a 4-dimensional Gaussian approximation to the posterior about
its peak $\hat{\vect{a}}$, one obtains
\begin{equation}
\rho \approx \frac{\mu_S}{L_0|S|\Delta A\Delta R}(2\pi)^2
L(\hat{\vect{a}})|\vect{C}(\hat{\vect{a}})|^{1/2},
\label{eq:rho}
\end{equation}
where the elements of (minus) the Hessian matrix $\mathbf{C}^{-1}$ are
given by (\ref{eq:error}) evaluated at the peak $\hat{\vect{a}}$.  The above
expression again ignores possible edge effects due to abrupt
truncation of the posterior by the priors.
 In Section~\ref{truncation} we give an account of how we solved this problem.
Using the expressions
(\ref{newlike}), (\ref{mfahat}) and \eqref{newlike2} for the likelihood, one finds (in logarithms)
\begin{equation}\label{eq:SelectionAprox}
\ln\rho \approx \ln \mu_S -\ln |S| + 2\ln(2\pi)-\ln (\Delta A\Delta
R) +
\tfrac{1}{2}\alpha(\hat{R})[\hat{A}(\hat{\vect{\underline{X}}},\hat{R})]^2
+ \tfrac{1}{2} \ln |\mathbf{C}(\hat{\vect{a}})|.
\end{equation}
Most importantly, since
\begin{equation}\label{eq:MuPropCond}
\mu_S \propto |S|,
\end{equation}
we see that $\ln \rho$ is independent of the size $|S|$ of the
region considered in the $(\vect{\underline{X}})$ - subspace,
provided $S$ encloses entirely (to a good approximation) the peak of
interest.
\subsection{Acceptance/rejection threshold}

Consider the variable $\nu$
\begin{equation}\label{eq:NormalAmpl1}
\nu\equiv\frac{\boldsymbol{d}^{T}\vect{M}^{-1}\boldsymbol{t}(\hat{a})}{\sqrt{\alpha}~\sigma^{2}}
,
\end{equation}
where $ \left < n_{i} n_{j} \right> = \sigma^{2} M_{ij} = N_{ij}$
and $\sqrt{\alpha}$ is the generalised signal-to-noise ratio for a
unit amplitude object (according to Savage \& Oliver
\cite{SavageOliver} notation).
This is the same as
\begin{equation}\label{eq:NormalAmpl2}
\nu=\frac{\boldsymbol{d}^{T}\vect{N}^{-1}\boldsymbol{t}(\hat{a})}{\sqrt{\boldsymbol{t}(\hat{a})^{T}\vect{N}^{-1}\boldsymbol{t}(\hat{a})}}
,
\end{equation}
which is the more common form one may find of $\nu \equiv$
``\emph{the normalized field amplitude}'' (L\'{o}pez-Caniego et al.
\cite{caniego}). ``\emph{The normalized field amplitude}'' is the
amplitude of a random stationary Gaussian field whose power spectrum
[$\mathcal{B(\eta)}$] satisfies the condition
$\intop_{\boldsymbol{\eta}}\mathcal{B}(\eta)
d\mathbf{\boldsymbol{\eta}}=1$,
ie, $\sigma=1$. Using Parseval theorem, it is straightforward to
show that the power spectrum of some background that was filtered by
an un-normalized matched filter, $\mathcal{M}(\eta)$, satisfies:

\begin{equation}\label{eq:NormalAmpl4}
\intop_{\boldsymbol{\eta}}\vect{M}(\eta)\,
d\mathbf{\boldsymbol{\eta}}\simeq\boldsymbol{t}(\hat{a})^{T}\vect{N}^{-1}\boldsymbol{t}(\hat{a}).
\end{equation}
Assuming that $\mu$
is neither a function of the position nor a function of the
considered area, this implies that $\mu$ must be a constant. These
conditions have been obtained using the principle of indifference
along with the restriction (\ref{eq:MuPropCond}). Integrating, one
obtains:

\begin{equation}\label{eq:MuTotalArea}
\mu_{S}=\frac{N~|s|}{\Delta_{s}},
\end{equation}
where $N$ is the expected total number of sources in the patch and
$\Delta_s$ is the total area of the patch.

Let us now recover the condition for symmetric loss, $\rho>1$,
together with the key result for model selection
(\ref{eq:ModelSelKeyEq}). Substituting (\ref{eq:MuTotalArea}) into
it and using expression (\ref{newlike}) we obtain:

\begin{equation}\label{eq:ModelAssSimpleCorr}
\nu>\frac{\hat{A}\sqrt{\alpha}}{2}+\frac{\mathcal{P}}{\hat{A}\sqrt{\alpha}}
\end{equation}
where $\mathcal{P}$ (the prior term) is:
\begin{equation}\label{eq:ModelAssPrior}
\mathcal{P}\equiv\ln\left(\frac{V_T}{N~2\pi}\right)+\ln\left(
(2\pi)^{-1}\left|\mathbf{C}(\hat{\boldsymbol{a}})\right|^{-\frac{1}{2}}
\right).
\end{equation}
where $V_{T}$ is the total parameter volume, ie, $V_{T} =
(A_{max}-A_{mim})\,(R_{max}- R_{min})\, \Delta_{s}$ and $N$ is the
expected number of sources in the map.
\begin{figure}
\begin{center}
\includegraphics[width=7cm]{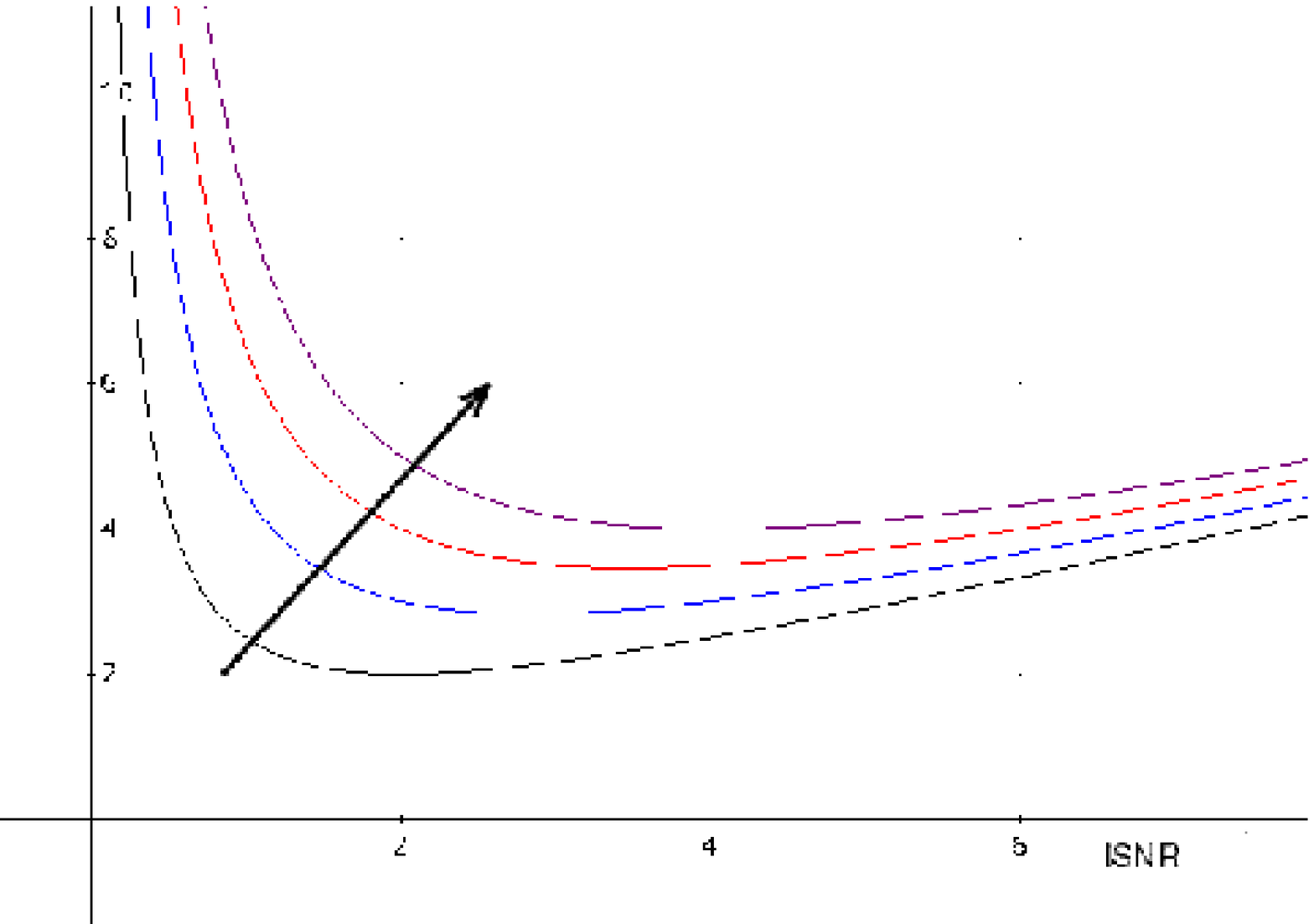}
\caption{Right hand side of inequality (\ref{eq:ModelAssSimpleCorr})
as function of ISNR $\equiv\hat{A}\sqrt{\alpha},$} The arrow shows
the direction of increasing $\mathcal{P}$
\label{fig:MinimumTresholdLevel}
\end{center}
\end{figure}

In Figure \ref{fig:MinimumTresholdLevel} we display the right hand
side of inequality (\ref{eq:ModelAssSimpleCorr}) as function of ISNR
$\equiv\hat{A}\sqrt{\alpha}$. We plot several curves for typical
values $\mathcal{P}$. Something which is immediately evident is that
each curve has a lower bound which depends on $\mathcal{P}$ which in
turn depends on the priors. Regardless of how small the peak is
there is always a lower bound for it to be considered a true
detection and not merely a noise fluctuation. This is to be expected
from an assessment condition which enforces a policy of robustness
against false detections (spurious). The threshold curve rises on
both directions: when the signal (peak) is too weak because there is
a strong possibility of being a noise fluctuation; and when the
signal is high because there is a lot of evidence supporting the
presence of an object. Hence, rising up the threshold assures an
extra level of security against spurious detections without
compromising the degree of detection.

Expanding $\mathcal{P}$ using the results from subsection
\ref{subsect:FisherAnalysis} and substituting them into
(\ref{eq:ModelAssPrior}) we obtain:

\begin{equation}\label{eq:ModelAssPriorExpanded}
\mathcal{P} =
\ln\left(\frac{V_T\sigma^4}{32\pi^2\sqrt{\mathcal{D}}\,\gamma N
\hat{A}^3\,\hat{R}} \right).
\end{equation}

\begin{figure}
\begin{center}
\includegraphics[width=7cm]{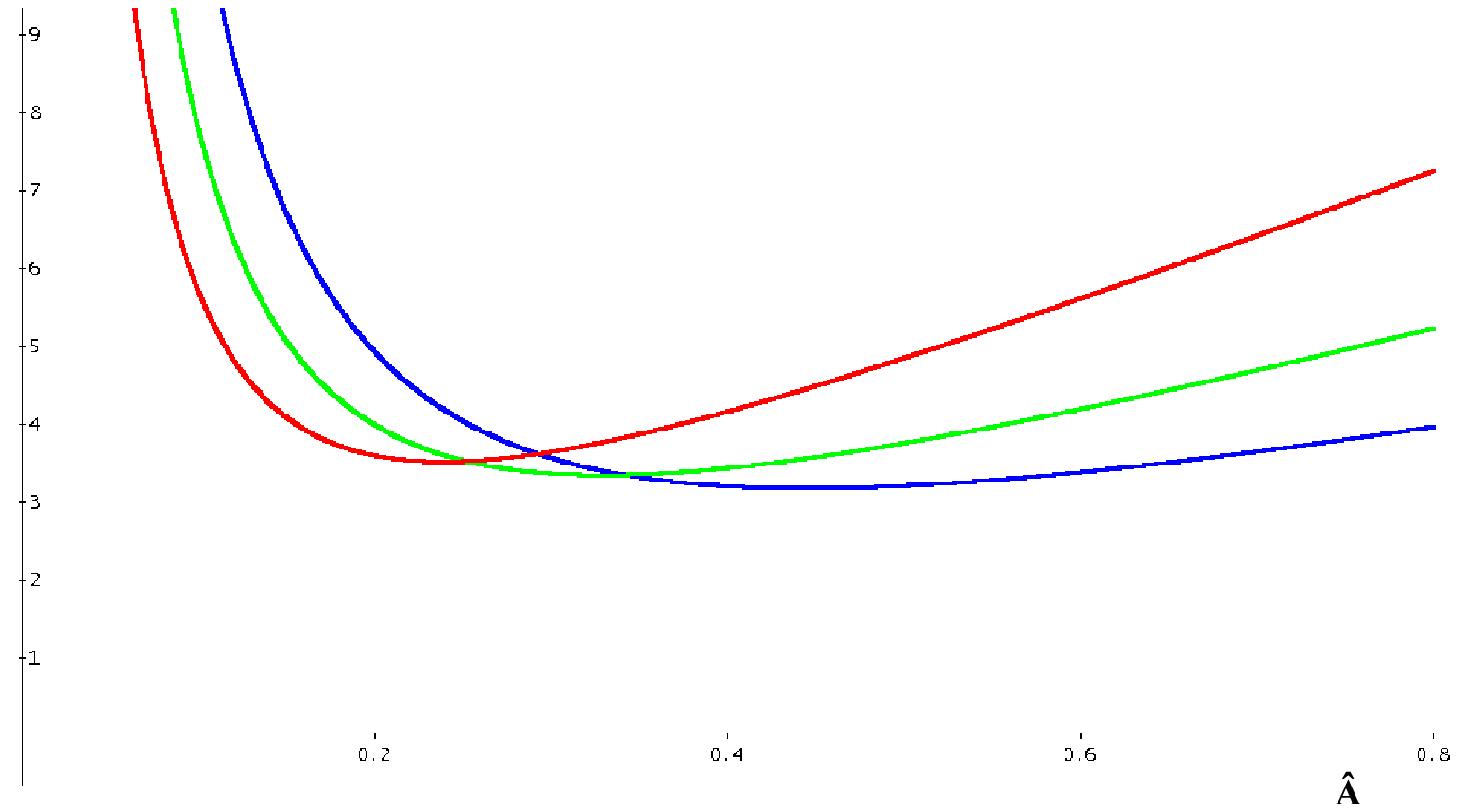}
\caption{Right hand side of inequality (\ref{eq:ModelAssSimpleCorr})
as function of "\emph{estimated source amplitude}" $\equiv\hat{A}$.
Three values of R were used (pixel units): 5 (blue), 7 (green), 10
(red).}\label{fig:MinimumTresholdEstAmpl}
\end{center}
\end{figure}

The curves in figure \ref{fig:MinimumTresholdEstAmpl} show the
dependency of the threshold for detection on the estimated amplitude
$\hat{A}\equiv$ ``estimated source amplitude'', of the putative
source under study. We used data that close mimics the example from
HM03: $V_T=81000~(5 \times 0.5 \times 32400)$, $N = 8$, $\sigma=1$,
$R\in[5,10]$ and $A\in[0.25, 0.75]$. We plot three curves for each
prior with $R$ taking the values
$\{5(\mbox{blue}),7(\mbox{green}),10(\mbox{red})\}$.
\subsection{Detection and characterisation robustness}
The Bayesian framework provides the necessary tools for both of the
primary actions of PowellSnakes: detection and characterization.
However, the impact of a mismatched prior parameter value has
considerably different consequences depending on what we are doing.
According to Jaynes \cite{jaynes}, the Bayesian inference rules
prescribe that when we have a sufficiently informative likelihood
the posterior is only slightly affected by the choice of the priors.
Hence, one may expect that a prior mismatch will only have a minimal
impact when we are dealing with bright sources. Our simulations are
consistent with these predictions. But when we are tackling faint
sources instead, the situation changes. Now the likelihood is no
longer very informative and the priors play a part. Although, when
doing parameter estimation and since we are using flat priors only,
they do not affect the parameter estimates. On the other hand,
something quite different happens when doing the detection step.
When performing model assessment the priors must be fully
normalized. Choosing a wrong prior parameter range can have an
important impact on the quality of the detection, either by
significantly increasing the number of false hits (spurious) or , on
the contrary, making the algorithm become "blind" to detecting
sources (misses). \footnote{In PowellSnakes (Version II) the flat
prior on the source amplitude $A$ has been replaced by a much more
realistic power law: $N(s) \propto s^{-\beta}$,~ and the extreme
sensitivity of the detection step upon the values of the prior
parameters was gone }

\subsection{Quality of the detection: An upper bound}
\label{subsect:UpperLimitISNR}

So far we have only discussed the condition on the detection related to the
tuning of the threshold for acceptance/rejection to minimize the
symmetric loss. We have said nothing about the level of success
achieved. In other words, we know our procedure is tuned in order to
minimize the number of undetected plus spurious sources, but we have
not yet discussed how well we can do. Following Van Trees
\cite{vantrees} the best a "\emph{detector}" can achieve is when we
have absolute prior certitude about the characteristics of the
detection process which is the same as knowing for sure the true
values for all parameters. This condition is also known as
"\emph{simple hypothesis test}" and translates into the following
prior:

\begin{equation}\label{eq:PriorSimpleHypoth}
\pi_0(\va)=\delta(a_1-a_{10})\delta(a_2-a_{20})\ldots\delta(a_i-a_{i0})
,
\end{equation}
where $\delta(x)$ is the Dirac delta function.
Recovering expression (\ref{eq:ModelSelKeyEq}), making $\mu_S =
1\Rightarrow \Pr(H_1)=\Pr(H_0) $, which stands for, no previous
information favours one model over the other $\equiv$ ``\emph{the
data must say it all}'', using the condition for symmetric loss
(\ref{eq:SymmLossCond}), we have that:
\begin{equation}\label{eq:LikeRatioSimple1}
\frac{L_1(\va_0)}{L_0}>1 .
\end{equation}
Taking logarithms on both sides of the inequality and using
expression (\ref{newlike}) we obtain:

\begin{equation}
-\tfrac{1}{2}\vect{s}(\vect{a}_0)^{\rm T}
\vect{N}^{-1}\vect{s}(\vect{a}_0) + \vect{d}^{\rm
T}\vect{N}^{-1}\vect{s}(\vect{a}_0)>0, \label{eq:LikeRatioSimple2}
\end{equation}
where $\vect{a}_0$ is the true parameter vector. The left hand side
of the above inequality is an estimator of $\frac{L_1(\va_0)}{L_0}$.
It is very easy to show that the likelihood ratio estimator
conditioned on ``\emph{there is no source}'',
$\frac{L_1(\va_0)}{L_0}_ {|H_0} \equiv  \Gamma_{|H_0}$
is Gaussian distributed with $N(-\tfrac{1}{2}ISNR_0^2,ISNR_0^2)$,
where $ISNR_0\equiv A_0\sqrt{\alpha_0}$. Hence
\begin{equation}\label{eq:LikeRatioSimple4}
\nu_{|H_0}\equiv\frac{\Gamma_{|H_0}+\tfrac{1}{2}ISNR_0^2}{ISNR_0}\sim
N(0,1) .
\end{equation}
Substituting $\nu_{|H_0}$ into inequality
(\ref{eq:LikeRatioSimple2}), we obtain
\begin{equation}
\nu_{|H_0}>\tfrac{1}{2} ISNR_0 . \label{eq:LikeRatioSimple5}
\end{equation}
Each time this inequality is satisfied we presume we are in the
presence of a source, a true detection. But in our assumptions we
have assumed the opposite. Hence, this is the condition for a
spurious detection. Therefore, the probability for the occurrence of
a spurious detection is:

\begin{equation}\label{eq:LikeRatioSimpleProba}
\Pr(\overbrace{H_1}|H_0) =
\frac{1-erf\left(\frac{\sqrt{2}~~ISNR_0}{4} \right)}{2} ,
\end{equation}
where $\overbrace{H_i}$ means ``\emph{once we have chosen } $H_i$''
and $erf(x)$ is the ``\emph{Gauss error function}''.
\begin{figure}
\begin{center}
\includegraphics[width=7cm]{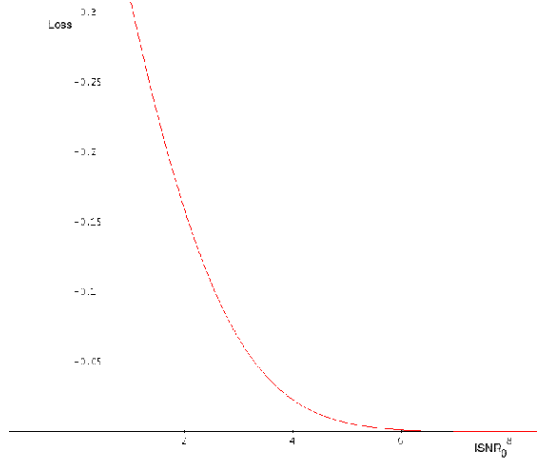}
\caption{Curve for the theoretical lower bound for
$\Pr(Err)\equiv$Loss as a function of $ISNR_0$}
\label{fig:TheoreticalLoss}
\end{center}
\end{figure}
Since the loss is symmetric, the probability for the other type of
error $\Pr(\overbrace{H_0}|H_1)$, the probability that an undetected
source occurs, should match that of the spurious one. Thus,

\begin{equation}\label{eq:LikeRatioSimpleProbaSym}
\Pr(\overbrace{H_0}|H_1)=\Pr(\overbrace{H_1}|H_0) =
\frac{1-erf\left(\frac{\sqrt{2}~~ISNR_0}{4} \right)}{2} .
\end{equation}
We are now in position to establish the lower bound for a detection.
 When performing the detection two independent and
exclusive types of errors can occur:

\begin{equation}\label{eq:LossMinBound1}
\begin{array}{llll}
\mbox{Spurious}     & \Rightarrow & \Pr(\overbrace{H_1}~H_0)=   & \Pr(\overbrace{H_1}|H_0)\Pr(H_0) \\
\mbox{Undetected}   & \Rightarrow & \Pr(\overbrace{H_0}~H_1)= &
\Pr(\overbrace{H_0}|H_1)\Pr(H_1) .
\end{array}
\end{equation}
The total probability for an error to occur is

\begin{equation}\label{eq:LossMinBound2}
\Pr(Err | \va_0) = \Pr(\overbrace{H_1}|H_0)\Pr(H_0) +
\Pr(\overbrace{H_0}|H_1)\Pr(H_1) ,
\end{equation}
where the conditioning in $\va_0$ explicity expresses the assumption of
perfect prior knowledge on the parameters values (``\emph{simple
hypothesis}''). Taking advantage of the symmetry
$\Pr(\overbrace{H_0}|H_1)=\Pr(\overbrace{H_1}|H_0)$ and the
normalization condition $\Pr(H_0) + \Pr(H_1) = 1$, we finally get

\begin{equation}\label{eq:LossMinBound3}
Loss_0\equiv \Pr(Err| \va_0) =
\frac{1-erf\left(\frac{\sqrt{2}~~ISNR_0}{4} \right)}{2} .
\end{equation}

Figure \ref{fig:TheoreticalLoss} shows the dependence of the loss
theoretical lower bound on $ISNR_0$.  One can see that $ISNR_0$
plays a pivotal role in defining an
upper level of the quality of the detection process.

In many situations the radius of the source intensity profile, as
recorded in the pixels map, may be considered known and constant
throughout the patch. One of these cases is that of considering
point shapeless sources. In this case one can further assume, with a
high degree of accuracy, that the pixelized intensity closely
follows the PSF of the antenna. Thus, considering a statistically
homogeneous background, the $ISNR_0$ becomes a function of the
source intensity only, $A_0$, as $\alpha_0 = \vt(R_0)^T
\vect{N}^{-1}\vt(R_0)$ may now be considered constant . This
separation allows us to make predictions about the minimum flux that
one can reliably detect taking into consideration our goals and the
experimental setup. Let us define ``\emph{Normalized Integrated
Signal to Noise Ratio}'' $\equiv NISNR$ which is the $ISNR$ of an
unit amplitude source template.

\begin{equation}\label{eq:NISNR}
NISNR \equiv\sqrt{\alpha} = \sqrt{\vt(R)^T\vect{N}^{-1}\vt(R)}
\end{equation}
This quantity is of considerable importance when establishing a
boundary on the source fluxes that can be reliably detected, as it
provides a scale of measure. One may think of $A$ (the source
amplitude) as being the ISNR measured in NISNR. For a background
correlation matrix considered diagonal (white noise, eq
\ref{eq:2.1}) and the Gaussian source template (\ref{objdef}), an
analytical evaluation of the $NISNR_0$ is possible:

\begin{equation}\label{eq:LossMinBoundWhiteNoiseAlpha}
NISNR_0=\frac{\sqrt{\pi}R_0}{\sigma} .
\end{equation}

In all interesting scenarios one rarely has an allowed parameter
domain which sums up to single point in parameter space
(``\emph{simple hypothesis}''). Now the question is how to define a
single number which somehow describes a representative figure for
the detection quality upper bound of the ensemble of ``\emph{simple
hypothesis}'' that constitute the volume of the parameter domain.  Within a Bayesian
framework we tackle this issue by
treating the simple hypothesis parameters as nuissance parameters and integrating them out.
Hence, our proposed solution is to average the ``\emph{simple
hypothesis}'' loss limit, $Loss_0$, over the properly normalized
probability priors for the parameters:

\begin{equation}\label{eq:LossMinBoundAverage1}
\langle Loss\rangle\equiv\Pr(Err) =
\intop_{\boldsymbol{\mathcal{V}}}\Pr(Err| \va) \pi(\va)~d^N\va,
\end{equation}
Substituting (\ref{eq:LossMinBound3}) into the equation (\ref{eq:LossMinBoundAverage1}) above, and
dropping the zero subscript once it no longer carries any special meaning,
we obtain:

\begin{equation}\label{eq:LossMinBoundAverage2}
\langle Loss\rangle =
\intop_{\boldsymbol{b}}\frac{1-erf\left(\frac{\sqrt{2}~~ISNR(\boldsymbol{b})}{4}
\right)}{2} ~\pi(\boldsymbol{b})~d^N\boldsymbol{b},
\end{equation}
where $\boldsymbol{b}$ stands for the entire parameter set except
the position coordinates. We are implicitly assuming a factorisable
parameter prior (see formula \ref{prior}). The above formula plays a
key role when defining the expected quality for a given detection
and measurement setup. If the predicted detection scenario has a low
average $ISNR$, we should never expect a good detection performance.
By experience we have verified that when tackling real situations
the expected loss is usually much larger than the predicted
theoretical bound.

When considering the case where the prior on the source amplitude,
$\pi(A)$, can be assumed flat on the region of detection,
and the source radius constant across the patch, to a good approximation an analytical evaluation of the above expression
is  possible:

\begin{equation}\label{eq:LossMinBoundAverageAnalyt}
\langle Loss\rangle
=\frac{1}{2(I_M-I_m)}\left[u\left(1-erf\left(\frac{\sqrt{2}~u}{4}\right)\right)-\sqrt{\frac{2}{\pi}}~e^{-u^2/8}\right]_{u=I_m}^{u=I_M},
\end{equation}
where $I_m\equiv$ ``minimum ISNR'' and $I_M\equiv$ ``maximum ISNR''.
\begin{figure}
\begin{center}
\includegraphics[width=7cm]{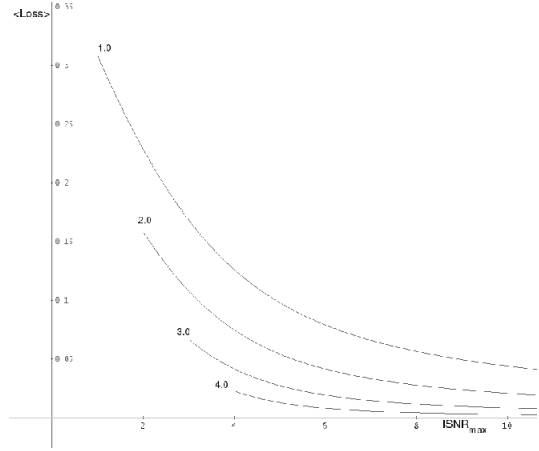}
\caption{Plot of equation (\ref{eq:LossMinBoundAverageAnalyt}). Each
curve represents a different value of $I_m$ shown close to it}
\label{fig:AverageLossAnalytical}
\end{center}
\end{figure}
From figure \ref{fig:AverageLossAnalytical} one realise the
importance of previewing the ISNR limits of our
measurement setup in order to define the expected upper bound on
the detection quality.

\subsection{Discussion on truncation of the posterior by the prior}
\label{truncation}

So far the procedure described (formula \ref{eq:SelectionAprox})
does not take into account the fact that the prior might abruptly
truncate the posterior before it falls (largely) to zero, and hence
may lead to an overestimation of the integral. The purpose of this
discussion is essentially directed towards the truncation of the
posterior which happens in the $\{A,R\}$ sub-space. When considering
the position sub-space,
the prior is independent of the area considered $|s|$, as far as
this fully encompasses the likelihood peak, which is always the case
when dealing with real scenarios.

When we were dealing with the source shape and the beam melted into
a single template of radius $R$ embedded into a background of white
noise only, the effect of truncation was already evident, especially
when were tackling cases of low and very low ISNR
($\equiv\hat{A}\sqrt{\alpha}$), but not extremely severe. This has
provided the rational behind the ``\emph{tuning parameter}'' see
below. The effect was not severe because the extent of the area
under the likelihood fitted fairly well inside the prior box because
the $A$ and $R$ parameters error bars were comparatively small.

\subsubsection{The tuning parameter $\vphi$}
\label{tuning}

When assuming flat priors for all the parameters, the priors become
constants and therefore can be taken out of the evidence integrals.
Thus, one may now use the approximate formula
(\ref{eq:SelectionAprox}) as a fast way of evaluating the evidence
integrals, as far as the truncation effects are kept under control.
But especially when dealing with faint sources we should expect
these effects to become significant and hence they must be taken
into consideration. Therefore, we added a tuning parameter, $\vphi$,
to correct for possible systematic deviations from the true value of
the integral, ie:

\begin{equation}\label{eq:TunParam_1}
 \frac{\vphi~(2 \pi) ^{M/2} |\vC({\hat{\va}})|^{1/2}} {V_{T}} L({\hat{\va}}) = \frac{\int_{\va} L(\va) d^{M} \va}{V_{T}}
\end{equation}
The new expression for the assessment level curve (right hand side
of inequality \ref{eq:ModelAssSimpleCorr})  which results from the
introduction of the ``\emph{tuning parameter}'', is just the old
form where we have only changed the $\mathcal{P}$ expression:

\begin{equation}\label{eq:ModelAssPriorExpanded}
\mathcal{P} =
\ln\left(\frac{V_T\sigma^4}{\vphi\,32\pi^2\sqrt{\mathcal{D}}\,\gamma
N \hat{A}^3\,\hat{R}} \right).
\end{equation}

\subsubsection{How to evaluate the parameter $\vphi$}

In the simple case of white noise it is possible to compute an average
value of the tuning parameter:

\begin{equation}\label{Eq:TunParEval1}
\left < \frac{\int_{V_{\va}} L(\va) d^{M} \va}{L_{0}} \right >
_{H_{1}} = \vphi~(2\pi)^{M/2} |\vC(\hat{\btheta})|^{1/2} \left <
\frac{L(\hat{\va})}{L_{0}} \right >_{H_{1}} ,
\end{equation}
where, $V_{\va} \equiv$ ``\emph{Parameter volume of a single
peak}''. Substituting expression (\ref{newlike}) into
(\ref{Eq:TunParEval1}) we get

\begin{equation}\label{Eq:TunParEval2}
\vphi = \frac{\int_{V_{\va}} \exp \left( \frac{\vs^{T}_{0} \vM^{-1}
\vs(\va)}{\sigma^{2}} - \frac{1}{2} \frac{\vs(\va)^{T} \vM ^{-1}
\vs(\va)}{\sigma^{2}} \right ) d^{4} \va} {(2 \pi) ^{2}
|\vC(\hat{\va})|^{1/2} \exp(\frac{1}{2} \hat{\alpha})} ,
\end{equation}
where ``Average value of the data under hypothesis $H_1 \equiv \left
< \vd \right >_{|H_1}$'' $~=~$ ``signal vector true value $\equiv
s_{0}$''.
The integral in the numerator is difficult to compute, but
restricting our analysis to assuming a white noise background only
$\vM^{-1} = I$, obviates a numerical evaluation of this integral.

\begin{equation}\label{Eq:TunParEval3}
\vphi = \frac{\int_{V_{\va}} \exp \left( \frac{\vs^{T}_{0}
\vs(\va)}{\sigma^{2}} - \frac{1}{2} \frac{\vs(\va)^{T}
\vs(\va)}{\sigma^{2}} \right ) d^{4} \va} {(2 \pi) ^{2}
|\vC(\hat{\va})|^{1/2} \exp(\frac{1}{2} \hat{\alpha})} .
\end{equation}
From expression (\ref{Eq:TunParEval3}), one sees that the value of
$\vphi$ depends on $\vs_0$, the true shape of the object. In
Table~\ref{tuning-I} we display several values of $\vphi$ which
result from the application of the above formula to typical
scenarios ($\mbox{ISNR}_0\equiv A_0 \sqrt{\alpha_0}$). The values we
 obtained for the ``tuning parameter'' are consistent with those
we first expected. When dealing with scenarios with faint
signals $\Rightarrow low~\left\langle\mbox{ISNR}_0 \right\rangle$ we
anticipate the parameter values being close to the priors inferior
bounds and at the same time large error bars are expected, hence the
likelihood truncation effects should make a significant contribution
to the value of the evidence integral. Thus, the low value of $\phi$
should not be a surprise. When the opposite happens
$\Rightarrow high~\left\langle\mbox{ISNR}_0 \right\rangle$ the
Gaussian approximation to the likelihood peak is no longer accurate
and the large $\phi$ is just a consequence of the wider tails of
true likelihood function.
\begin{table}
\caption{Tuning parameter $\vphi$ as function of
$\left\langle\mbox{ISNR}_0\right\rangle$}
\begin{center}
\begin{tabular}{|c|l|} \hline
$\left\langle\mbox{ISNR}_0
\right\rangle$~is small ( $\leq  5$ )&  $\vphi \in (0.66 , 1]$  \\
\hline $5 < $~$\left\langle\mbox{ISNR}_0 \right\rangle$~$< 7$ &
$\vphi \approx$ 1 \\ \hline $\left\langle\mbox{ISNR}_0
\right\rangle$  is large ( $\geq 7$ ) & $\vphi \in (1 , 1.33]$ \\
\hline
\end{tabular}
\end{center}
\label{tuning-I}
\end{table}


\subsubsection{Impact of $\vphi$ on the performance of the algorithm}

 Using a series of simulations (see section \ref{results}) for different detection scenarios, we estimated the impact of the tuning parameter, $\vphi$, on the performance of the algorithm. We concluded that:

\begin{itemize}
\item The algorithm is not very sensitive to changes of $\vphi$.

\item The differences in performance never exceeded $1.5 \%$  when compared with the fiducial value $\vphi = 1$.

\item The largest difference occurred when we dealt with very small values of $\left\langle\mbox{ISNR}_0 \right\rangle(<4)$,  or very large $( > 8 )$ .

\item  And most important of all: ''\emph{The exact values of $\vphi$ always produced the best results}''.

\end{itemize}
Therefore the golden rule displayed in Table~\ref{table:tuning}
should be applied.

\begin{table}
\caption{Tuning parameter $\vphi$; The golden rule}
\begin{center}
\begin{tabular}{|c|l|} \hline
$\left\langle\mbox{ISNR}_0 \right\rangle \leq 5$&                    use     $\vphi = 0.75$\\
$ 5 < \left\langle\mbox{ISNR}_0 \right\rangle < 7$&                             use  $\vphi = 1.00$\\
$\left\langle\mbox{ISNR}_0 \right\rangle \geq 7$&
use    $\vphi = 1.15$\\ \hline
\end{tabular}
\end{center}
\label{table:tuning}
\end{table}


\section{Introducing ``colour''} \label{sect:IntroColour}

When dealing with real astronomical detection problems, the
background diffuse component of the maps is not usually dominated by
the instrumental noise. Nowadays, the most common scenario is the
one where the largest signal component of the background has an
astronomical origin: The ubiquitous CMB and the galactic foregrounds
(free-free, dust, synchrotron). Currently we are focusing on the CMB
especially because it can be considered statistically homogeneous
and isotropic across the entire sphere with great accuracy
(assumptions of our simplified model). The CMB radiation field has
already an intrinsic coherence length ($\sim10'$), which taking into
consideration the current surveys resolution (usually lower than
$1'$ per pixel), may be hardly considered white. In addition, as it
is collected by an antenna with a finite aperture (usually several
pixels), whose net effect, in Fourier space, is that of a low pass
filter, an extra degree of correlation is injected into the
background. Hence, only an algorithm which appropriately deals with
a background whose power spectrum is not flat could possibly cope
with such a setup. The power spectrum of the simplified background
model we proposed (see \ref{eq:IntColourPSpect1}) is not yet fully
realistic. However, despite of its simplicity, this model, already
introduces the main characteristics of this background. This allow
us a direct comparison between the model predictions and the
simulations results. When dealing with a background assumed to be
generated by a stationary Gaussian process but now with an arbitrary
coherence length, the condition \eqref{eq:2.1} on the inverse
correlation matrix no longer holds. Nevertheless, despite the fact
the correlation matrix is no longer diagonal, if the condition of
statistical homogeneity still holds, one may always assume the
background correlation matrix to be circulant. Taking advantage of
the circulant property of the correlation matrix and transforming to
Fourier space, the problem significantly simplifies as the power
spectrum of a circulant correlation matrix is always diagonal. This
is not a severe limitation because considering patches of small size
only, the condition for statistical homogeneity is a good
approximation. We are assuming not only statistical homogeneity but
isotropy as well. Most of the formulas presented here will hold even
if we drop the condition for statistical isotropy. We shall ignore
this fact and we will use the statistical isotropy condition
throughout the remaining text. We do so because it gives us the
opportunity for significant simplifications without restricting the
opportunity for real applications, as this condition holds pretty
well when tackling the great majority of the problems of interest.
Expressing equation \eqref{eq:1.2} (multiplied by -1)  in Fourier
space one obtains
\begin{equation}
\ln\left(L_{H_{1}}\right)=c^{'}+\intop_{\boldsymbol{\eta}}\frac{\tilde{d}(\boldsymbol{\eta})\tilde{\tau}(\boldsymbol{-\eta})}{\mathcal{B}(\eta)}\,
d\mathbf{\boldsymbol{\eta}}-\frac{1}{2}\left(\intop_{\boldsymbol{\eta}}\frac{\tilde{\tau}(\boldsymbol{\eta})\tilde{\tau}(\boldsymbol{-\eta})}{\mathcal{B}(\eta)}\,
d\mathbf{\boldsymbol{\eta}}+\intop_{\boldsymbol{\eta}}\frac{\tilde{d}(\boldsymbol{\eta})\tilde{d}(\boldsymbol{-\eta})}{\mathcal{B}(\eta)}\,
d\mathbf{\boldsymbol{\eta}}\right),\label{eq:33.1}\end{equation}
where the usual $\mathbf{\boldsymbol{k}} = 2\pi  \mathbf{\boldsymbol{\eta}}$.
In a similar way for the likelihood $L_{H_{0}}$ of the background
\begin{equation}\label{eq:ColorLikeH0}
\ln\left(L_{H_{0}}\right)=c^{'}-\frac{1}{2}\left(\intop_{\boldsymbol{\eta}}\frac{\tilde{d}(\boldsymbol{\eta})\tilde{d}(\boldsymbol{-\eta})}{\mathcal{B}(\eta)}\,
d\mathbf{\boldsymbol{\eta}} \right),
\end{equation}
where each of the symbols $\tilde{\tau}$, $\tilde{d}$ are the
Fourier transforms of $\tau$ and $d$ respectively,
and $\mathcal{B}(\eta)$ is the normalized Power Spectrum of the
background, ie, $ \intop_{\boldsymbol{\eta}}\mathcal{B}(\eta) d\mathbf{\boldsymbol{\eta}}=1 \Leftrightarrow \sigma=1$
In what follows the symbols with a tilde on top mean the Fourier
transform of the original symbol (without the tilde). We are
assuming that a pre-normalization of the map's pixels has already
been done by dividing the value of each map's pixels by the map rms
value ($\equiv\sigma$), which implies a value of $\sigma=1$ for the
resulting map.

Let us now introduce, for the first time, correlation effects in the
background diffuse component. This correlation is to be expected not
only because it already exists in the original astronomical
background (CMB), and in the diffuse foregrounds (galactic
components), but also due to the effect of the antenna PSF. We are
considering now a simplified model for the autocorrelation function
of the background, after being passed through the antenna. This
model is of the form:

\begin{equation}\label{eq:IntColourCorrFunct1}
\left\langle \vx_j\vx_i\right\rangle_{b} \equiv
\boldsymbol{n}_{b}(|\vx_j-\vx_i|) =
e^{-\frac{1}{2}\frac{(\vx_j-\vx_i)^2}{s_0^2}},
\end{equation}
where $s_0$ is a measure of the ``\emph{coherence}''  length of the
Gaussian isotropic homogeneous stationary background field (see
Goodman \cite{Goodman})\footnote{There is not a complete consensus
about the exact definition of ``\emph{coherence length}''. We shall
follow Goodman's definition, ``\emph{coherence length}'' $\equiv
\sqrt{\mbox{coherence area}}$, which applied to our particular form
of the autocorrelation function gives $\sqrt{\pi} s_0$}. The
autocorrelation function is already properly normalized as required
by our previous assumptions, ie, $ \sigma^2_{b} = \left\langle \vx_i\vx_i\right\rangle_{b} =\boldsymbol{n}_{b}(0)= 1$
Transforming to Fourier space and using the Wiener-Khinchin theorem
(Goodman \cite{Goodman}) we get:

\begin{equation}\label{eq:IntColourPSpect1}
\mathcal{B}_{b}(\eta)= \frac{1}{2\pi\eta_0^2}~
e^{-\frac{1}{2}\left(\frac{\eta}{\eta_0}\right)^2} ,
\end{equation}
where $\mathcal{B}(\eta)$ is the ``\emph{Power spectrum}'' and
$\eta_0 = \left(2\pi s_0\right)^{-1}$.

Together with the astronomical components one should always expect
the presence of ``\emph{pixel noise}''. In our signal model,
``\emph{pixel noise}'' stands for all the noise components resulting
from the instrumental setup. We are assuming the ``\emph{pixel
noise}'' to be independent from pixel to pixel, hence configuring a
white noise process. We also assume that the``\emph{pixel noise}'' is uncorrelated with all
the astronomical components, thus allowing us to write the total
background power spectrum as the sum of these two components:

\begin{equation}\label{eq:IntColourPSpect2}
\mathcal{B}(\eta) = w_B
\frac{1}{2\pi\eta_0^2}~e^{-\frac{1}{2}\left(\frac{\eta}{\eta_0}\right)^2}
+ w_I
\end{equation}
where $\mathcal{B}_I(\eta)=1$ (white noise process) and we
added the constants $\left\{w_I,w_B\right\}$ in order to allow the
signal components to be present in different amounts, hence
encompassing a broader range of possible scenarios. These constants
must satisfy the normalization condition $w_I+w_B=1$ as the total
power spectrum must integrate to unity.

\subsection{ISNR - the symmetric loss lower
bound}\label{subsect:ColourSymLoss}

\begin{figure}
\begin{center}
\includegraphics[width=7cm]{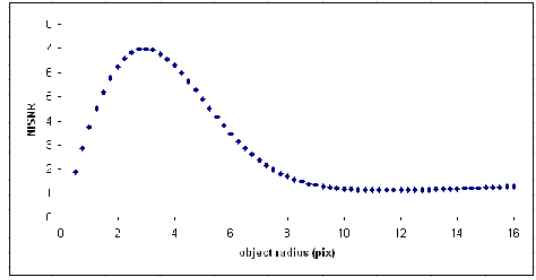}
\caption{NISNR ($\sigma$ units) vs ``source radius'' $R$ (pix) ;
$s_0$ = 11.57 pix, $w_I=1/5$, $w_B=4/5$ } \label{fig:ColorNISNR}
\end{center}
\end{figure}

Let us begin by expressing in Fourier space, the equivalent representation of the
ISNR, using the appropriate form of the Parseval theorem:

\begin{equation}\label{eq:ColorISNR_def}
ISNR(\va)^2=\tau(\va)\vN^{-1}\tau(\va)=\intop_{\boldsymbol{\eta}}\frac{\tilde{\tau}(\boldsymbol{\eta},\boldsymbol{a})\tilde{\tau}(\boldsymbol{-\eta},\boldsymbol{a})}{\mathcal{B}(\eta)}\,
d\mathbf{\boldsymbol{\eta}} .
\end{equation}
First we will handle the simplest possible case where the
radii of the sources are constant. We start by evaluating the
$NISNR$:

\begin{equation}\label{eq:ColorNISNR_def}
NISNR(\va)=\sqrt{\intop_{\boldsymbol{\eta}}\frac{\tilde{t}(\boldsymbol{\eta},\boldsymbol{a})\tilde{t}(\boldsymbol{-\eta},\boldsymbol{a})}{\mathcal{B}(\eta)}\,
d\mathbf{\boldsymbol{\eta}}}
\end{equation}
as this will provide us with a metric to predict how deep we can go in
the flux scale without compromising the detection quality.
Substituting the expression (\ref{eq:IntColourPSpect2}) for the
power spectrum and the unity amplitude Gaussian template
(\ref{objdef}) into (\ref{eq:ColorNISNR_def}) and evaluating the
integral, (which has numerical solution only) and using typical values
for the sources and background parameters we obtain the function plotted in figure
\ref{fig:ColorNISNR}.

When we were dealing with a white noise only background the $NISNR$
was proportional to the radii of the sources (see formula
\ref{eq:LossMinBoundWhiteNoiseAlpha}). However when we introduce the
correlation, a completely different curve results. Now, the
dependence of the $NISNR$ on the object radius stops being linear
and a complex shape emerges.

\subsection{The ``\emph{whitening}'' equalizer}

A quantity which plays a central role in the detection process is
the likelihood ratio of the competing models (see formulas \ref{eq:33.1} and
\ref{eq:ColorLikeH0}). After taking logarithms one obtains:

\begin{equation}\label{eq:WhiteEquLikRatio0}
\ln\left(\frac{L_{H_{1}}}{L_{H_{0}}}\right)=\intop_{\boldsymbol{\eta}}\frac{\tilde{d}(\boldsymbol{\eta})\tilde{\tau}(\boldsymbol{-\eta})}{\mathcal{B}(\eta)}\,
d\mathbf{\boldsymbol{\eta}}-\frac{1}{2}\intop_{\boldsymbol{\eta}}\frac{\tilde{\tau}(\boldsymbol{\eta})\tilde{\tau}(\boldsymbol{-\eta})}{\mathcal{B}(\eta)}\,
d\mathbf{\boldsymbol{\eta}}
\end{equation}
Now let us assume that our source template may be modelled by the
following expression which is a generalization of  equation
\eqref{objdef}:

\begin{equation}
\tilde{\tau}(\boldsymbol{\eta})=\tilde{h}(b_{\beta},\boldsymbol{\eta})\,
e^{-i2\pi\boldsymbol{\eta}.\boldsymbol{r}_{0}},\label{eq:WhiteEquTemplate}
\end{equation}
where $b_{\beta}\equiv\mbox{all variables but position }$,
$i=\sqrt{-1}$, $\boldsymbol{r}_{0}\equiv(x_{0},y_{0})$ and
$\tilde{h}(b_{\beta},\boldsymbol{\eta})$ is the Fourier transform of
the source template centred in the origin. Substituting into
(\ref{eq:WhiteEquLikRatio0}) we get

\begin{equation}\label{eq:WhiteEquLikRatio1}
\ln\left(\frac{L_{H_{1}}}{L_{H_{0}}}\right)=\mathcal{F}^{-1}\left\{\frac{\tilde{d}(\boldsymbol{\eta})\tilde{h}(b_{\beta},\boldsymbol{\eta})}{\mathcal{B}(\eta)}
\right\}_{\boldsymbol{r}_{0}}-\frac{1}{2}\intop_{\boldsymbol{\eta}}\frac{\tilde{h}(b_{\beta},\boldsymbol{\eta})^2}{\mathcal{B}(\eta)}\,d\boldsymbol{\eta}
,
\end{equation}
where $\mathcal{F}^{-1}$ is the inverse bi-dimensional Fourier
transform. One should already be familiar with this expression: the
second term is $1/2$ of the $ISNR$ expressed in Fourier space, which
doesn't depend on the position coordinates, and the first, in the
argument of the inverse Fourier transform, is the ``\emph{Linear
matched filter}'' applied to the map, where:

\begin{equation}\label{eq:WhiteEquMatchFilt}
\mbox{``\emph{Linear matched filter}``}\equiv
\frac{\tilde{h}(b_{\beta},\boldsymbol{\eta})}{\alpha
\mathcal{B}(\eta)} .
\end{equation}
Further we can read this expression in a somewhat different manner.
 Let us write the power spectrum as the product of
its square root (the expected amplitude spectrum), ie
$\mathcal{B}(\eta) =
\sqrt{\mathcal{B}(\eta)}\sqrt{\mathcal{B}(\eta)}$.
Substituting into the argument of the Fourier transform (first term
of \ref{eq:WhiteEquLikRatio1}) we have

\begin{equation}\label{eq:WhiteEquSpliting}
\frac{\tilde{d}(\boldsymbol{\eta})}{\sqrt{\mathcal{B}(\eta)}}\frac{\tilde{h}(b_{\beta},\boldsymbol{\eta})}{\sqrt{\mathcal{B}(\eta)}}
.
\end{equation}
Evaluating the square of the first term of
(\ref{eq:WhiteEquSpliting}) ensemble average, ie its power
spectrum we obtain:

\begin{equation}\label{eq:WhiteEquSplitPSpect}
\left\langle\frac{\tilde{d}(\boldsymbol{\eta})^2}{\mathcal{B}(\eta)}\right\rangle=1
.
\end{equation}
Hence, after applying the linear filter $\mathcal{B}(\eta)^{-1/2}$
we removed the ``\emph{colour}'' from the background
transforming it into a ``\emph{white}'' Gaussian random field. This
is the reason why one calls the linear matched filter  a ``\emph{whitening
equalizer}''. Let us name the resulting map after being
``\emph{whitened}'':

\begin{equation}\label{eq:WhiteEquWhitedMap}
\mbox{\emph{``whitened map''}}\equiv
\tilde{\omega}(\boldsymbol{\eta})\equiv
\frac{\tilde{d}(\boldsymbol{\eta})}{\sqrt{\mathcal{B}(\eta)}} .
\end{equation}
Of course, the sources buried in the map become ``\emph{whitened}''
as well. In fact the second term of (\ref{eq:WhiteEquSpliting}) is
nothing else but  a whitened source:

\begin{equation}\label{eq:WhiteEquWhitedSource}
\mbox{\emph{``whitened source''}}\equiv
\tilde{\psi}(b_{\beta},\boldsymbol{\eta})\equiv
\frac{\tilde{h}(b_{\beta},\boldsymbol{\eta})}{\sqrt{\mathcal{B}(\eta)}}
.
\end{equation}
Rewriting formula (\ref{eq:WhiteEquLikRatio1}) using these new
definitions

\begin{equation}\label{eq:WhiteEquWhiteEquiv}
\ln\left(\frac{L_{H_{1}}}{L_{H_{0}}}\right)=\mathcal{F}^{-1}\left\{\tilde{\omega}(\boldsymbol{\eta})\tilde{\psi}(b_{\beta},\boldsymbol{\eta})
\right\}_{\boldsymbol{r}_{0}}-\frac{1}{2}\intop_{\boldsymbol{\eta}}\tilde{\psi}(b_{\beta},\boldsymbol{\eta})^2\,d\boldsymbol{\eta}
.
\end{equation}
It can be easily verified that this formula, where we used the
``\emph{whitened}'' counterparts of our entities, reduces  to the
familiar white-noise likelihood ratio (formulas
\ref{eq:LikeRatioSimple1} and \ref{eq:LikeRatioSimple2}) expressed
for convenience in Fourier space. Hence, everything we said for the
white noise case applies here, even when we ought to consider the
effects of correlation in the background, as long as we use the
``\emph{whitened}'' entities instead.
\begin{figure}
\begin{center}
\includegraphics[width=7cm]{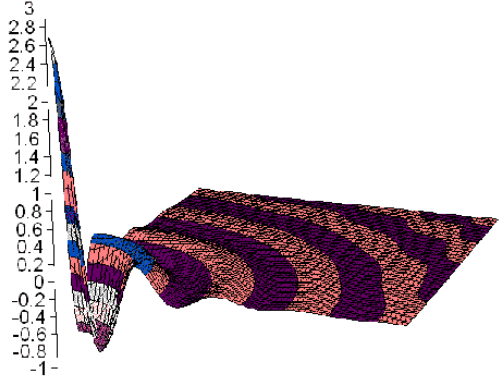}
\caption{``\emph{whitened source}'' $R\ll s_0$ : radius $R=3.4$ ,
$NISNR\approx 23.31$ ; Background parameters $s_0$ = 11.57 ,
$w_I=0.01$, $w_B=0.99$; all distances in pixels}
\label{fig:WhitenedSourceHigh}
\end{center}
\end{figure}
\begin{figure}
\begin{center}
\includegraphics[width=7cm]{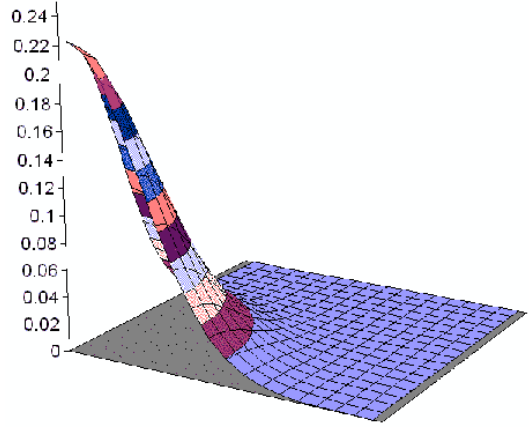}
\caption{``\emph{whitened source}'' $R\approx s_0$ : radius $R=3.4$
, $NISNR\approx 1.47$ ; Background parameters $s_0$ = 2.0 ,
$w_I=1/5$, $w_B=4/5$; all distances in pixels}
\label{fig:WhitenedSourceLow}
\end{center}
\end{figure}

\subsection{Fisher Analysis in Fourier space}\label{subsect:FisherColour}

Proceeding with the same line of reasoning as that initiated in subsection \ref{subsect:FisherAnalysis}, but this time
expressing $L_{H_1}$ using Fourier space (\ref{eq:33.1}), we start
by splitting the evaluation of the Fisher matrix into three different cases:

\begin{itemize}
\item The first case refers to those coefficients whose differentiation variables
$b_{i}$ are not positional. We shall collectively name them as
$\equiv\boldsymbol{C}$ .\begin{equation} \frac{\partial^{2}\left\{
-\ln\left(\mathit{L}_{H_{1}}\right)\right\} }{\partial b_{i}\partial
b_{j}}=-\intop_{\boldsymbol{\eta}}\mathcal{B}(\eta)^{-1}\:\frac{\partial\:\tilde{h}(b_{\beta},\boldsymbol{\eta})}{\partial
b_{i}}\,.\,\frac{\partial\:\tilde{h}(b_{\beta},\boldsymbol{\eta})}{\partial
b_{j}}\, d\mathbf{\boldsymbol{\eta}} .\label{eq:3.4}\end{equation}
These elements of the $\boldsymbol{F}$ matrix will be generally
$\neq0$ .
\item The second case deals with those coefficients where one of the differentiation
variables belongs to set $b_{i}$ and the other to $x_{i}$ , the
position set.\begin{equation} \frac{\partial^{2}\left\{
-\ln\left(\mathit{L}_{H_{1}}\right)\right\} }{\partial b_{i}\partial
x_{k}}=i2\pi\intop_{\boldsymbol{\eta}}\eta_{x_{k}}\:\mathcal{B}(\eta)^{-1}\:\frac{\partial\:\tilde{h}(b_{\beta},-\boldsymbol{\eta})}{\partial
b_{i}}\,.\,\tilde{h}(b_{\beta},\boldsymbol{\eta})\,
d\mathbf{\boldsymbol{\eta}}. \label{eq:3.5}\end{equation} One may
rewrite the expression \eqref{eq:3.5} taking advantage of the
symmetries of the expressions involved:
\begin{equation}
=-4\pi\intop_{\boldsymbol{\eta}>0}\eta_{x_{k}}\:\mathcal{B}(\eta)^{-1}\:
Im\left[\frac{\partial\:\tilde{h^{*}}(b_{\beta},\boldsymbol{\eta})}{\partial b_{i}}\,.\,\tilde{h}(b_{\beta},\boldsymbol{\eta})\right]\,
d\mathbf{\boldsymbol{\eta}},\label{eq:3.6}\end{equation} where
$\boldsymbol{\eta}>0$ means that the integral extends to all the
Fourier modes which have $\eta_{x_{k}}>0$ (half of the Fourier
plane). A necessary condition for those class of coefficients being
equal to 0 is $\tilde{h}(b_{\beta},\boldsymbol{\eta})$ being
real. On the other hand, $h(b_{\beta},\boldsymbol{r})$ having
reflection symmetry on both axes, is a sufficient condition for that
to happen, which is the case for our current model (formula
\ref{eq:1.3}).
\item The third case is when we are dealing with the position parameters
only.\begin{equation} \frac{\partial^{2}\left\{
-\ln\left(\mathit{L}_{H_{1}}\right)\right\} }{\partial x_{k}\partial
x_{j}}=4\pi^{2}\intop_{\boldsymbol{\eta}}\eta_{x_{k}}\eta_{x_{j}}\:\mathcal{B}(\eta)^{-1}\:\tilde{h^{2}}(b_{\beta},\boldsymbol{\eta})\,
d\mathbf{\boldsymbol{\eta}} .\label{eq:3.7}\end{equation} As the
example given above, if $h(b_{\beta},\boldsymbol{r})$ has reflection
symmetry on both axes, this is a sufficient condition for (where we
are ignoring the trivial case $h(b_{\beta},\boldsymbol{r})=0$
)\begin{equation} \frac{\partial^{2}\left\{
-\ln\left(\mathit{L}_{H_{1}}\right)\right\} }{\partial x_{k}\partial
x_{j}}\equiv\left\{ \begin{array}{cc}
=0 & k\neq j\\
\neq0 & k=j .\end{array}\right.\label{eq:3.8}\end{equation} to
happen. This third set of the Fisher coefficients will be named as
$\equiv P$.
\end{itemize}
Ordering the parameters by collecting the position parameters
$x_{k}$ in the upper left corner, the Fisher matrix will then
exhibit a structure of the form displayed ($\boxtimes\equiv$ in
general a non zero value).

\begin{equation}
F_{ij}=\left[\begin{array}{cc}
\begin{array}{ccc}
\boxtimes &  & 0\\
 & P\\
0 &  & \boxtimes\end{array} & 0\\
0 & \left[\begin{array}{ccccc}
\boxtimes &  & \ldots &  & \boxtimes\\
 & \ddots\\
\vdots &  & C &  & \vdots\\
 &  &  & \ddots\\
\boxtimes &  & \cdots &  &
\boxtimes\end{array}\right]\end{array}\right]\label{eq:3.9}
\end{equation}
One may take advantage of the depicted symmetry of the Fisher matrix
when trying to compute its inverse.

\begin{equation}
F_{ij}^{-1}=\left[\begin{array}{ccc}
F_{xx}^{-1} & 0 & 0\\
0 & F_{yy}^{-1} & 0\\
0 & 0 & \left[\boldsymbol{C}^{-1}\right]\end{array}\right]
\label{eq:3.10}
\end{equation}
As an illustrative example and again, with the help of formula
\eqref{eq:2.9} , one may explicitly write the error bars on the
position parameters

\begin{equation}
\Delta X=\Delta
Y=\frac{1}{2\pi^{3/2}\sqrt{\intop_{0}^{\infty}\eta^{3}\mathcal{B}(\eta)^{-1}\:\tilde{h^{2}}(b_{\beta},\eta)\,
d\eta}},\label{eq:3.11}\end{equation} where we have employed the
circular symmetry of proposed template \eqref{eq:1.3}.

\subsection{Extending to real scenarios}
So far we have only considered an extremely simplified model. Our
purpose was two fold:
\begin{itemize}
\item Make the problem practical in terms of required resources.
\item Avoid obscuring and cluttering the fundamentals with complex "\emph{decoration}" details.
\end{itemize}
However, we should always keep in mind that the experimental setup
we wish to address in the future, the Planck Surveyor satellite, and
the object of its study, the full sky, adds other significant
challenges as well. Three of them are worth of further discussion:

\begin{description}
\item[\textbf{\emph{The in-homogeneity of the statistical properties of the
background:}}]
When introducing the galactic emission which is mainly limited to a
narrow band around the galactic plane, a severe breakdown on the
condition of statistical homogeneity occurs. Moreover the
instrumental pixel noise is never completely homogeneous. As result
of the scan strategy there are zones of the sphere which will be
sampled a greater number of times than others, increasing the
integration time and leading to lower noise levels than those of the
areas where beam spends less time. Using small patches ($\sim$
128/256 x 128/256 pixels) and re-computing the background for each
one of them will decrease this problem to a level which can be
safely ignored. Though, it must be noted this is not a limitation of
the Bayesian framework which, on theoretical grounds, is prepared
for dealing with in-homogeneous backgrounds equally well. We only
take advantage of the circulant properties of the covariance matrix,
transforming to Fourier space as an implementation technique which
allows us a much simpler and fast solution to our problem rending it
practical and efficient. However, the same set of assumptions is
required in the derivation of the optimal linear filters together
with an implicit assumption of Gaussianity (HM03).

\item[\textbf{\emph{The non-Gaussianity of the galactic diffuse foregrounds:}}]
The galactic diffuse fields, for instances dust, which can show
considerable values of non-Gaussianity, as result of the Central
Limit Theorem, only introduce extremely small distortions on the
Gaussian distribution of the Fourier modes. The non-Gaussian part of
the field brings in no more than small correlations between the
different Fourier modes which can be securely ignored when compared
for instances with the potentially much more serious non-homogeneity
of the background (HM03) (Rocha et al.\cite{GracaNGSims}).
\item[\textbf{\emph{The "confusion" noise as result of the contribution to the
background of the faint unresolved point sources:}}] This is the
dominant component of the background of the Planck high frequency (
545 and 857 GHz) maps on high galactic latitudes. This component may
be modelled as a Poissonian sampling process with a constant rate
source. Any realistic background model should take this into
account.
\end{description}



\section{Results}
\label{results}

We tested the presented ideas performing an extensive set of
simulations using two different toy models and different detection
scenarios (the latter solely for the white noise only model (see
subsection \ref{subsect:ResWhiteOnly})). In both cases we used a
Gaussian source template (\ref{objdef}). The parameters for each
source were drawn from uniform distributions with bounds given in
table \ref{table:models}. The patches are square grids of 200 x 200
pixels with a border of $2\times$ the largest source radius, in
pixels. The sources were spread uniformly throughout the patch. Care
was taken to avoid source clustering and the presence of sources in
the patch's borders. We imposed a safe distance between sources by
defining a ``\emph{bounding}'' square, centred on the source with
side $= \gamma \times 2 \times R$ pixels wide, where $R\equiv$
``\emph{Radius of object}'' (see table \ref{table:models} for the
values of $\gamma$). We forbade the overlapping of these
``\emph{bounding}'' squares.

\subsection{Performance of the algorithms}

We evaluated the performance of the different algorithms by
computing the ``\emph{Total Error}'' defined as:
\[ \text{`\emph{Total Error}' = Percentage of `\emph{undetected sources}'+ Percentage of \emph{`spurious detections}'}\]
where:
\begin{itemize}
\item A 'undetected source'  happens  when: a  simulated object is not recognized by the algorithm as a source.
\item  A 'spurious detection'  happens when:
\begin{itemize}
\item An object is detected
\item There is no simulated object such that the entire set of its parameters belongs to the volume defined by the estimated parameters of the detected object $\pm 3 \sigma_{p}$.
\end{itemize}
\item The `\emph{Percentage}' is computed dividing the quantities by the total number of sources in the
patch.
\end{itemize}
The algorithm that minimizes the `\emph{Total error}' is the one
that performs better.

\subsection{White noise background}\label{subsect:ResWhiteOnly}

In Table \ref{table:res-priors} we show  the performance of
PowellSnakes, where, $\phi$ is the tuning parameter.
The sources are imbedded into a background which is a stationary
Gaussian random field with a coherence length much smaller than one
pixel, usually known as Gaussian white noise process.

Two scenarios were tested:
\begin{itemize}
\item HMcL (Hobson-McLachlan). With this example we tried to mimic, as exactly as possible, the
HM03 ``\textsf{Toy problem 4.3}''
\item LISNR (Low Integrated Signal to Noise Ratio). This example was
inspired in another from HM03, where a SZ example was
given, but we have only retained the range of the $ISNR$ involved.
We did not try to simulate the SZ case at all as we have always
worked with the Gaussian source template.

Two situations were investigated:
\begin{itemize}
\item With a fixed number of sources per patch: ``\textbf{fix}''
\item With a variable number of sources per patch: ``\textbf{variable}''.
In  this case the number of sources for each patch was drawn from an
uniform distribution whose minimum limit was zero and the maximum
was the value displayed in table \ref{table:models} ($N_{objs}$)
\end{itemize}
\end{itemize}
Table \ref{table:models} has a summary of the parameters' values,
background and foregrounds (sources), for each of the simulated
examples. Figure \ref{fig:LikeWhiteNoise} represents a typical
likelihood manifold for this case ($A$ and $R$ dimensions suppressed).

Each run (map complete cleanup) takes less than a second ($\sim $ 2
each second) on a regular PC (Pentium IV, 2.39 GHz). A considerable
effort to optimize the code was made. An intelligent management of
previously computed values was essential in achieving such an high
performance. Provisions for parallelizing the code were made but not
yet employed.

\begin{table}
\begin{center}
\begin{tabular}{|c|c|c|c|}
\hline Type of simulation & LISNR-16-fix & LISNR-16-var& HMcL \\
\hline $\left\langle ISNR \right\rangle$ & 3.84 & 3.62 & 4.68 \\
\hline $\phi$ & 0.75 & 0.66 & 0.66 \\
\hline $N_ {simul}$ & 5000 & 1000& 10000 \\
\hline $\%$ detections & 67.41 \% & 56.41 \% & 82.95 \% \\
\hline $\%$ spurious & 9.60 \% & 8.62 \% & 8.19 \% \\
\hline Total error & 42.19 \% & 52.20 \% & 25.15 \%
\end{tabular}
\end{center}
\caption{Summary of the results from the simulations with a white
noise background.$N_{\text{simul}}$ is the number of patches which
have been simulated, $\phi$ is the tuning parameter}
\label{table:res-priors}
\end{table}

\begin{table}
\begin{center}
\begin{tabular}{|c|c|c|c|} \hline
\multicolumn{4}{|c|}{Detection models}\\ \hline &LISNR-16-var &
LISNR-16-fix & Hobson-McLachlan \footnote{astro-ph/0204457 (HMcL)}\\
\hline $R_{max}$ (pixels)  & 5.4 & 5.4 & 10 \\ \hline $R_{min}$
(pixels) & 3.21& 3.21 & 5 \\ \hline $A_{max}$ & 0.76 & 0.8 & 1 \\
\hline $A_{min}$  & 0.2 & 0.22 & 0.5 \\ \hline $\sigma$ & 1 & 1 & 2
\\ \hline $N_{objs} $& 16 (variable) & 16 (fixed) & 8 (variable)
\\ \hline $ \gamma$ & 1.5 & 1.5 & 2.5 \\ \hline
\end{tabular}
\end{center}
\caption{Detection models parameters, where the $R_{max}$ and
$R_{min}$ are the maximum and minimum radius respectively; and
$A_{max}$ and $A_{min}$ are the maximum and minimum amplitude
respectively; The radii units are pixels and the source amplitudes
units are given in the same unit as $\sigma$} \label{table:models}
\end{table}%

\subsubsection{Discussion}

One of our major goals when we started this project was making
Bayesian methods run fast enough in order to allow us to collect
significant statistics figures by performing massive simulations in
a reasonable amount of time. We have fulfilled our goal, as you may
verify in table \ref{table:res-priors}, several thousands of
simulations were performed in each proposed scenario making our
figures statistically sounding. Our results are consistent with the
theoretical framework presented here. Examples with lower ISNR show
higher levels of ``\emph{Total error}'' as expect and significantly
higher than the lower bound (see subsection
\ref{subsect:UpperLimitISNR}). Cases with a fixed number of sources
per patch always show a lower error. This should not be a surprise
as in our priors we have assumed a constant expected number of
sources per patch and the lower the dispersion in the number of
sources the better the prior fitted the data, hence the higher
performance.
Worth of a reference is the ``\emph{Tuning parameter}'' (subsection
\ref{truncation}). In table \ref{table:res-priors} we are only
showing results achieved using the  ``\emph{Tuning parameter}''
values which performed better. All the presented examples are low
ISNR cases, thus we have only employed values lower than 1 (see
table \ref{table:res-priors}). These ``\emph{best choice}'' values
are consistent with our predictions.


\subsection{``Coloured'' background}\label{subsect:ResColour}

\begin{figure}
\begin{center}
\includegraphics[width=7cm]{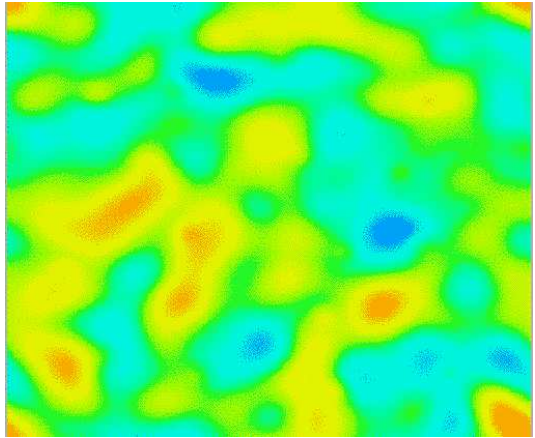}
\caption{``\emph{Coloured}'' background - astronomical component;
Power spectrum was modelled using formula
(\ref{eq:IntColourPSpect1}); $s_0 = 11.57 \text{pixels}$}
\label{fig:CMBBackground}
\end{center}
\end{figure}

\begin{figure}
\begin{center}
\includegraphics[width=7cm]{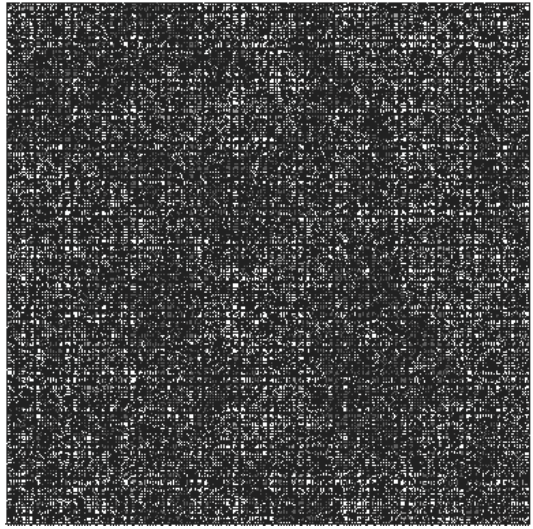}
\caption{``\emph{Coloured}'' background - complete; astronomical
component $10~\mu K $ , pixel noise $20~\mu K \Rightarrow w_I=4/5,
w_B=1/5$} \label{fig:ColouredBackground}
\end{center}
\end{figure}

\begin{figure}
\begin{center}
\includegraphics[width=7cm]{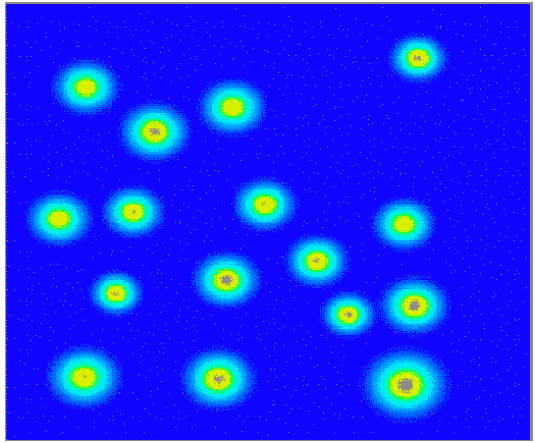}
\caption{ The simulated sources. The shape of the sources was
assumed Gaussian (\ref{objdef}); all the parameters, $\left\{A,R,X,Y
\right\}$, were drawn for uniform distributions (check the text for
details)} \label{fig:Sources}
\end{center}
\end{figure}

Our arrangement for the ``\emph{coloured}'' background simulations
followed the same guidelines as those for the white noise only
case. A greater care concerning the realism of the setup was taken
into consideration. We used a pixel size of $1'\times1'$, and small
patches of 200 x 200 pixels ($ \sim 3.33^{\circ} \times
3.33^{\circ}$). The background consisted of two uncorrelated
components:
\begin{itemize}
\item A stationary Gaussian random field with a Gaussian power spectrum.
The power spectrum, already smoothed by the antenna, was modelled by
formula (\ref{eq:IntColourPSpect1}) with a coherence length scale
parameter of $s_0=11.57$ pixels  (see figure
\ref{fig:CMBBackground})
\item Pixel noise with a ``\emph{white}'' profile with constant level across the map
\item The CMB RMS level used was $10 \mu K $ and the pixel noise $20 \mu K$ which together implied $w_I=4/5$ and
$w_B=1/5$ (see expression \ref{eq:IntColourPSpect2}, figure
\ref{fig:ColouredBackground} )
\end{itemize}
The simulated sources had amplitudes ranging from 0.9 to 1.0, in
normalized units (normalized by the total rms noise level). The
radii of the sources were inside the interval [3.2, 5.0] pixels. The
whole setup results in an average $\left\langle ISNR \right\rangle $
of $\simeq 3.84$. In each simulated map we put 16 objects with their
parameters drawn from uniform distributions (whose limits are given
above) . Following the same prescription as for the white noise only
case, we prevented the sources from forming clusters and being
placed on the patch borders (see figure \ref{fig:Sources}) . Table
\ref{table:res_colour} contains the results obtained. We have
performed a total of 5000 simulations.

\begin{figure}
\begin{center}
\includegraphics[width=7cm]{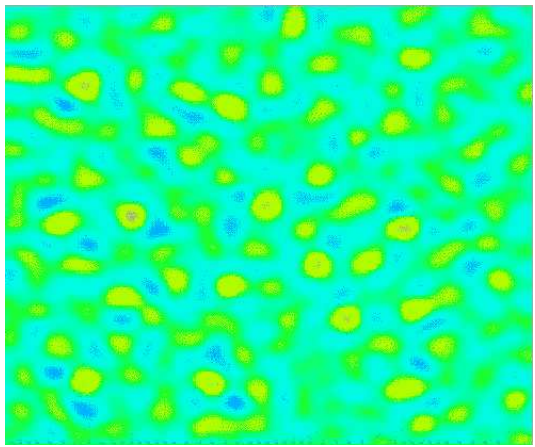}
\caption{Likelihood ratio $\frac{L_{H_1}}{L_{H_0}}$ ($A$ and $R$
dimensions suppressed) top view; The sources are the same as in
figure \ref{fig:Sources}; the background parameters are those from
the simulations: $s_0=11.57
~\text{pixels},~w_I=4/5,~w_B=1/5,\left\langle NISNR \right\rangle
\sim 3.84$} \label{sim:filter}
\end{center}
\end{figure}

\subsubsection{Discussion}

When we introduced correlation in the background, the likelihood
manifold has now considerably changed from the white noise only
example. When keeping $\left\{A, R\right\}$ constant, looking at the
position subspace one may see that the likelihood maxima are now
extremely narrow and high and they have become surrounded by
subsidiary peaks of considerable height (see figure
\ref{fig:LikelihoodColourBaseDetail}).
\begin{figure}
\begin{center}
\includegraphics[width=7cm]{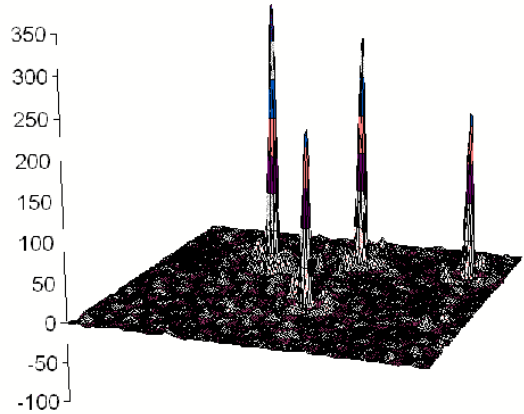}
\caption{``\emph{Coloured background}'' - The likelihood ratio
$\frac{L_{H_1}}{L_{H_0}}$ manifold position subspace. Example where
the instrumental noise is very low; background parameters:
$s_0=11.57 ~\text{pixels},~w_I=0.01,~w_B=0.99,~NISNR \sim 23.31$}
\label{fig:LikelihoodColourBase}
\end{center}
\end{figure}
This scenario is even more acute when the proportion of instrumental
noise becomes lower in relation to the ``\emph{coloured}''
component. In figure (\ref{fig:LikelihoodColourBase}) we depict an
exaggerated case (for clarity) where the instrumental noise is only
$10\%$ of the astronomical component. Nevertheless we should stress
that current detection instrumentations are already getting very
close to these values. The Planck mission LFI instrument (Planck
``\textit{Blue book}'' \cite{bluebook}) has a previewed average
ratio of $\sim20\%$ between the astronomical components and the
detectors instrumental noise. Though a more balanced proportion
should be expected on the majority of existing instrumental setups.
When dealing with such scenarios the search for the likelihood
maxima becomes harder. The likelihood peaks being thin and
completely surrounded by a very high number of other fake peaks are
extremely difficult to find. One now needs to unfold a much higher
number of ``\emph{snakes}'', typically four times the number of the
white noise only case. The use of the ``\emph{jump procedure}'' is
now mandatory in order to pass through the large accumulation of
fake peaks which encircles the real ones.
Since the likelihood peaks are now extremely thin, the error bars on
the source localization are now extremely small, usually much less
than a pixel. This is the equivalent of assuming that the estimated
values for the position parameters perfectly match the true ones.
When we start employing the Bayesian procedure for validation of the
detection we should replace the flat position prior (see
\ref{eq:E1PeakPrior}) by a Dirac delta centred on the values
previously estimated (see \ref{eq:PriorSimpleHypoth}) for the
position of the source:

\begin{equation}\label{eq:PriorSimpleHypoth}
\pi_0(\va)=\frac{\delta(x-x_{0})\delta(y-y_{0})}{\Delta  A\Delta R}
\end{equation}

As a final note on the speed of the code execution we must note the
substantial decrease of performance when dealing with the
``\emph{coloured background}''. The average time for completing a
full patch is now about 8 seconds which is about an order of
magnitude more than in the previous case. Despite this significant
reduction, our initial purpose (performing massive amount of
simulations) has not been compromised as one can still undertake
thousands of simulations in a reasonable amount of time.

A summary of the sequence of steps performed by  PowellSnakes, Version I (the version here exposed), is given in Appendix A.



\section{Conclusions}\label{sect:conclusions}

The detection and characterisation of discrete objects is a common
problem in many areas of astrophysics and cosmology.  When
performing this task most methods assume a smoothly varying
background with a characteristic scale or coherence length much
larger then the scale of the discrete objects under scrutiny.
However high-resolution observations of the cosmic microwave
background, CMB, pose a challenge to such assumptions. CMB radiation
has a coherence scale length of the order of $\sim 10$ arcmin, close
to that of the objects of interest such as  extragalactic `point'
sources or the Sunyaev-Zel'dovitch effect in galaxy clusters.   In
addition the instrumental noise levels can be greater than the
amplitude of the discrete objects. The common approach for dealing
with such difficulties is to apply a matched filter to the initial
map and instead analyse the filtered map. These approaches have
reported good performances. However the  filtering process is only
optimal among the limited class of linear filters and is dissociated
from the subsequent object detection step of selection performed on
the filtered maps. Hobson \& McLachlan (\cite{mikecharlie}; HM03)
were the first to introduce  a Bayesian approach to the detection
and characterization of discrete objects in a diffuse background.
Its implementation was performed using Monte-Carlo Markov chain
(MCMC) sampling from the posterior which was very computationally
expensive. This prompted us to
 explore a new, fast method for performing Bayesian object detection in which sampling is replaced by multiple local
maximisation of the posterior, and the evaluation of errors and
Bayesian evidence values is performed by making a Gaussian
approximation to the posterior at each peak. In this paper we
presented such a novel approach, PowellSnakes.
We start by pre-filtering the map and explicitly show that for a
given value of $R$ (radius of the source), peaks in the filtered
field correspond to peaks in the log-likelihood considered as
function of the position of the sources. We replace the standard
form of the likelihood for the object parameters by an alternative
exact form that is much quicker to evaluate. Next we launch $N$
downhill minimizations in the 2-dimensional ($X$,$Y$) using the
Powell algorithm. In order to avoid local maxima in the likelihood
surface we devised the so called `jump procedure', a  new concept
borrowed from the realm of quantum mechanics. Next we perform
multiple Powell minimizations in the full 4-dimensional space
($X$,$Y$,$A$,$R$), starting each minimization from the positions
found in the previous step. To evaluate the performance of an
algorithm we introduced here for the first time the concept of
`symmetric loss' within the context of the proposed subjects. The
Peak-based prior, $P$, was proposed and an upper bound on the
quality of a detection presented. We showed that the quantity,
`Normalised Integrated Signal to Noise Ratio',  NISNR , plays an
important role when establishing a boundary on the source fluxes
that can be reliably detected. We included a discussion on the
truncation of the posterior by the prior and proposed a novel
solution by introducing the `tuning parameter', $\phi$.
 We studied in detail the consequences of adding correlation into the diffuse background which makes the substructure where the sources are imbedded.
 As the Linear filter works as a whitening equalizer, the coloured noise case reduces to the white noise as long as we use the `whitened' map and sources.
Using simulations of several typical scenarios we showed that our
approach performs very well on both white and coloured background.
We found that our results are consistent with our theoretical
framework. For the white noise background, cases with a lower,
`Integrated Signal to Noise Ratio', ISNR, exhibit higher levels of
`Total Error' as expected. Cases with a fixed number of  sources per
patch show a lower error.
The number of spurious detections increases with this prior however
the number of detected sources increases by a larger amount
resulting in a lower `Total error'. When we introduce colour in the
background the likelihood changed considerably: the maxima in the
position subspace are now extremely narrow and high surrounded by
subsidiary peaks of considerable height. The search for the
likelihood maxima becomes harder: one has to resort to a larger
number of `snakes'. The usage of the jump procedure becomes
mandatory.
In this case the errors in the source localization are very small,
usually much less than a pixel. This means that we can replace the
flat position prior by a Dirac delta function centred on the values
previously estimated.

Furthermore this approach yields a speed-up over sampling-based
methods of many 100s, making the computational complexity of the
approach comparable to that of linear filtering methods. Here we
presented  PowellSnakes, in its  first incarnation:`PowellSnakes I'.
An account of an updated implementation, `PowellSnakes II'  will be
published shortly.  The application of the method to realistic
simulated Planck observations will be presented in a forthcoming
publication.

\section*{Acknowledgements}

PC thanks the Cavendish Astrophysics Group of the University of Cambridge for support and hospitality during the progression of this work.
GR acknowledges support from the US Planck Project, which is funded by the NASA Science Mission Directorate.
GR would like to aknowledge useful discussions with Krzystof G\'{o}rsky and Charles Lawrence.

\section{Appendix A}

PowellSnakes is a new fast Bayesian approach for the detection of
discrete objects immersed in a diffuse background.  This new method
speeds up traditional Bayesian techniques by:

\begin{itemize}
\item replacing the standard form of the likelihood for the parameters
characterizing the discrete objects by an alternative exact form
that is much quicker to evaluate;

\item using a simultaneous multiple
minimization code based on Powell's direction set algorithm to
locate rapidly the local maxima in the posterior; and

\item deciding
whether each located posterior peak corresponds to a real object by
performing a Bayesian model selection using an approximate evidence
value based on a local Gaussian approximation to the peak. The
construction of this Gaussian approximation also provides the
covariance matrix of the uncertainties in the derived parameter
values for the object in question.

This new approach provides a
speed up in performance by a factor of `hundreds' as compared to
existing Bayesian source extraction methods that use MCMC to explore
the parameter space, such as that presented by Hobson \& McLachlan
\cite{mikecharlie}. The method can be implemented in either real and
Fourier space. In the case of objects embedded in a homogeneous
random field, working in Fourier space provides a further speed up
that takes advantage of the fact that the correlation matrix of the
background is circulant.
Its performance is found to be comparable if not better to that of
frequentist techniques such as applying optimal and wavelet filters
. Furthermore PowellSnakes has the advantage of consistently
defining  the threshold for acceptance/rejection based on priors,
the same cannot be said of the frequentist methods.

\end{itemize}

Let's start by recapitulating that the {\it Powell's} method is the
prototype of  "direction-set methods" for maximization or
minimization of functions. These are the methods to apply when you
cannot easily compute derivatives. However this method requires a
one-dimensional minimization sub-algorithm, eg. {\it Brent's}
method. The {\it Brent's} method is an interpolation scheme whereby
you alternate between a parabolic step and golden sections. It is
used in one-dimensional  minimization problems without calculation
of the derivative. (for more details see Numerical Recipes
\cite{NR}, pg 389, 395, 406).

\noindent
The steps followed in PowellSnakes, Version I, are:
\begin{enumerate}
\item Compute the classical linear matched filter $ \hat{A} = \frac{t(\vx)^{T} N^{-1} \vd}{t(\vx)^{T} N^{-1} t(\vx)}$ for the whole patch in the Fourier domain - the denominator is constant and needs to be evaluated only once which can be done in Fourier space as well
\item Compute all the constants
\item Fourier transform the map - a 4Mpix map takes at most 3 secs (FFTW)
\item Filter the map ($\simeq 2$ secs)
\item transform back to the map space using the inverse fourier transform ($\simeq 3$ secs)
\item Evaluate the $ln (\frac{P(H_{1}| \vd)}{P(H_{0}| \vd)}) $ map ($\simeq 2$ secs)
\item Find the peaks of  $ln (\frac{P(H_{1}| \vd)}{P(H_{0}| \vd)}) $ field - check if they are higher than a certain acceptance level: find the peaks using a Powell minimizer with jumps. The evaluation of the objective function is fast - use a kind of evaluation cache (pick a value from a precalculated array).\\
However  we still need to find the optimal radius, positions and amplitude in real space. When using the matched filter an average value for the radius of the source has been used and the values of the peaks are only approximate.  Now we need to find the optimal parameters.
  \item Error bars are calculated from approximate versions of the analytic formulae (while MCMC provides these errors automatically)
 \item The stop criteria - no stop criteria - however there is a way of ensuring that no peak is le
 ft behind - we proceed as follows: first of all we can have an idea of the area (not the volume) under the likelihood peaks.  Let's consider now a imaginary pixel of the size of this area. Split the map into this imaginary pixels and start a PowellSnakes search in the center of each one of them. It is unlikely to miss any of the significant likelihood maxima (specially if we use the jump technique). One might think this process to be time consuming but in fact this is not the case since the objective function is pre-calculated.
\end{enumerate}

\subsection{Characteristics}

Let's now characterize PowellSnakes in terms of its speed and sensitivity.

\subsubsection{Speed}

The PowellSnakes algorithm is a two step process, in the first part of the algorithm we proceed as follows:

\begin {enumerate}
\item First the map is filtered with a linear matched filter given by:
\begin{equation}
\hat{A}(\vX,R) = \frac{\vt^{\rm T}(\vX,R)\vN^{-1}\vd}
{{\vt}^{\rm T}(\vX,R)\vN^{-1}\vt(\vX,R)}.
\label{mfahat2}
\end{equation}
 The evaluation is done in Fourier space using the average radius of the possible range of radiae. The map is Fourier transformed again back to real space. If the radiae of the objects were all equal, and if we knew their true value, the resulted filtered field would be proportional to the likelihood field with the absolute maxima on top of the objects. When the available range of radiae is small, filtering with the `average' matched filter will not produce absolute maxima but in the great majority of the situations a local peak is created over the objects.
\item Next we search this map with the Powell minimizer in two dimensions (the position) with the jumps switched on. This is an extremely fast operation because the objective function is already pre-calculated, and we just need to pick up a value from an array, or better, compute a bilinear interpolation.
\end{enumerate}
In the second part of the algorithm:
\begin{enumerate}
\item  Now we already have a good idea of where to look further:  Around the maxima of the likelihood restricted to the subspace of the position of the objects. In step 1) we produced a `whitened' version of the map, and we `whitened' the test objects, at the same time. At this point we switch off the jumps, and launch `PowellSnakes' in all  four dimensions (position, radius and amplitude), taking as starting points the previously found maxima restricted to the position subspace. Use the complete expression for the likelihood (no analytical shortcuts).
\item Once Powell minimizer finishes, pick up the optimal values for the parameters and test them for acceptance/rejection. If and only if the maximum is accepted as `good', subtract the object from the map. (This is different from the our old solution: sometimes, when we subtracted a rejected peak in the neighbourhood of a true but still undetected object, we ended up damaging the good peak. Unfortunately most of the times that is enough to prevent the real object to be detected. Right now we only subtract accepted peaks to avoid multiple detections of the same object. We think we should try to avoid the subtraction at all, since there is always the risk that a spurious detection and consequent subtraction will damage a good peak.)
\end{enumerate}

\subsubsection{Sensitivity}

Now, let's consider the Sensitivity:
\begin{enumerate}
\item Analytical solutions are in general great, however when noise is taken into consideration the functions become hard to handle. Furthermore most of the theorems don't apply in this case. Therefore one should try to avoid analytical solutions.
\item Sometimes the estimated values of the parameters fall off the allowed range used to define the Bayesian criterion for acceptance. Therefore  even if the error bars allow them, we don't accept them instead send them for testing (accept/reject), or reject them immediately. Rather you should attach them a very low probability and let the computation continue.
\item PowellSnakes depends on several parameters in order to optimise its performance / sensitivity: ie, the number of launched "snakes", average number of sources per patch, etc.
Since the detection conditions (background, noise, signal-to-noise
ratio, etc.) can exhibit some drastically changes across the sphere,
these parameters should be occasionally re-adjusted. In its current
implementation PowellSnakes allows the definition of sphere zones.
Within each zone a different set of these parameters may be loaded
from the parameter file in order to tune its sensitivity /
performance.
\item Do not expect miracles: If the NISNR is low you cannot detect reliably.
\end{enumerate}

\subsubsection{Other technical details}

\begin{enumerate}
\item The structure of the program was designed to be ready to take full advantage of a multiprocessor system. The performance will scale linearly with the number of processors.
\item Implemented in ISO C++.
\item Completely built on templates. This feature allows you to consider the precision you want.
\item Designed as a `framework', ie you do not need to understand how PowellSnakes works to use it. Inherit from its classes, customize it, and PowellSnakes will call your code when needed.
\end{enumerate}



\label{lastpage}
\end{document}

%% file: rgb.tex
\definecolor{AliceBlue}{rgb}{0.94,0.97,1.00}
\definecolor{AntiqueWhite1}{rgb}{1.00,0.94,0.86}
\definecolor{AntiqueWhite2}{rgb}{0.93,0.87,0.80}
\definecolor{AntiqueWhite3}{rgb}{0.80,0.75,0.69}
\definecolor{AntiqueWhite4}{rgb}{0.55,0.51,0.47}
\definecolor{AntiqueWhite}{rgb}{0.98,0.92,0.84}
\definecolor{BlanchedAlmond}{rgb}{1.00,0.92,0.80}
\definecolor{BlueViolet}{rgb}{0.54,0.17,0.89}
\definecolor{CadetBlue1}{rgb}{0.60,0.96,1.00}
\definecolor{CadetBlue2}{rgb}{0.56,0.90,0.93}
\definecolor{CadetBlue3}{rgb}{0.48,0.77,0.80}
\definecolor{CadetBlue4}{rgb}{0.33,0.53,0.55}
\definecolor{CadetBlue}{rgb}{0.37,0.62,0.63}
\definecolor{CornflowerBlue}{rgb}{0.39,0.58,0.93}
\definecolor{DarkBlue}{rgb}{0.00,0.00,0.55}
\definecolor{DarkCyan}{rgb}{0.00,0.55,0.55}
\definecolor{DarkGoldenrod1}{rgb}{1.00,0.73,0.06}
\definecolor{DarkGoldenrod2}{rgb}{0.93,0.68,0.05}
\definecolor{DarkGoldenrod3}{rgb}{0.80,0.58,0.05}
\definecolor{DarkGoldenrod4}{rgb}{0.55,0.40,0.03}
\definecolor{DarkGoldenrod}{rgb}{0.72,0.53,0.04}
\definecolor{DarkGray}{rgb}{0.66,0.66,0.66}
\definecolor{DarkGreen}{rgb}{0.00,0.39,0.00}
\definecolor{DarkGrey}{rgb}{0.66,0.66,0.66}
\definecolor{DarkKhaki}{rgb}{0.74,0.72,0.42}
\definecolor{DarkMagenta}{rgb}{0.55,0.00,0.55}
\definecolor{DarkOliveGreen1}{rgb}{0.79,1.00,0.44}
\definecolor{DarkOliveGreen2}{rgb}{0.74,0.93,0.41}
\definecolor{DarkOliveGreen3}{rgb}{0.64,0.80,0.35}
\definecolor{DarkOliveGreen4}{rgb}{0.43,0.55,0.24}
\definecolor{DarkOliveGreen}{rgb}{0.33,0.42,0.18}
\definecolor{DarkOrange1}{rgb}{1.00,0.50,0.00}
\definecolor{DarkOrange2}{rgb}{0.93,0.46,0.00}
\definecolor{DarkOrange3}{rgb}{0.80,0.40,0.00}
\definecolor{DarkOrange4}{rgb}{0.55,0.27,0.00}
\definecolor{DarkOrange}{rgb}{1.00,0.55,0.00}
\definecolor{DarkOrchid1}{rgb}{0.75,0.24,1.00}
\definecolor{DarkOrchid2}{rgb}{0.70,0.23,0.93}
\definecolor{DarkOrchid3}{rgb}{0.60,0.20,0.80}
\definecolor{DarkOrchid4}{rgb}{0.41,0.13,0.55}
\definecolor{DarkOrchid}{rgb}{0.60,0.20,0.80}
\definecolor{DarkRed}{rgb}{0.55,0.00,0.00}
\definecolor{DarkSalmon}{rgb}{0.91,0.59,0.48}
\definecolor{DarkSeaGreen1}{rgb}{0.76,1.00,0.76}
\definecolor{DarkSeaGreen2}{rgb}{0.71,0.93,0.71}
\definecolor{DarkSeaGreen3}{rgb}{0.61,0.80,0.61}
\definecolor{DarkSeaGreen4}{rgb}{0.41,0.55,0.41}
\definecolor{DarkSeaGreen}{rgb}{0.56,0.74,0.56}
\definecolor{DarkSlateBlue}{rgb}{0.28,0.24,0.55}
\definecolor{DarkSlateGray1}{rgb}{0.59,1.00,1.00}
\definecolor{DarkSlateGray2}{rgb}{0.55,0.93,0.93}
\definecolor{DarkSlateGray3}{rgb}{0.47,0.80,0.80}
\definecolor{DarkSlateGray4}{rgb}{0.32,0.55,0.55}
\definecolor{DarkSlateGray}{rgb}{0.18,0.31,0.31}
\definecolor{DarkSlateGrey}{rgb}{0.18,0.31,0.31}
\definecolor{DarkTurquoise}{rgb}{0.00,0.81,0.82}
\definecolor{DarkViolet}{rgb}{0.58,0.00,0.83}
\definecolor{DeepPink1}{rgb}{1.00,0.08,0.58}
\definecolor{DeepPink2}{rgb}{0.93,0.07,0.54}
\definecolor{DeepPink3}{rgb}{0.80,0.06,0.46}
\definecolor{DeepPink4}{rgb}{0.55,0.04,0.31}
\definecolor{DeepPink}{rgb}{1.00,0.08,0.58}
\definecolor{DeepSkyBlue1}{rgb}{0.00,0.75,1.00}
\definecolor{DeepSkyBlue2}{rgb}{0.00,0.70,0.93}
\definecolor{DeepSkyBlue3}{rgb}{0.00,0.60,0.80}
\definecolor{DeepSkyBlue4}{rgb}{0.00,0.41,0.55}
\definecolor{DeepSkyBlue}{rgb}{0.00,0.75,1.00}
\definecolor{DimGray}{rgb}{0.41,0.41,0.41}
\definecolor{DimGrey}{rgb}{0.41,0.41,0.41}
\definecolor{DodgerBlue1}{rgb}{0.12,0.56,1.00}
\definecolor{DodgerBlue2}{rgb}{0.11,0.53,0.93}
\definecolor{DodgerBlue3}{rgb}{0.09,0.45,0.80}
\definecolor{DodgerBlue4}{rgb}{0.06,0.31,0.55}
\definecolor{DodgerBlue}{rgb}{0.12,0.56,1.00}
\definecolor{FloralWhite}{rgb}{1.00,0.98,0.94}
\definecolor{ForestGreen}{rgb}{0.13,0.55,0.13}
\definecolor{GhostWhite}{rgb}{0.97,0.97,1.00}
\definecolor{GreenYellow}{rgb}{0.68,1.00,0.18}
\definecolor{HotPink1}{rgb}{1.00,0.43,0.71}
\definecolor{HotPink2}{rgb}{0.93,0.42,0.65}
\definecolor{HotPink3}{rgb}{0.80,0.38,0.56}
\definecolor{HotPink4}{rgb}{0.55,0.23,0.38}
\definecolor{HotPink}{rgb}{1.00,0.41,0.71}
\definecolor{IndianRed1}{rgb}{1.00,0.42,0.42}
\definecolor{IndianRed2}{rgb}{0.93,0.39,0.39}
\definecolor{IndianRed3}{rgb}{0.80,0.33,0.33}
\definecolor{IndianRed4}{rgb}{0.55,0.23,0.23}
\definecolor{IndianRed}{rgb}{0.80,0.36,0.36}
\definecolor{LavenderBlush1}{rgb}{1.00,0.94,0.96}
\definecolor{LavenderBlush2}{rgb}{0.93,0.88,0.90}
\definecolor{LavenderBlush3}{rgb}{0.80,0.76,0.77}
\definecolor{LavenderBlush4}{rgb}{0.55,0.51,0.53}
\definecolor{LavenderBlush}{rgb}{1.00,0.94,0.96}
\definecolor{LawnGreen}{rgb}{0.49,0.99,0.00}
\definecolor{LemonChiffon1}{rgb}{1.00,0.98,0.80}
\definecolor{LemonChiffon2}{rgb}{0.93,0.91,0.75}
\definecolor{LemonChiffon3}{rgb}{0.80,0.79,0.65}
\definecolor{LemonChiffon4}{rgb}{0.55,0.54,0.44}
\definecolor{LemonChiffon}{rgb}{1.00,0.98,0.80}
\definecolor{LightBlue1}{rgb}{0.75,0.94,1.00}
\definecolor{LightBlue2}{rgb}{0.70,0.87,0.93}
\definecolor{LightBlue3}{rgb}{0.60,0.75,0.80}
\definecolor{LightBlue4}{rgb}{0.41,0.51,0.55}
\definecolor{LightBlue}{rgb}{0.68,0.85,0.90}
\definecolor{LightCoral}{rgb}{0.94,0.50,0.50}
\definecolor{LightCyan1}{rgb}{0.88,1.00,1.00}
\definecolor{LightCyan2}{rgb}{0.82,0.93,0.93}
\definecolor{LightCyan3}{rgb}{0.71,0.80,0.80}
\definecolor{LightCyan4}{rgb}{0.48,0.55,0.55}
\definecolor{LightCyan}{rgb}{0.88,1.00,1.00}
\definecolor{LightGoldenrod1}{rgb}{1.00,0.93,0.55}
\definecolor{LightGoldenrod2}{rgb}{0.93,0.86,0.51}
\definecolor{LightGoldenrod3}{rgb}{0.80,0.75,0.44}
\definecolor{LightGoldenrod4}{rgb}{0.55,0.51,0.30}
\definecolor{LightGoldenrodYellow}{rgb}{0.98,0.98,0.82}
\definecolor{LightGoldenrod}{rgb}{0.93,0.87,0.51}
\definecolor{LightGray}{rgb}{0.83,0.83,0.83}
\definecolor{LightGreen}{rgb}{0.56,0.93,0.56}
\definecolor{LightGrey}{rgb}{0.83,0.83,0.83}
\definecolor{LightPink1}{rgb}{1.00,0.68,0.73}
\definecolor{LightPink2}{rgb}{0.93,0.64,0.68}
\definecolor{LightPink3}{rgb}{0.80,0.55,0.58}
\definecolor{LightPink4}{rgb}{0.55,0.37,0.40}
\definecolor{LightPink}{rgb}{1.00,0.71,0.76}
\definecolor{LightSalmon1}{rgb}{1.00,0.63,0.48}
\definecolor{LightSalmon2}{rgb}{0.93,0.58,0.45}
\definecolor{LightSalmon3}{rgb}{0.80,0.51,0.38}
\definecolor{LightSalmon4}{rgb}{0.55,0.34,0.26}
\definecolor{LightSalmon}{rgb}{1.00,0.63,0.48}
\definecolor{LightSeaGreen}{rgb}{0.13,0.70,0.67}
\definecolor{LightSkyBlue1}{rgb}{0.69,0.89,1.00}
\definecolor{LightSkyBlue2}{rgb}{0.64,0.83,0.93}
\definecolor{LightSkyBlue3}{rgb}{0.55,0.71,0.80}
\definecolor{LightSkyBlue4}{rgb}{0.38,0.48,0.55}
\definecolor{LightSkyBlue}{rgb}{0.53,0.81,0.98}
\definecolor{LightSlateBlue}{rgb}{0.52,0.44,1.00}
\definecolor{LightSlateGray}{rgb}{0.47,0.53,0.60}
\definecolor{LightSlateGrey}{rgb}{0.47,0.53,0.60}
\definecolor{LightSteelBlue1}{rgb}{0.79,0.88,1.00}
\definecolor{LightSteelBlue2}{rgb}{0.74,0.82,0.93}
\definecolor{LightSteelBlue3}{rgb}{0.64,0.71,0.80}
\definecolor{LightSteelBlue4}{rgb}{0.43,0.48,0.55}
\definecolor{LightSteelBlue}{rgb}{0.69,0.77,0.87}
\definecolor{LightYellow1}{rgb}{1.00,1.00,0.88}
\definecolor{LightYellow2}{rgb}{0.93,0.93,0.82}
\definecolor{LightYellow3}{rgb}{0.80,0.80,0.71}
\definecolor{LightYellow4}{rgb}{0.55,0.55,0.48}
\definecolor{LightYellow}{rgb}{1.00,1.00,0.88}
\definecolor{LimeGreen}{rgb}{0.20,0.80,0.20}
\definecolor{MediumAquamarine}{rgb}{0.40,0.80,0.67}
\definecolor{MediumBlue}{rgb}{0.00,0.00,0.80}
\definecolor{MediumOrchid1}{rgb}{0.88,0.40,1.00}
\definecolor{MediumOrchid2}{rgb}{0.82,0.37,0.93}
\definecolor{MediumOrchid3}{rgb}{0.71,0.32,0.80}
\definecolor{MediumOrchid4}{rgb}{0.48,0.22,0.55}
\definecolor{MediumOrchid}{rgb}{0.73,0.33,0.83}
\definecolor{MediumPurple1}{rgb}{0.67,0.51,1.00}
\definecolor{MediumPurple2}{rgb}{0.62,0.47,0.93}
\definecolor{MediumPurple3}{rgb}{0.54,0.41,0.80}
\definecolor{MediumPurple4}{rgb}{0.36,0.28,0.55}
\definecolor{MediumPurple}{rgb}{0.58,0.44,0.86}
\definecolor{MediumSeaGreen}{rgb}{0.24,0.70,0.44}
\definecolor{MediumSlateBlue}{rgb}{0.48,0.41,0.93}
\definecolor{MediumSpringGreen}{rgb}{0.00,0.98,0.60}
\definecolor{MediumTurquoise}{rgb}{0.28,0.82,0.80}
\definecolor{MediumVioletRed}{rgb}{0.78,0.08,0.52}
\definecolor{MidnightBlue}{rgb}{0.10,0.10,0.44}
\definecolor{MintCream}{rgb}{0.96,1.00,0.98}
\definecolor{MistyRose1}{rgb}{1.00,0.89,0.88}
\definecolor{MistyRose2}{rgb}{0.93,0.84,0.82}
\definecolor{MistyRose3}{rgb}{0.80,0.72,0.71}
\definecolor{MistyRose4}{rgb}{0.55,0.49,0.48}
\definecolor{MistyRose}{rgb}{1.00,0.89,0.88}
\definecolor{NavajoWhite1}{rgb}{1.00,0.87,0.68}
\definecolor{NavajoWhite2}{rgb}{0.93,0.81,0.63}
\definecolor{NavajoWhite3}{rgb}{0.80,0.70,0.55}
\definecolor{NavajoWhite4}{rgb}{0.55,0.47,0.37}
\definecolor{NavajoWhite}{rgb}{1.00,0.87,0.68}
\definecolor{NavyBlue}{rgb}{0.00,0.00,0.50}
\definecolor{OldLace}{rgb}{0.99,0.96,0.90}
\definecolor{OliveDrab1}{rgb}{0.75,1.00,0.24}
\definecolor{OliveDrab2}{rgb}{0.70,0.93,0.23}
\definecolor{OliveDrab3}{rgb}{0.60,0.80,0.20}
\definecolor{OliveDrab4}{rgb}{0.41,0.55,0.13}
\definecolor{OliveDrab}{rgb}{0.42,0.56,0.14}
\definecolor{OrangeRed1}{rgb}{1.00,0.27,0.00}
\definecolor{OrangeRed2}{rgb}{0.93,0.25,0.00}
\definecolor{OrangeRed3}{rgb}{0.80,0.22,0.00}
\definecolor{OrangeRed4}{rgb}{0.55,0.15,0.00}
\definecolor{OrangeRed}{rgb}{1.00,0.27,0.00}
\definecolor{PaleGoldenrod}{rgb}{0.93,0.91,0.67}
\definecolor{PaleGreen1}{rgb}{0.60,1.00,0.60}
\definecolor{PaleGreen2}{rgb}{0.56,0.93,0.56}
\definecolor{PaleGreen3}{rgb}{0.49,0.80,0.49}
\definecolor{PaleGreen4}{rgb}{0.33,0.55,0.33}
\definecolor{PaleGreen}{rgb}{0.60,0.98,0.60}
\definecolor{PaleTurquoise1}{rgb}{0.73,1.00,1.00}
\definecolor{PaleTurquoise2}{rgb}{0.68,0.93,0.93}
\definecolor{PaleTurquoise3}{rgb}{0.59,0.80,0.80}
\definecolor{PaleTurquoise4}{rgb}{0.40,0.55,0.55}
\definecolor{PaleTurquoise}{rgb}{0.69,0.93,0.93}
\definecolor{PaleVioletRed1}{rgb}{1.00,0.51,0.67}
\definecolor{PaleVioletRed2}{rgb}{0.93,0.47,0.62}
\definecolor{PaleVioletRed3}{rgb}{0.80,0.41,0.54}
\definecolor{PaleVioletRed4}{rgb}{0.55,0.28,0.36}
\definecolor{PaleVioletRed}{rgb}{0.86,0.44,0.58}
\definecolor{PapayaWhip}{rgb}{1.00,0.94,0.84}
\definecolor{PeachPuff1}{rgb}{1.00,0.85,0.73}
\definecolor{PeachPuff2}{rgb}{0.93,0.80,0.68}
\definecolor{PeachPuff3}{rgb}{0.80,0.69,0.58}
\definecolor{PeachPuff4}{rgb}{0.55,0.47,0.40}
\definecolor{PeachPuff}{rgb}{1.00,0.85,0.73}
\definecolor{PowderBlue}{rgb}{0.69,0.88,0.90}
\definecolor{RosyBrown1}{rgb}{1.00,0.76,0.76}
\definecolor{RosyBrown2}{rgb}{0.93,0.71,0.71}
\definecolor{RosyBrown3}{rgb}{0.80,0.61,0.61}
\definecolor{RosyBrown4}{rgb}{0.55,0.41,0.41}
\definecolor{RosyBrown}{rgb}{0.74,0.56,0.56}
\definecolor{RoyalBlue1}{rgb}{0.28,0.46,1.00}
\definecolor{RoyalBlue2}{rgb}{0.26,0.43,0.93}
\definecolor{RoyalBlue3}{rgb}{0.23,0.37,0.80}
\definecolor{RoyalBlue4}{rgb}{0.15,0.25,0.55}
\definecolor{RoyalBlue}{rgb}{0.25,0.41,0.88}
\definecolor{SaddleBrown}{rgb}{0.55,0.27,0.07}
\definecolor{SandyBrown}{rgb}{0.96,0.64,0.38}
\definecolor{SeaGreen1}{rgb}{0.33,1.00,0.62}
\definecolor{SeaGreen2}{rgb}{0.31,0.93,0.58}
\definecolor{SeaGreen3}{rgb}{0.26,0.80,0.50}
\definecolor{SeaGreen4}{rgb}{0.18,0.55,0.34}
\definecolor{SeaGreen}{rgb}{0.18,0.55,0.34}
\definecolor{SkyBlue1}{rgb}{0.53,0.81,1.00}
\definecolor{SkyBlue2}{rgb}{0.49,0.75,0.93}
\definecolor{SkyBlue3}{rgb}{0.42,0.65,0.80}
\definecolor{SkyBlue4}{rgb}{0.29,0.44,0.55}
\definecolor{SkyBlue}{rgb}{0.53,0.81,0.92}
\definecolor{SlateBlue1}{rgb}{0.51,0.44,1.00}
\definecolor{SlateBlue2}{rgb}{0.48,0.40,0.93}
\definecolor{SlateBlue3}{rgb}{0.41,0.35,0.80}
\definecolor{SlateBlue4}{rgb}{0.28,0.24,0.55}
\definecolor{SlateBlue}{rgb}{0.42,0.35,0.80}
\definecolor{SlateGray1}{rgb}{0.78,0.89,1.00}
\definecolor{SlateGray2}{rgb}{0.73,0.83,0.93}
\definecolor{SlateGray3}{rgb}{0.62,0.71,0.80}
\definecolor{SlateGray4}{rgb}{0.42,0.48,0.55}
\definecolor{SlateGray}{rgb}{0.44,0.50,0.56}
\definecolor{SlateGrey}{rgb}{0.44,0.50,0.56}
\definecolor{SpringGreen1}{rgb}{0.00,1.00,0.50}
\definecolor{SpringGreen2}{rgb}{0.00,0.93,0.46}
\definecolor{SpringGreen3}{rgb}{0.00,0.80,0.40}
\definecolor{SpringGreen4}{rgb}{0.00,0.55,0.27}
\definecolor{SpringGreen}{rgb}{0.00,1.00,0.50}
\definecolor{SteelBlue1}{rgb}{0.39,0.72,1.00}
\definecolor{SteelBlue2}{rgb}{0.36,0.67,0.93}
\definecolor{SteelBlue3}{rgb}{0.31,0.58,0.80}
\definecolor{SteelBlue4}{rgb}{0.21,0.39,0.55}
\definecolor{SteelBlue}{rgb}{0.27,0.51,0.71}
\definecolor{VioletRed1}{rgb}{1.00,0.24,0.59}
\definecolor{VioletRed2}{rgb}{0.93,0.23,0.55}
\definecolor{VioletRed3}{rgb}{0.80,0.20,0.47}
\definecolor{VioletRed4}{rgb}{0.55,0.13,0.32}
\definecolor{VioletRed}{rgb}{0.82,0.13,0.56}
\definecolor{WhiteSmoke}{rgb}{0.96,0.96,0.96}
\definecolor{YellowGreen}{rgb}{0.60,0.80,0.20}
\definecolor{aliceblue}{rgb}{0.94,0.97,1.00}
\definecolor{antiquewhite}{rgb}{0.98,0.92,0.84}
\definecolor{aquamarine1}{rgb}{0.50,1.00,0.83}
\definecolor{aquamarine2}{rgb}{0.46,0.93,0.78}
\definecolor{aquamarine3}{rgb}{0.40,0.80,0.67}
\definecolor{aquamarine4}{rgb}{0.27,0.55,0.45}
\definecolor{aquamarine}{rgb}{0.50,1.00,0.83}
\definecolor{azure1}{rgb}{0.94,1.00,1.00}
\definecolor{azure2}{rgb}{0.88,0.93,0.93}
\definecolor{azure3}{rgb}{0.76,0.80,0.80}
\definecolor{azure4}{rgb}{0.51,0.55,0.55}
\definecolor{azure}{rgb}{0.94,1.00,1.00}
\definecolor{beige}{rgb}{0.96,0.96,0.86}
\definecolor{bisque1}{rgb}{1.00,0.89,0.77}
\definecolor{bisque2}{rgb}{0.93,0.84,0.72}
\definecolor{bisque3}{rgb}{0.80,0.72,0.62}
\definecolor{bisque4}{rgb}{0.55,0.49,0.42}
\definecolor{bisque}{rgb}{1.00,0.89,0.77}
\definecolor{black}{rgb}{0.00,0.00,0.00}
\definecolor{blanchedalmond}{rgb}{1.00,0.92,0.80}
\definecolor{blue1}{rgb}{0.00,0.00,1.00}
\definecolor{blue2}{rgb}{0.00,0.00,0.93}
\definecolor{blue3}{rgb}{0.00,0.00,0.80}
\definecolor{blue4}{rgb}{0.00,0.00,0.55}
\definecolor{blueviolet}{rgb}{0.54,0.17,0.89}
\definecolor{blue}{rgb}{0.00,0.00,1.00}
\definecolor{brown1}{rgb}{1.00,0.25,0.25}
\definecolor{brown2}{rgb}{0.93,0.23,0.23}
\definecolor{brown3}{rgb}{0.80,0.20,0.20}
\definecolor{brown4}{rgb}{0.55,0.14,0.14}
\definecolor{brown}{rgb}{0.65,0.16,0.16}
\definecolor{burlywood1}{rgb}{1.00,0.83,0.61}
\definecolor{burlywood2}{rgb}{0.93,0.77,0.57}
\definecolor{burlywood3}{rgb}{0.80,0.67,0.49}
\definecolor{burlywood4}{rgb}{0.55,0.45,0.33}
\definecolor{burlywood}{rgb}{0.87,0.72,0.53}
\definecolor{cadetblue}{rgb}{0.37,0.62,0.63}
\definecolor{chartreuse1}{rgb}{0.50,1.00,0.00}
\definecolor{chartreuse2}{rgb}{0.46,0.93,0.00}
\definecolor{chartreuse3}{rgb}{0.40,0.80,0.00}
\definecolor{chartreuse4}{rgb}{0.27,0.55,0.00}
\definecolor{chartreuse}{rgb}{0.50,1.00,0.00}
\definecolor{chocolate1}{rgb}{1.00,0.50,0.14}
\definecolor{chocolate2}{rgb}{0.93,0.46,0.13}
\definecolor{chocolate3}{rgb}{0.80,0.40,0.11}
\definecolor{chocolate4}{rgb}{0.55,0.27,0.07}
\definecolor{chocolate}{rgb}{0.82,0.41,0.12}
\definecolor{coral1}{rgb}{1.00,0.45,0.34}
\definecolor{coral2}{rgb}{0.93,0.42,0.31}
\definecolor{coral3}{rgb}{0.80,0.36,0.27}
\definecolor{coral4}{rgb}{0.55,0.24,0.18}
\definecolor{coral}{rgb}{1.00,0.50,0.31}
\definecolor{cornflowerblue}{rgb}{0.39,0.58,0.93}
\definecolor{cornsilk1}{rgb}{1.00,0.97,0.86}
\definecolor{cornsilk2}{rgb}{0.93,0.91,0.80}
\definecolor{cornsilk3}{rgb}{0.80,0.78,0.69}
\definecolor{cornsilk4}{rgb}{0.55,0.53,0.47}
\definecolor{cornsilk}{rgb}{1.00,0.97,0.86}
\definecolor{cyan1}{rgb}{0.00,1.00,1.00}
\definecolor{cyan2}{rgb}{0.00,0.93,0.93}
\definecolor{cyan3}{rgb}{0.00,0.80,0.80}
\definecolor{cyan4}{rgb}{0.00,0.55,0.55}
\definecolor{cyan}{rgb}{0.00,1.00,1.00}
\definecolor{darkblue}{rgb}{0.00,0.00,0.55}
\definecolor{darkcyan}{rgb}{0.00,0.55,0.55}
\definecolor{darkgoldenrod}{rgb}{0.72,0.53,0.04}
\definecolor{darkgray}{rgb}{0.66,0.66,0.66}
\definecolor{darkgreen}{rgb}{0.00,0.39,0.00}
\definecolor{darkgrey}{rgb}{0.66,0.66,0.66}
\definecolor{darkkhaki}{rgb}{0.74,0.72,0.42}
\definecolor{darkmagenta}{rgb}{0.55,0.00,0.55}
\definecolor{darkolive}{rgb}{0.33,0.42,0.18}
\definecolor{darkorange}{rgb}{1.00,0.55,0.00}
\definecolor{darkorchid}{rgb}{0.60,0.20,0.80}
\definecolor{darkred}{rgb}{0.55,0.00,0.00}
\definecolor{darksalmon}{rgb}{0.91,0.59,0.48}
\definecolor{darksea}{rgb}{0.56,0.74,0.56}
\definecolor{darkslate}{rgb}{0.18,0.31,0.31}
\definecolor{darkslate}{rgb}{0.18,0.31,0.31}
\definecolor{darkslate}{rgb}{0.28,0.24,0.55}
\definecolor{darkturquoise}{rgb}{0.00,0.81,0.82}
\definecolor{darkviolet}{rgb}{0.58,0.00,0.83}
\definecolor{deeppink}{rgb}{1.00,0.08,0.58}
\definecolor{deepsky}{rgb}{0.00,0.75,1.00}
\definecolor{dimgray}{rgb}{0.41,0.41,0.41}
\definecolor{dimgrey}{rgb}{0.41,0.41,0.41}
\definecolor{dodgerblue}{rgb}{0.12,0.56,1.00}
\definecolor{firebrick1}{rgb}{1.00,0.19,0.19}
\definecolor{firebrick2}{rgb}{0.93,0.17,0.17}
\definecolor{firebrick3}{rgb}{0.80,0.15,0.15}
\definecolor{firebrick4}{rgb}{0.55,0.10,0.10}
\definecolor{firebrick}{rgb}{0.70,0.13,0.13}
\definecolor{floralwhite}{rgb}{1.00,0.98,0.94}
\definecolor{forestgreen}{rgb}{0.13,0.55,0.13}
\definecolor{gainsboro}{rgb}{0.86,0.86,0.86}
\definecolor{ghostwhite}{rgb}{0.97,0.97,1.00}
\definecolor{gold1}{rgb}{1.00,0.84,0.00}
\definecolor{gold2}{rgb}{0.93,0.79,0.00}
\definecolor{gold3}{rgb}{0.80,0.68,0.00}
\definecolor{gold4}{rgb}{0.55,0.46,0.00}
\definecolor{goldenrod1}{rgb}{1.00,0.76,0.15}
\definecolor{goldenrod2}{rgb}{0.93,0.71,0.13}
\definecolor{goldenrod3}{rgb}{0.80,0.61,0.11}
\definecolor{goldenrod4}{rgb}{0.55,0.41,0.08}
\definecolor{goldenrod}{rgb}{0.85,0.65,0.13}
\definecolor{gold}{rgb}{1.00,0.84,0.00}
\definecolor{gray0}{rgb}{0.00,0.00,0.00}
\definecolor{gray100}{rgb}{1.00,1.00,1.00}
\definecolor{gray10}{rgb}{0.10,0.10,0.10}
\definecolor{gray11}{rgb}{0.11,0.11,0.11}
\definecolor{gray12}{rgb}{0.12,0.12,0.12}
\definecolor{gray13}{rgb}{0.13,0.13,0.13}
\definecolor{gray14}{rgb}{0.14,0.14,0.14}
\definecolor{gray15}{rgb}{0.15,0.15,0.15}
\definecolor{gray16}{rgb}{0.16,0.16,0.16}
\definecolor{gray17}{rgb}{0.17,0.17,0.17}
\definecolor{gray18}{rgb}{0.18,0.18,0.18}
\definecolor{gray19}{rgb}{0.19,0.19,0.19}
\definecolor{gray1}{rgb}{0.01,0.01,0.01}
\definecolor{gray20}{rgb}{0.20,0.20,0.20}
\definecolor{gray21}{rgb}{0.21,0.21,0.21}
\definecolor{gray22}{rgb}{0.22,0.22,0.22}
\definecolor{gray23}{rgb}{0.23,0.23,0.23}
\definecolor{gray24}{rgb}{0.24,0.24,0.24}
\definecolor{gray25}{rgb}{0.25,0.25,0.25}
\definecolor{gray26}{rgb}{0.26,0.26,0.26}
\definecolor{gray27}{rgb}{0.27,0.27,0.27}
\definecolor{gray28}{rgb}{0.28,0.28,0.28}
\definecolor{gray29}{rgb}{0.29,0.29,0.29}
\definecolor{gray2}{rgb}{0.02,0.02,0.02}
\definecolor{gray30}{rgb}{0.30,0.30,0.30}
\definecolor{gray31}{rgb}{0.31,0.31,0.31}
\definecolor{gray32}{rgb}{0.32,0.32,0.32}
\definecolor{gray33}{rgb}{0.33,0.33,0.33}
\definecolor{gray34}{rgb}{0.34,0.34,0.34}
\definecolor{gray35}{rgb}{0.35,0.35,0.35}
\definecolor{gray36}{rgb}{0.36,0.36,0.36}
\definecolor{gray37}{rgb}{0.37,0.37,0.37}
\definecolor{gray38}{rgb}{0.38,0.38,0.38}
\definecolor{gray39}{rgb}{0.39,0.39,0.39}
\definecolor{gray3}{rgb}{0.03,0.03,0.03}
\definecolor{gray40}{rgb}{0.40,0.40,0.40}
\definecolor{gray41}{rgb}{0.41,0.41,0.41}
\definecolor{gray42}{rgb}{0.42,0.42,0.42}
\definecolor{gray43}{rgb}{0.43,0.43,0.43}
\definecolor{gray44}{rgb}{0.44,0.44,0.44}
\definecolor{gray45}{rgb}{0.45,0.45,0.45}
\definecolor{gray46}{rgb}{0.46,0.46,0.46}
\definecolor{gray47}{rgb}{0.47,0.47,0.47}
\definecolor{gray48}{rgb}{0.48,0.48,0.48}
\definecolor{gray49}{rgb}{0.49,0.49,0.49}
\definecolor{gray4}{rgb}{0.04,0.04,0.04}
\definecolor{gray50}{rgb}{0.50,0.50,0.50}
\definecolor{gray51}{rgb}{0.51,0.51,0.51}
\definecolor{gray52}{rgb}{0.52,0.52,0.52}
\definecolor{gray53}{rgb}{0.53,0.53,0.53}
\definecolor{gray54}{rgb}{0.54,0.54,0.54}
\definecolor{gray55}{rgb}{0.55,0.55,0.55}
\definecolor{gray56}{rgb}{0.56,0.56,0.56}
\definecolor{gray57}{rgb}{0.57,0.57,0.57}
\definecolor{gray58}{rgb}{0.58,0.58,0.58}
\definecolor{gray59}{rgb}{0.59,0.59,0.59}
\definecolor{gray5}{rgb}{0.05,0.05,0.05}
\definecolor{gray60}{rgb}{0.60,0.60,0.60}
\definecolor{gray61}{rgb}{0.61,0.61,0.61}
\definecolor{gray62}{rgb}{0.62,0.62,0.62}
\definecolor{gray63}{rgb}{0.63,0.63,0.63}
\definecolor{gray64}{rgb}{0.64,0.64,0.64}
\definecolor{gray65}{rgb}{0.65,0.65,0.65}
\definecolor{gray66}{rgb}{0.66,0.66,0.66}
\definecolor{gray67}{rgb}{0.67,0.67,0.67}
\definecolor{gray68}{rgb}{0.68,0.68,0.68}
\definecolor{gray69}{rgb}{0.69,0.69,0.69}
\definecolor{gray6}{rgb}{0.06,0.06,0.06}
\definecolor{gray70}{rgb}{0.70,0.70,0.70}
\definecolor{gray71}{rgb}{0.71,0.71,0.71}
\definecolor{gray72}{rgb}{0.72,0.72,0.72}
\definecolor{gray73}{rgb}{0.73,0.73,0.73}
\definecolor{gray74}{rgb}{0.74,0.74,0.74}
\definecolor{gray75}{rgb}{0.75,0.75,0.75}
\definecolor{gray76}{rgb}{0.76,0.76,0.76}
\definecolor{gray77}{rgb}{0.77,0.77,0.77}
\definecolor{gray78}{rgb}{0.78,0.78,0.78}
\definecolor{gray79}{rgb}{0.79,0.79,0.79}
\definecolor{gray7}{rgb}{0.07,0.07,0.07}
\definecolor{gray80}{rgb}{0.80,0.80,0.80}
\definecolor{gray81}{rgb}{0.81,0.81,0.81}
\definecolor{gray82}{rgb}{0.82,0.82,0.82}
\definecolor{gray83}{rgb}{0.83,0.83,0.83}
\definecolor{gray84}{rgb}{0.84,0.84,0.84}
\definecolor{gray85}{rgb}{0.85,0.85,0.85}
\definecolor{gray86}{rgb}{0.86,0.86,0.86}
\definecolor{gray87}{rgb}{0.87,0.87,0.87}
\definecolor{gray88}{rgb}{0.88,0.88,0.88}
\definecolor{gray89}{rgb}{0.89,0.89,0.89}
\definecolor{gray8}{rgb}{0.08,0.08,0.08}
\definecolor{gray90}{rgb}{0.90,0.90,0.90}
\definecolor{gray91}{rgb}{0.91,0.91,0.91}
\definecolor{gray92}{rgb}{0.92,0.92,0.92}
\definecolor{gray93}{rgb}{0.93,0.93,0.93}
\definecolor{gray94}{rgb}{0.94,0.94,0.94}
\definecolor{gray95}{rgb}{0.95,0.95,0.95}
\definecolor{gray96}{rgb}{0.96,0.96,0.96}
\definecolor{gray97}{rgb}{0.97,0.97,0.97}
\definecolor{gray98}{rgb}{0.98,0.98,0.98}
\definecolor{gray99}{rgb}{0.99,0.99,0.99}
\definecolor{gray9}{rgb}{0.09,0.09,0.09}
\definecolor{gray}{rgb}{0.75,0.75,0.75}
\definecolor{green1}{rgb}{0.00,1.00,0.00}
\definecolor{green2}{rgb}{0.00,0.93,0.00}
\definecolor{green3}{rgb}{0.00,0.80,0.00}
\definecolor{green4}{rgb}{0.00,0.55,0.00}
\definecolor{greenyellow}{rgb}{0.68,1.00,0.18}
\definecolor{green}{rgb}{0.00,1.00,0.00}
\definecolor{grey0}{rgb}{0.00,0.00,0.00}
\definecolor{grey100}{rgb}{1.00,1.00,1.00}
\definecolor{grey10}{rgb}{0.10,0.10,0.10}
\definecolor{grey11}{rgb}{0.11,0.11,0.11}
\definecolor{grey12}{rgb}{0.12,0.12,0.12}
\definecolor{grey13}{rgb}{0.13,0.13,0.13}
\definecolor{grey14}{rgb}{0.14,0.14,0.14}
\definecolor{grey15}{rgb}{0.15,0.15,0.15}
\definecolor{grey16}{rgb}{0.16,0.16,0.16}
\definecolor{grey17}{rgb}{0.17,0.17,0.17}
\definecolor{grey18}{rgb}{0.18,0.18,0.18}
\definecolor{grey19}{rgb}{0.19,0.19,0.19}
\definecolor{grey1}{rgb}{0.01,0.01,0.01}
\definecolor{grey20}{rgb}{0.20,0.20,0.20}
\definecolor{grey21}{rgb}{0.21,0.21,0.21}
\definecolor{grey22}{rgb}{0.22,0.22,0.22}
\definecolor{grey23}{rgb}{0.23,0.23,0.23}
\definecolor{grey24}{rgb}{0.24,0.24,0.24}
\definecolor{grey25}{rgb}{0.25,0.25,0.25}
\definecolor{grey26}{rgb}{0.26,0.26,0.26}
\definecolor{grey27}{rgb}{0.27,0.27,0.27}
\definecolor{grey28}{rgb}{0.28,0.28,0.28}
\definecolor{grey29}{rgb}{0.29,0.29,0.29}
\definecolor{grey2}{rgb}{0.02,0.02,0.02}
\definecolor{grey30}{rgb}{0.30,0.30,0.30}
\definecolor{grey31}{rgb}{0.31,0.31,0.31}
\definecolor{grey32}{rgb}{0.32,0.32,0.32}
\definecolor{grey33}{rgb}{0.33,0.33,0.33}
\definecolor{grey34}{rgb}{0.34,0.34,0.34}
\definecolor{grey35}{rgb}{0.35,0.35,0.35}
\definecolor{grey36}{rgb}{0.36,0.36,0.36}
\definecolor{grey37}{rgb}{0.37,0.37,0.37}
\definecolor{grey38}{rgb}{0.38,0.38,0.38}
\definecolor{grey39}{rgb}{0.39,0.39,0.39}
\definecolor{grey3}{rgb}{0.03,0.03,0.03}
\definecolor{grey40}{rgb}{0.40,0.40,0.40}
\definecolor{grey41}{rgb}{0.41,0.41,0.41}
\definecolor{grey42}{rgb}{0.42,0.42,0.42}
\definecolor{grey43}{rgb}{0.43,0.43,0.43}
\definecolor{grey44}{rgb}{0.44,0.44,0.44}
\definecolor{grey45}{rgb}{0.45,0.45,0.45}
\definecolor{grey46}{rgb}{0.46,0.46,0.46}
\definecolor{grey47}{rgb}{0.47,0.47,0.47}
\definecolor{grey48}{rgb}{0.48,0.48,0.48}
\definecolor{grey49}{rgb}{0.49,0.49,0.49}
\definecolor{grey4}{rgb}{0.04,0.04,0.04}
\definecolor{grey50}{rgb}{0.50,0.50,0.50}
\definecolor{grey51}{rgb}{0.51,0.51,0.51}
\definecolor{grey52}{rgb}{0.52,0.52,0.52}
\definecolor{grey53}{rgb}{0.53,0.53,0.53}
\definecolor{grey54}{rgb}{0.54,0.54,0.54}
\definecolor{grey55}{rgb}{0.55,0.55,0.55}
\definecolor{grey56}{rgb}{0.56,0.56,0.56}
\definecolor{grey57}{rgb}{0.57,0.57,0.57}
\definecolor{grey58}{rgb}{0.58,0.58,0.58}
\definecolor{grey59}{rgb}{0.59,0.59,0.59}
\definecolor{grey5}{rgb}{0.05,0.05,0.05}
\definecolor{grey60}{rgb}{0.60,0.60,0.60}
\definecolor{grey61}{rgb}{0.61,0.61,0.61}
\definecolor{grey62}{rgb}{0.62,0.62,0.62}
\definecolor{grey63}{rgb}{0.63,0.63,0.63}
\definecolor{grey64}{rgb}{0.64,0.64,0.64}
\definecolor{grey65}{rgb}{0.65,0.65,0.65}
\definecolor{grey66}{rgb}{0.66,0.66,0.66}
\definecolor{grey67}{rgb}{0.67,0.67,0.67}
\definecolor{grey68}{rgb}{0.68,0.68,0.68}
\definecolor{grey69}{rgb}{0.69,0.69,0.69}
\definecolor{grey6}{rgb}{0.06,0.06,0.06}
\definecolor{grey70}{rgb}{0.70,0.70,0.70}
\definecolor{grey71}{rgb}{0.71,0.71,0.71}
\definecolor{grey72}{rgb}{0.72,0.72,0.72}
\definecolor{grey73}{rgb}{0.73,0.73,0.73}
\definecolor{grey74}{rgb}{0.74,0.74,0.74}
\definecolor{grey75}{rgb}{0.75,0.75,0.75}
\definecolor{grey76}{rgb}{0.76,0.76,0.76}
\definecolor{grey77}{rgb}{0.77,0.77,0.77}
\definecolor{grey78}{rgb}{0.78,0.78,0.78}
\definecolor{grey79}{rgb}{0.79,0.79,0.79}
\definecolor{grey7}{rgb}{0.07,0.07,0.07}
\definecolor{grey80}{rgb}{0.80,0.80,0.80}
\definecolor{grey81}{rgb}{0.81,0.81,0.81}
\definecolor{grey82}{rgb}{0.82,0.82,0.82}
\definecolor{grey83}{rgb}{0.83,0.83,0.83}
\definecolor{grey84}{rgb}{0.84,0.84,0.84}
\definecolor{grey85}{rgb}{0.85,0.85,0.85}
\definecolor{grey86}{rgb}{0.86,0.86,0.86}
\definecolor{grey87}{rgb}{0.87,0.87,0.87}
\definecolor{grey88}{rgb}{0.88,0.88,0.88}
\definecolor{grey89}{rgb}{0.89,0.89,0.89}
\definecolor{grey8}{rgb}{0.08,0.08,0.08}
\definecolor{grey90}{rgb}{0.90,0.90,0.90}
\definecolor{grey91}{rgb}{0.91,0.91,0.91}
\definecolor{grey92}{rgb}{0.92,0.92,0.92}
\definecolor{grey93}{rgb}{0.93,0.93,0.93}
\definecolor{grey94}{rgb}{0.94,0.94,0.94}
\definecolor{grey95}{rgb}{0.95,0.95,0.95}
\definecolor{grey96}{rgb}{0.96,0.96,0.96}
\definecolor{grey97}{rgb}{0.97,0.97,0.97}
\definecolor{grey98}{rgb}{0.98,0.98,0.98}
\definecolor{grey99}{rgb}{0.99,0.99,0.99}
\definecolor{grey9}{rgb}{0.09,0.09,0.09}
\definecolor{grey}{rgb}{0.75,0.75,0.75}
\definecolor{honeydew1}{rgb}{0.94,1.00,0.94}
\definecolor{honeydew2}{rgb}{0.88,0.93,0.88}
\definecolor{honeydew3}{rgb}{0.76,0.80,0.76}
\definecolor{honeydew4}{rgb}{0.51,0.55,0.51}
\definecolor{honeydew}{rgb}{0.94,1.00,0.94}
\definecolor{hotpink}{rgb}{1.00,0.41,0.71}
\definecolor{indianred}{rgb}{0.80,0.36,0.36}
\definecolor{ivory1}{rgb}{1.00,1.00,0.94}
\definecolor{ivory2}{rgb}{0.93,0.93,0.88}
\definecolor{ivory3}{rgb}{0.80,0.80,0.76}
\definecolor{ivory4}{rgb}{0.55,0.55,0.51}
\definecolor{ivory}{rgb}{1.00,1.00,0.94}
\definecolor{khaki1}{rgb}{1.00,0.96,0.56}
\definecolor{khaki2}{rgb}{0.93,0.90,0.52}
\definecolor{khaki3}{rgb}{0.80,0.78,0.45}
\definecolor{khaki4}{rgb}{0.55,0.53,0.31}
\definecolor{khaki}{rgb}{0.94,0.90,0.55}
\definecolor{lavenderblush}{rgb}{1.00,0.94,0.96}
\definecolor{lavender}{rgb}{0.90,0.90,0.98}
\definecolor{lawngreen}{rgb}{0.49,0.99,0.00}
\definecolor{lemonchiffon}{rgb}{1.00,0.98,0.80}
\definecolor{lightblue}{rgb}{0.68,0.85,0.90}
\definecolor{lightcoral}{rgb}{0.94,0.50,0.50}
\definecolor{lightcyan}{rgb}{0.88,1.00,1.00}
\definecolor{lightgoldenrod}{rgb}{0.93,0.87,0.51}
\definecolor{lightgoldenrod}{rgb}{0.98,0.98,0.82}
\definecolor{lightgray}{rgb}{0.83,0.83,0.83}
\definecolor{lightgreen}{rgb}{0.56,0.93,0.56}
\definecolor{lightgrey}{rgb}{0.83,0.83,0.83}
\definecolor{lightpink}{rgb}{1.00,0.71,0.76}
\definecolor{lightsalmon}{rgb}{1.00,0.63,0.48}
\definecolor{lightsea}{rgb}{0.13,0.70,0.67}
\definecolor{lightsky}{rgb}{0.53,0.81,0.98}
\definecolor{lightslate}{rgb}{0.47,0.53,0.60}
\definecolor{lightslate}{rgb}{0.47,0.53,0.60}
\definecolor{lightslate}{rgb}{0.52,0.44,1.00}
\definecolor{lightsteel}{rgb}{0.69,0.77,0.87}
\definecolor{lightyellow}{rgb}{1.00,1.00,0.88}
\definecolor{limegreen}{rgb}{0.20,0.80,0.20}
\definecolor{linen}{rgb}{0.98,0.94,0.90}
\definecolor{magenta1}{rgb}{1.00,0.00,1.00}
\definecolor{magenta2}{rgb}{0.93,0.00,0.93}
\definecolor{magenta3}{rgb}{0.80,0.00,0.80}
\definecolor{magenta4}{rgb}{0.55,0.00,0.55}
\definecolor{magenta}{rgb}{1.00,0.00,1.00}
\definecolor{maroon1}{rgb}{1.00,0.20,0.70}
\definecolor{maroon2}{rgb}{0.93,0.19,0.65}
\definecolor{maroon3}{rgb}{0.80,0.16,0.56}
\definecolor{maroon4}{rgb}{0.55,0.11,0.38}
\definecolor{maroon}{rgb}{0.69,0.19,0.38}
\definecolor{mediumaquamarine}{rgb}{0.40,0.80,0.67}
\definecolor{mediumblue}{rgb}{0.00,0.00,0.80}
\definecolor{mediumorchid}{rgb}{0.73,0.33,0.83}
\definecolor{mediumpurple}{rgb}{0.58,0.44,0.86}
\definecolor{mediumsea}{rgb}{0.24,0.70,0.44}
\definecolor{mediumslate}{rgb}{0.48,0.41,0.93}
\definecolor{mediumspring}{rgb}{0.00,0.98,0.60}
\definecolor{mediumturquoise}{rgb}{0.28,0.82,0.80}
\definecolor{mediumviolet}{rgb}{0.78,0.08,0.52}
\definecolor{midnightblue}{rgb}{0.10,0.10,0.44}
\definecolor{mintcream}{rgb}{0.96,1.00,0.98}
\definecolor{mistyrose}{rgb}{1.00,0.89,0.88}
\definecolor{moccasin}{rgb}{1.00,0.89,0.71}
\definecolor{navajowhite}{rgb}{1.00,0.87,0.68}
\definecolor{navyblue}{rgb}{0.00,0.00,0.50}
\definecolor{navy}{rgb}{0.00,0.00,0.50}
\definecolor{oldlace}{rgb}{0.99,0.96,0.90}
\definecolor{olivedrab}{rgb}{0.42,0.56,0.14}
\definecolor{orange1}{rgb}{1.00,0.65,0.00}
\definecolor{orange2}{rgb}{0.93,0.60,0.00}
\definecolor{orange3}{rgb}{0.80,0.52,0.00}
\definecolor{orange4}{rgb}{0.55,0.35,0.00}
\definecolor{orangered}{rgb}{1.00,0.27,0.00}
\definecolor{orange}{rgb}{1.00,0.65,0.00}
\definecolor{orchid1}{rgb}{1.00,0.51,0.98}
\definecolor{orchid2}{rgb}{0.93,0.48,0.91}
\definecolor{orchid3}{rgb}{0.80,0.41,0.79}
\definecolor{orchid4}{rgb}{0.55,0.28,0.54}
\definecolor{orchid}{rgb}{0.85,0.44,0.84}
\definecolor{palegoldenrod}{rgb}{0.93,0.91,0.67}
\definecolor{palegreen}{rgb}{0.60,0.98,0.60}
\definecolor{paleturquoise}{rgb}{0.69,0.93,0.93}
\definecolor{paleviolet}{rgb}{0.86,0.44,0.58}
\definecolor{papayawhip}{rgb}{1.00,0.94,0.84}
\definecolor{peachpuff}{rgb}{1.00,0.85,0.73}
\definecolor{peru}{rgb}{0.80,0.52,0.25}
\definecolor{pink1}{rgb}{1.00,0.71,0.77}
\definecolor{pink2}{rgb}{0.93,0.66,0.72}
\definecolor{pink3}{rgb}{0.80,0.57,0.62}
\definecolor{pink4}{rgb}{0.55,0.39,0.42}
\definecolor{pink}{rgb}{1.00,0.75,0.80}
\definecolor{plum1}{rgb}{1.00,0.73,1.00}
\definecolor{plum2}{rgb}{0.93,0.68,0.93}
\definecolor{plum3}{rgb}{0.80,0.59,0.80}
\definecolor{plum4}{rgb}{0.55,0.40,0.55}
\definecolor{plum}{rgb}{0.87,0.63,0.87}
\definecolor{powderblue}{rgb}{0.69,0.88,0.90}
\definecolor{purple1}{rgb}{0.61,0.19,1.00}
\definecolor{purple2}{rgb}{0.57,0.17,0.93}
\definecolor{purple3}{rgb}{0.49,0.15,0.80}
\definecolor{purple4}{rgb}{0.33,0.10,0.55}
\definecolor{purple}{rgb}{0.63,0.13,0.94}
\definecolor{red1}{rgb}{1.00,0.00,0.00}
\definecolor{red2}{rgb}{0.93,0.00,0.00}
\definecolor{red3}{rgb}{0.80,0.00,0.00}
\definecolor{red4}{rgb}{0.55,0.00,0.00}
\definecolor{red}{rgb}{1.00,0.00,0.00}
\definecolor{rosybrown}{rgb}{0.74,0.56,0.56}
\definecolor{royalblue}{rgb}{0.25,0.41,0.88}
\definecolor{saddlebrown}{rgb}{0.55,0.27,0.07}
\definecolor{salmon1}{rgb}{1.00,0.55,0.41}
\definecolor{salmon2}{rgb}{0.93,0.51,0.38}
\definecolor{salmon3}{rgb}{0.80,0.44,0.33}
\definecolor{salmon4}{rgb}{0.55,0.30,0.22}
\definecolor{salmon}{rgb}{0.98,0.50,0.45}
\definecolor{sandybrown}{rgb}{0.96,0.64,0.38}
\definecolor{seagreen}{rgb}{0.18,0.55,0.34}
\definecolor{seashell1}{rgb}{1.00,0.96,0.93}
\definecolor{seashell2}{rgb}{0.93,0.90,0.87}
\definecolor{seashell3}{rgb}{0.80,0.77,0.75}
\definecolor{seashell4}{rgb}{0.55,0.53,0.51}
\definecolor{seashell}{rgb}{1.00,0.96,0.93}
\definecolor{sienna1}{rgb}{1.00,0.51,0.28}
\definecolor{sienna2}{rgb}{0.93,0.47,0.26}
\definecolor{sienna3}{rgb}{0.80,0.41,0.22}
\definecolor{sienna4}{rgb}{0.55,0.28,0.15}
\definecolor{sienna}{rgb}{0.63,0.32,0.18}
\definecolor{skyblue}{rgb}{0.53,0.81,0.92}
\definecolor{slateblue}{rgb}{0.42,0.35,0.80}
\definecolor{slategray}{rgb}{0.44,0.50,0.56}
\definecolor{slategrey}{rgb}{0.44,0.50,0.56}
\definecolor{snow1}{rgb}{1.00,0.98,0.98}
\definecolor{snow2}{rgb}{0.93,0.91,0.91}
\definecolor{snow3}{rgb}{0.80,0.79,0.79}
\definecolor{snow4}{rgb}{0.55,0.54,0.54}
\definecolor{snow}{rgb}{1.00,0.98,0.98}
\definecolor{springgreen}{rgb}{0.00,1.00,0.50}
\definecolor{steelblue}{rgb}{0.27,0.51,0.71}
\definecolor{tan1}{rgb}{1.00,0.65,0.31}
\definecolor{tan2}{rgb}{0.93,0.60,0.29}
\definecolor{tan3}{rgb}{0.80,0.52,0.25}
\definecolor{tan4}{rgb}{0.55,0.35,0.17}
\definecolor{tan}{rgb}{0.82,0.71,0.55}
\definecolor{thistle1}{rgb}{1.00,0.88,1.00}
\definecolor{thistle2}{rgb}{0.93,0.82,0.93}
\definecolor{thistle3}{rgb}{0.80,0.71,0.80}
\definecolor{thistle4}{rgb}{0.55,0.48,0.55}
\definecolor{thistle}{rgb}{0.85,0.75,0.85}
\definecolor{tomato1}{rgb}{1.00,0.39,0.28}
\definecolor{tomato2}{rgb}{0.93,0.36,0.26}
\definecolor{tomato3}{rgb}{0.80,0.31,0.22}
\definecolor{tomato4}{rgb}{0.55,0.21,0.15}
\definecolor{tomato}{rgb}{1.00,0.39,0.28}
\definecolor{turquoise1}{rgb}{0.00,0.96,1.00}
\definecolor{turquoise2}{rgb}{0.00,0.90,0.93}
\definecolor{turquoise3}{rgb}{0.00,0.77,0.80}
\definecolor{turquoise4}{rgb}{0.00,0.53,0.55}
\definecolor{turquoise}{rgb}{0.25,0.88,0.82}
\definecolor{violetred}{rgb}{0.82,0.13,0.56}
\definecolor{violet}{rgb}{0.93,0.51,0.93}
\definecolor{wheat1}{rgb}{1.00,0.91,0.73}
\definecolor{wheat2}{rgb}{0.93,0.85,0.68}
\definecolor{wheat3}{rgb}{0.80,0.73,0.59}
\definecolor{wheat4}{rgb}{0.55,0.49,0.40}
\definecolor{wheat}{rgb}{0.96,0.87,0.70}
\definecolor{whitesmoke}{rgb}{0.96,0.96,0.96}
\definecolor{white}{rgb}{1.00,1.00,1.00}
\definecolor{yellow1}{rgb}{1.00,1.00,0.00}
\definecolor{yellow2}{rgb}{0.93,0.93,0.00}
\definecolor{yellow3}{rgb}{0.80,0.80,0.00}
\definecolor{yellow4}{rgb}{0.55,0.55,0.00}
\definecolor{yellowgreen}{rgb}{0.60,0.80,0.20}
\definecolor{yellow}{rgb}{1.00,1.00,0.00}